\RequirePackage{booktabs} 
\documentclass[pdflatex,sn-mathphys-ay]{sn-jnl}

\usepackage{subfiles}
\usepackage{graphicx}%
\usepackage{multirow}%
\usepackage{mathtools}
\usepackage{amsmath,amssymb,amsfonts}%
\usepackage{amsthm}%
\usepackage{newtxmath}
\usepackage{mathrsfs}%
\usepackage[title]{appendix}%
\usepackage{xcolor}%
\usepackage{textcomp}%
\usepackage{manyfoot}%
\usepackage{booktabs}%
\usepackage[linesnumbered,ruled,vlined]{algorithm2e}
\usepackage{algpseudocode}%
\usepackage{nicematrix}
\usepackage{listings}%
\usepackage{caption}
\usepackage{subcaption}%
\usepackage{float}

\usepackage{enumitem}

\usepackage{placeins} 

\usepackage{tikz}
\tikzstyle{block}=[draw opacity=0.7,line width=1.4cm]
\usetikzlibrary{positioning, arrows, shapes, fit, calc}

\let\oldnl\nl
\newcommand{\nonl}{\renewcommand{\nl}{\let\nl\oldnl}} 

\usepackage{comment}

\theoremstyle{thmstyleone}%
\newtheorem{theorem}{Theorem}
\newtheorem{proposition}[theorem]{Proposition}%

\AfterEndEnvironment{manprop}{\noindent\ignorespaces}

\newtheorem{mylemma}{Lemma}

\AfterEndEnvironment{manlemma}{\noindent\ignorespaces}

\AfterEndEnvironment{mancoro}{\noindent\ignorespaces}

\newenvironment{mypf}[2]{\proof[\textbf{Proof of {#1} {#2}.}]}{\endproof}
\AfterEndEnvironment{mypf}{\noindent\ignorespaces}

\newtheorem{mydef}{Definition}

\AfterEndEnvironment{mandef}{\noindent\ignorespaces}

\newtheorem{myexample}{Example}
\newenvironment{manexample}[1]
    {\renewcommand\themyexample{#1}\myexample}{\endmyexample}
\AfterEndEnvironment{manexample}{\noindent\ignorespaces}

\newtheorem{myremark}{Remark}
\newenvironment{manremark}[1]
    {\renewcommand\themyremark{#1}\myremark}{\endmyremark}
\AfterEndEnvironment{manremark}{\noindent\ignorespaces}

\theoremstyle{thmstyletwo}%
\newtheorem{example}{Example}%
\newtheorem{remark}{Remark}%

\theoremstyle{thmstylethree}%
\newtheorem{definition}{Definition}%

\algnewcommand\Input{\State \textbf{Input:\ }}
\algnewcommand\Output{\State \textbf{Output:\ }}

\DeclareMathOperator{\normal}{N}
\DeclareMathOperator{\tstu}{t}
\DeclareMathOperator{\Bin}{Bin}
\DeclareMathOperator{\Pois}{Poisson}
\DeclareMathOperator{\Betatext}{Beta}
\DeclareMathOperator{\Unif}{Uniform}

\DeclareMathOperator{\Ga}{Gamma}
\DeclareMathOperator{\MultN}{Multinomial}
\DeclareMathOperator{\pool}{pool}
\DeclareMathOperator{\meld}{meld}
\DeclareMathOperator{\logit}{logit}
\DeclareMathOperator{\surv}{sur}
\DeclareMathOperator{\imm}{imm}
\DeclareMathOperator{\for}{for}
\DeclareMathOperator{\andd}{and}
\DeclareMathOperator{\with}{with}
\DeclareMathOperator{\where}{where}

\DeclareMathOperator{\thh}{th}

\newcommand{\bbInd}{\varmathbb{1}}
\newcommand{\bsM}{\boldsymbol{M}}
\newcommand{\bsQ}{\boldsymbol{Q}}
\newcommand{\bsw}{\boldsymbol{w}}
\newcommand{\bsx}{\boldsymbol{x}}

\newcommand{\bsY}{\boldsymbol{Y}}
\newcommand{\calC}{\mathcal{C}}

\newcommand{\bsalpha}{\boldsymbol{\alpha}}
\newcommand{\bsphi}{\boldsymbol{\phi}}
\newcommand{\bspsi}{\boldsymbol{\psi}}
\newcommand{\bstheta}{\boldsymbol{\theta}}

\newcommand{\lr}{\left(}
\newcommand{\rr}{\right)}
\newcommand{\ls}{\left[}
\newcommand{\rs}{\right]}
\newcommand{\lb}{\left\lbrace}
\newcommand{\rb}{\right\rbrace}

\newcommand{\smcsq}{\text{SMC}^2}
\newcommand{\tth}{\text{th}}
\newcommand{\smc}{\text{\tiny SMC}}

\definecolor{safegreen}{HTML}{388E3C}

\begin{document}

\title[Article Title]{Chained Markov melding using divide and conquer sequential Monte Carlo}


\author*{\fnm{Yixuan} \sur{Liu}}

\author{\fnm{Robert J. B.} \sur{Goudie}}


\affil{\orgdiv{MRC Biostatistics Unit}, \orgname{University of Cambridge},
\orgaddress{%
\city{Cambridge},
\country{UK}}}




\abstract{Specifying a full Bayesian model that integrates multiple data sources can be challenging. One natural approach is to specify each individual model separately and join them afterwards. This is the approach adopted in Markov melding. However, when adjacent submodels share common quantities, as in \textit{chained Markov melding}, posterior inference can be challenging for existing MCMC-based approaches. In this paper, we propose a new multi-stage sampler for chained Markov models involving an arbitrary number of submodels. The proposed sampler adopts a divide-and-conquer sequential Monte Carlo approach for the tree-structured model that fits naturally with the structure of chained Markov melding. The resulting multi-stage sampler provides a flexible alternative for sampling from complex joint models, as its separate sampling scheme for different submodels avoids the need for directly sampling from the full model. We demonstrate applications of the sampler through two examples. The first is a toy example involving 11 submodels of various types. The second example considers an ecologically integrated population model that combines multiple datasets to estimate immigration and reproduction rates.}




\maketitle

\section{Introduction}\label{se1}

The incorporation of evidence from multiple data sources, which are often different in size and complexity, is an important challenge in many areas of statistical analysis. 
Bayesian approaches are an intuitive approach to this problem because they naturally accommodate the synthesis of information through the posterior distribution, ensuring that uncertainties inherent in each data source are coherently propagated to the final inference.
Numerous studies have adopted such an approach, contributing either methodological developments or approaches to specific real-world problems. For example, \citet{ades2003}, \citet{ades2006}, \citet{lunn2013} and \citet{presanis2014} all applied hierarchical models to statistically synthesise multiple data sources, and in biostatistics there is a growing interest in using Bayesian methods to join longitudinal and survival models \citep{ibrahim2001, guo2004, lawrence-gould2015, mauff2020, chen2025}. Real-world challenges addressed involving multiple data sources include: combining survey and population data to increase the precision of estimation of Hispanic fertility rates in the US \citep{rendall2009}; using a Bayesian joint model to improve the estimation of survival times of HIV/AIDS patients in Brazil \citep{martins2016}; and developing a hierarchical model to integrate testing data and self-reported questionnaire data to estimate positivity for severe acute respiratory syndrome coronavirus 2 \citep{donnat2020}. Futhermore, in ecology, \textit{integrated population models} (IPMs) are an analytical framework that enables the joint analysis of demographic and survey data to estimate \textit{population dynamics}, such as immigration and reproduction rates, and Bayesian approaches to IPMs have become popular \citep{king2008, abadi2010a, abadi2010b, rhodes2011, woodworth2017, finke2019}. 

These approaches to fully integrating evidence sources through a joint Bayesian model are often computationally demanding. Consequently, researchers frequently resort to ad-hoc approximations. A pervasive example of such an approach is `plug-in' estimation with a two-stage design, in which in stage one one model is used to estimate the posterior for a parameter \(\phi\) that appears in both models. This is then summarised by a point estimate \(\hat{\phi}\), such as the maximum a posteriori estimate \citep{alvares2023, alvares2025}, the maximum likelihood estimate \citep{murawska2012}, or simply the mean or median \citep{leiva-yamaguchi2021}. This fixed value is then plugged into the second model in stage two, treating \(\hat{\phi}\) as a known constant. While computationally convenient, this stepwise approach will underestimate uncertainty, yielding artificially narrow credible intervals that fail to reflect the true uncertainty of the system.

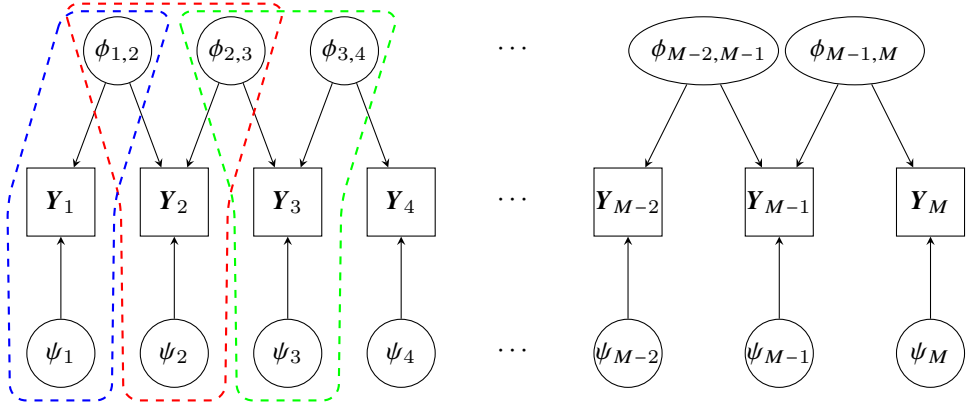
\begin{figure}
    \centering
    \usetikzlibrary{calc}
    \begin{tikzpicture}
        \node[draw=black, minimum width=.9cm, minimum height=.9cm] (y1) at (0,0) {$\bsY_{1}$};
        \node[ellipse, draw=black, minimum width=.9cm, minimum height=.9cm] (psi1) at (0,-2) {$\psi_{1}$};
        \node[draw=black, minimum width = .9cm, minimum height=.9cm] (y2) at (1.5,0) {$\bsY_{2}$};
        \node[ellipse, draw=black, minimum width=.9cm, minimum height=.9cm] (psi2) at (1.5,-2) {$\psi_{2}$};
        \node[draw=black, minimum width=.9cm, minimum height=.9cm] (y3) at (3,0) {$\bsY_{3}$};
        \node[ellipse, draw=black, minimum width=.9cm, minimum height=.9cm] (psi3) at (3,-2) {$\psi_{3}$};
        \node[draw=black, minimum width=.9cm, minimum height=.9cm] (y4) at (4.5, 0) {$\bsY_{4}$};
        \node[ellipse, draw=black, minimum width=.9cm, minimum height=.9cm] (psi4) at (4.5, -2) {$\psi_{4}$};
        \node[] at (6,0) {$\cdots$};
        \node[] at (6,-2) {$\cdots$};
        \node[draw=black, minimum size = .9cm, inner sep=0pt] (ym2) at (7.5,0) {$\bsY_{M-2}$};
        \node[circle, draw=black, minimum size=.9cm, inner sep=0pt] (psim2) at (7.5,-2) {$\psi_{M-2}$};
        \node[draw=black, minimum size = .9cm, inner sep=0pt] (ym1) at (9.5,0) {$\bsY_{M-1}$};
        \node[circle, draw=black, minimum size=.9cm, inner sep=0pt] (psim1) at (9.5,-2) {$\psi_{M-1}$};
        \node[draw=black, minimum width = .9cm, minimum height=.9cm, inner sep=0pt] (ym) at (11.5,0) {$\bsY_{M}$};
        \node[ellipse, draw=black, minimum size=.9cm, inner sep=0pt] (psim) at (11.5,-2) {$\psi_{M}$};

        \node[ellipse, draw=black, minimum width=.9cm, minimum height=.9cm, inner sep=0pt] (phi12) at (.75, 2) {$\phi_{1,2}$};
        \node[ellipse, draw=black, minimum width=.9cm, minimum height=.9cm, inner sep=0pt] (phi23) at (2.25, 2) {$\phi_{2,3}$};
        \node[ellipse, draw=black, minimum width=.9cm, minimum height=.9cm, inner sep=0pt] (phi34) at (3.75, 2) {$\phi_{3,4}$};
        \node[] at (6,2) {$\cdots$};
        \node[ellipse, draw=black, text width=1.4cm, align=center, minimum height=.9cm, inner sep=0pt] (phim21) at (8.5, 2) {$\phi_{M-2,M-1}$};
        \node[ellipse, draw=black, text width=1.3cm, align=center, minimum height=.9cm, inner sep=0pt] (phim1) at (10.5, 2) {$\phi_{M-1,M}$};

        \draw[-stealth] (psi1) -- (y1);
        \draw[-stealth] (psi2) -- (y2);
        \draw[-stealth] (psi3) -- (y3);
        \draw[-stealth] (psi4) -- (y4);
        \draw[-stealth] (psim2) -- (ym2);
        \draw[-stealth] (psim1) -- (ym1);
        \draw[-stealth] (psim) -- (ym);

        \draw[-stealth] (phi12) -- (y1);
        \draw[-stealth] (phi12) -- (y2);
        \draw[-stealth] (phi23) -- (y2);
        \draw[-stealth] (phi23) -- (y3);
        \draw[-stealth] (phi34) -- (y3);
        \draw[-stealth] (phi34) -- (y4);
        \draw[-stealth] (phim21) -- (ym2);
        \draw[-stealth] (phim21) -- (ym1);
        \draw[-stealth] (phim1) -- (ym1);
        \draw[-stealth] (phim1) -- (ym);

        \draw[red, dashed, thick, rounded corners=6pt]
          ($ (phi12.north west) + (-0.40, 0.3) $) --
          ($ (y2.west) + (-0.25, 0) $) --
          ($ (psi2.south west) + (-0.35,-0.30) $) --
          ($ (psi2.south east) + ( 0.35,-0.30) $) --
          ($ (y2.east) + (0.25, 0) $) -- 
          ($ (phi23.north east) + (0.4, 0.3) $) --
          cycle;

        \draw[blue, dashed, thick, rounded corners=6pt]
          ($ (phi12.north west) + (-0.4, 0.2) $) -- 
          ($ (y1.west) + (-0.25, 0) $) --
          ($ (psi1.south west) + (-0.35,-0.3) $) -- 
          ($ (psi1.south east) + (0.35,-0.3) $) -- 
          ($ (y1.east) + (0.25,0) $) --
          ($ (phi12.north east) + (0.4,0.2) $) -- cycle;

        \draw[green, dashed, thick, rounded corners=6pt]
          ($ (phi23.north west) + (-0.4, 0.2) $) -- 
          ($ (y3.west) + (-0.25, 0) $) --
          ($ (psi3.south west) + (-0.35,-0.3) $) -- 
          ($ (psi3.south east) + (0.35,-0.3) $) -- 
          ($ (y3.east) + (0.25,0) $) --
          ($ (phi34.north east) + (0.4,0.2) $) -- cycle;

    \end{tikzpicture}
    \caption{DAG for a chain model involving $M$ submodels. Squares denote observable random variables, and circles denote parameters. Arrows indicate dependencies between nodes in the model. The other DAGs in this article use the same representation. The blue, red and green dashed shapes represent Submodel 1, Submodel 2 and Submodel 3, respectively.}
    \label{Fig_chain_dag}
\end{figure}

Fully Bayesian inference addresses this deficiency, but both specifying and fitting a suitable fully Bayesian joint model is challenging.
An appealing and convenient approach to addressing this challenge is to specify submodels for each of the sources and combine them thereafter.
However, mathematically it is not immediately clear how to integrate two submodels that share a parameter but were defined independently. \cite{goudie2019} proposed \textit{Markov melding} as a generic solution to this problem. Markov melding integrates the submodels by defining a joint prior on the common quantity \(\phi\), with the (potentially different) priors on the common quantity combined by `pooling'. This approach has been extended by \cite{manderson2023} to settings in which a `chain' of models is combined. The \textit{directed acyclic graph} (DAG) of a general chain structure model is presented in Figure \ref{Fig_chain_dag}. In brief, any two adjacent submodels in a chain share a common parameter. For example, in Figure \ref{Fig_chain_dag}, submodels 1 (the blue dashed shape) and 2 (the red dashed shape) share $\phi_{1,2}$, and submodels 2 and 3 (the green dashed shape) share $\phi_{2,3}$. More details about this model will be provided in Section \ref{se2.1}. Markov melding has been adopted in several substantive application areas, including integrated population models in ecology \citep{vaneeMeldingWildlifeSurveys2023, vaneeMeldedIntegratedPopulation2025} and data integration for modelling infectious diseases \citep{nicholsonInteroperabilityStatisticalModels2022, birrell2025}.

A challenge with this approach is that inference for the joint Markov melding model may be computationally difficult.
The ideal computational approach for the Markov melding model would reflect the modular nature of the original submodels, and allow a sequential inference approach in which evidence from each submodel is gradually incorporated. 
\cite{goudie2019} proposed a `multi-stage' Markov chain Monte Carlo (MCMC) algorithm that enables this workflow, and this algorithm was refined by \cite{manderson2022} when the (often implicit) priors on the common quantity \(\phi\) are difficult to evaluate. For the chain Markov melding case, \cite{manderson2023} proposed extensions to handle the chain structure of the graph. However, these algorithms remain fragile and, as noted by \cite{manderson2023}, the current methods do not scale well to longer chains involving more models.

In this paper, we propose replacing the two-stage MCMC algorithm of \citet{manderson2023} with \textit{Divide-and-conquer sequential Monte Carlo} (D\&C-SMC) \citep{lindsten2017, kuntz2024}, a class of sequential Monte Carlo (SMC) algorithms that adopt a divide-and-conquer approach and which extends the applicability of SMC from sequential models (such as hidden Markov models) to tree-structured sequential models. This approach is well-suited to sampling from hierarchical or graphical models where sub-posteriors can be estimated in parallel and then systematically merged. In empirical studies, D\&C-SMC has been shown to outperform traditional approaches, such as MCMC, providing more accurate posterior expectation and normalising constant approximations. Furthermore, the method enables components to be sampled in parallel, facilitating a reduction in computational time. 
We show that the chain Markov melding model can be converted into a tree-structured model. 
We illustrate our approach on a simulation study involving 11 submodels of various types and a case study on an ecologically IPM.

The remainder of the paper is organised as follows. Section \ref{se2} reviews the chained Markov melding and D\&C-SMC methods. In Section \ref{se3}, we introduce the proposed sampler, which combines these two methods for chained Markov models. In particular, Section \ref{se3.1} considers the case of three submodels, and Section \ref{se3.2} is used to describe the extension of the sampler to the case of an arbitrary number of submodels. Section \ref{se3.3} discusses an extension of the sampler incorporating the $\smcsq$ algorithm. The simulation study is presented in Section \ref{se4}, and the integrated population model (IPM) example is described in Section \ref{se5}.

\section{Background}\label{se2}

In this section, we first review the existing Markov melding methodology for joining submodels into a single joint Bayesian model.
We then review D\&C-SMC, which we will adopt in Section~\ref{se3} for improving inference of Markov melding models.
Throughout the paper, we let $p$ be either a probability mass function for discrete random variables or a probability density function for continuous random variables.

\subsection{Markov melding}\label{se2.1}

\subsubsection{Common parameter Markov melding}

Markov melding \citep{goudie2019} was originally proposed for the scenario of combining submodels that all share a single common quantity $\phi$.
Suppose we have $M$ submodels, indexed by $m = 1, \dots, M$, each of which involve the common quantity $\phi$, submodel-specific quantities $\psi_m$, and submodel-specific observable quantities $\bsY_{m}$.
Each submodel has joint density $p_{m}(\phi, \psi_{m}, \bsY_{m})$ for $m=1,...,M$.
Markov melding provides a method for combining these $M$ separate submodels into a single joint model $p_{\meld}(\phi,\psi_{1},...,\psi_{M},\bsY_{1},...,\bsY_{M})$.

Combining submodels requires particular care when the priors for $\phi$ in the $M$ submodels are not all identical.
Suppose the marginal prior for $\phi$ is $p_{m}(\phi)$ under submodel $m = 1, \dots, M$.
Markov melding pools these $M$ marginal priors $p_1(\phi), \dots, p_M(\phi)$ using a pooling function $g$, which combines the information of all marginal priors to form a single pooled prior
\begin{equation}\label{eq:sigle-pooling}
p_{\pool}(\phi):=g(p_{1}(\phi),...,p_{M}(\phi)).
\end{equation}
Using this pooled prior, the joint model under Markov melding is then \begin{align}\label{eq:single-meld-joint}
    p_{\meld}(\phi,\psi_{1},...,\psi_{M},\bsY_{1},...,\bsY_{M})=p_{\pool}(\phi)\prod_{m=1}^{M}\frac{p_{m}(\phi,\psi_{m},\bsY_{m})}{p_{m}(\phi)}.
\end{align} \\

\subsubsection{Chained Markov melding}
\citet{manderson2023} generalised Markov melding to the situation in which the submodels do not have a single common quantity $\phi$, but instead are linked in a chain structure.
For example, with $M = 3$, suppose that Submodels 1 and 2 share a common quantity $\phi_{1, 2}$ and Submodels 2 and 3 share a different common quantity $\phi_{2, 3}$.
In general, as illustrated in Figure~\ref{Fig_chain_dag}, let $\phi_{m-1,m}$ be the common parameter of $(m-1)^{\tth}$ and $m^{\tth}$ submodels, and let $\psi_{m}$ be submodel-specific parameter that belongs only to the $m^{\text{th}}$ submodel.  
Let $\bstheta_{m}=(\phi_{m-1,m},\phi_{m,m+1},\psi_{m})$ be the corresponding parameter space for the $m^{\text{th}}$ submodel, with $\bstheta_{1}=(\phi_{1,2},\psi_{1})$ for the $1^{\text{st}}$ submodel and $\theta_{M}=(\phi_{M-1,M},\psi_{M})$ for the $M^{\text{th}}$ submodel.
Denote the probability function of $(\bstheta_{m}, \bsY_{m})$ for the $m^{\thh}$ submodel by $p_{m}(\bstheta_{m}, \bsY_{m})$. For convenience, denote the collections of all $\phi_{m}$, $\psi_{m}$ and $\bsY_{m}$ by $\bsphi=(\phi_{1,2},...,\phi_{M-1,M})$, $\bspsi=(\psi_{1},...,\psi_{M})$ and $\bsY=(\bsY_{1},...,\bsY_{M})$ respectively. 

Generalising Markov melding for chains requires a joint pooled prior for $\bsphi$. \citet{manderson2023} extended the idea of pooled priors to this setting as \begin{align}\label{eq:chained-pooling}
    p_{\pool}(\bsphi)=g(p_{1}(\phi_{1,2}),p_{2}(\phi_{1,2},\,\phi_{2,3}),\, ...,\, p_{M}(\phi_{M-1,M})).
\end{align} Note that \eqref{eq:chained-pooling} is a generic function that does not require $\phi_{m}\in\bsphi, i=1,...,M$ to be independent. Using this pooled prior, the joint Markov melding model for the chain of submodels is \begin{align}\label{eq:chained-meld-joint}
    p_{\meld}(\bsphi,\bspsi,\bsY)&=p_{\pool}(\bsphi)\frac{p_{1}(\phi_{1,2},\,\psi_{1},\,\bsY_{1})}{p_{1}(\phi_{1,2})}\frac{p_{M}(\phi_{M-1,M},\,\psi_{M},\,\bsY_{M})}{p_{M}(\phi_{M-1,M})}\nonumber\\
    &\qquad\qquad\times\prod_{m=2}^{M-1}\lr\frac{p_{m}(\phi_{m-1,m},\,\phi_{m,m+1},\,\psi_{m},\,\bsY_{m})}{p_{m}(\phi_{m-1,m},\,\phi_{m,m+1})}\rr.
\end{align}
We focus on this general case, since 
\eqref{eq:single-meld-joint} is clearly a special case.

\subsubsection{Pooling priors}
Several pooling priors have been proposed \citep{goudie2019, manderson2023}, but in this manuscript, we focus on the \textit{logarithmic pooling}. The basic logarithmically pooled prior \citep{goudie2019}, corresponding to \eqref{eq:sigle-pooling}, is defined as \begin{align}\label{eq:basic-log-pooling}
    p_{\pool,\log}(\phi)=\frac{1}{K_{\log}(\lambda)}\prod_{m=1}^{M}p_{m}(\phi)^{\lambda_{m}},\quad K_{\log}(\lambda)=\int\prod_{m=1}^{M}p_{m}(\phi)^{\lambda_{m}}d\phi,
\end{align}  
where $\lambda=(\lambda_{1},...,\lambda_{M})$ are nonnegative weights \citep{ohagan2006, genest1986}. \cite{manderson2023} extended this pooled prior for the chain model \begin{align}\label{eq:chained-log-pooling}
    &p_{\pool,\log}(\bsphi)=\frac{1}{K_{\log}(\lambda)}p_{1}(\phi_{1,2})^{\lambda_{1}}\prod_{m=2}^{M-1}\lr p_{m}(\phi_{m-1,m},\phi_{m,m+1})^{\lambda_{m}}\rr p_{M}(\phi_{M-1,M})^{\lambda_{M}},
\end{align} where $K_{\log}(\lambda)$ is the
normalising constant and $\lambda=(\lambda_{1},...,\lambda_{M})$ are nonnegative weights with $\sum_{m=1}^{M}\lambda_{m}\geq1$. A special case of the log pooling is \textit{product-of-expert (PoE) pooling} \citep{hinton2002} where $\lambda_{1}=\cdots=\lambda_{M}$.

\subsubsection{Computation}\label{se2.1.4}
To draw samples from the posterior distribution of \eqref{eq:chained-meld-joint} of the combined model for $M=3$, \citet{manderson2023} proposed a two-stage \textit{parallel sampler}, which can be regarded as a chain-model-based generalisation of the multi-stage samplers in \citet{lunn2013}, \citet{goudie2019} and \citet{hooten2021}%
.
To facilitate the staged computation, we decompose the pooled prior into \begin{align}\label{eq:pooled-prior-decomp-M3}
    p_{\pool}(\phi_{1,2},\phi_{2,3})=p_{\pool,1}(\phi_{1,2})p_{\pool,2}(\phi_{1,2},\phi_{2,3})p_{\pool,3}(\phi_{2,3}).
\end{align} This assumption can always be trivially satisfied by specifying improper and/or flat distributions for $p_{\pool,1}(\phi_{1,2})$ and $p_{\pool,2}(\phi_{2,3})$. However, it may preferable to retain the original priors in each submodel by choosing $p_{\pool,1}(\phi_{1,2})=p_{1}(\phi_{1,2})$ and $p_{\pool,2}(\phi_{2,3})=p_{2}(\phi_{2,3})$, and then making appropriate adjustments to $p_{\pool,2}(\phi_{1,2},\phi_{2,3})$.

\begin{figure}
    \centering
    \begin{tikzpicture}
        \node[dashed, draw=red, minimum width=7cm, minimum height=1.5cm, inner sep=0pt] (p2) at (0,0) {$p_{\meld,2}(\textcolor{red}{\phi_{1,2}},\textcolor{red}{\phi_{2,3}},\psi_{2}\mid\bsY_{2})$};
        \node[red] () at (0,.6) {$s_{2}$};

        \node[dashed, draw=blue, minimum width=4cm, minimum height=1cm, inner sep=0pt] (p1) at (-4,2.5) {$p_{\meld,1}(\textcolor{blue}{\phi_{1,2}},\psi_{1}\mid\bsY_{1})$};
        \node[blue] () at (-2.8,2.15) {$s_{1}$};
        \node[dashed, draw=blue, minimum width=4cm, minimum height=1cm, inner sep=0pt] (p3) at (4,2.5) {$p_{\meld,3}(\textcolor{blue}{\phi_{2,3}},\psi_{2}\mid\bsY_{3})$};
        \node[blue] () at (2.8,2.15) {$s_{1}$};

        \draw[-stealth] (-3,2) -- (-2.2,.7);
        \draw[-stealth] (3,2) -- (2.2,.7);
    \end{tikzpicture}
\caption{Diagram illustrating the sampling process of the two-stage parallel sampler. 
The computational stage is denoted by $s_{t}$, with $t=1,2$ and the densities shown reflect the additional terms considered at that stage.
The arrows represent the merging direction. The common parameters are highlighted.}
    \label{Fig_two_stage_orig}
\end{figure}
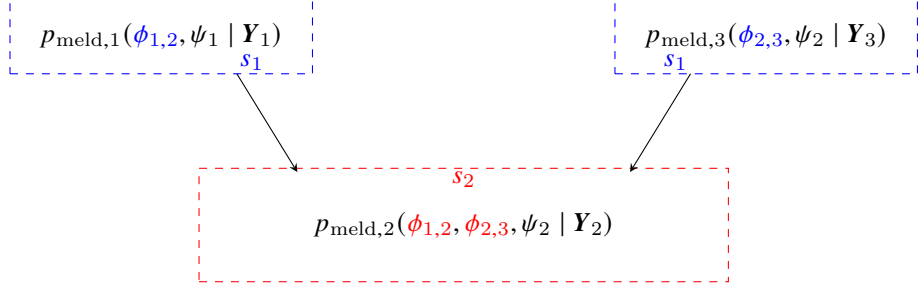

Figure \ref{Fig_two_stage_orig} demonstrates how the sampler works. 
In stage one ($s_{1}$), the sampler draws samples of $\phi_{1,2}$ and $\phi_{2,3}$ by targeting the joint subposterior \begin{align}\label{eq:meld-subposterior-1-3}
    &p_{\meld,1,3}(\phi_{1,2},\phi_{2,3},\psi_{1},\psi_{2}|\bsY_{1},\bsY_{2})\nonumber\\
    &\qquad\qquad\propto p_{\pool,1}(\phi_{1,2})\frac{p_{1}(\phi_{1,2},\psi_{1},\bsY_{1})}{p_{1}(\phi_{1,2})}p_{\pool,3}(\phi_{2,3})\frac{p_{3}(\phi_{2,3},\psi_{3},\bsY_{3})}{p_{3}(\phi_{2,3})}.
\end{align} The above equation indicates that the sampling processes for the two submodels are independent, so the subposteriors can be separately targeted as  \begin{align}\label{eq:meld-posterior-sub1}
    p_{\meld,1}(\phi_{1,2},\psi_{1}|\bsY_{1})\propto p_{\pool,1}(\phi_{1,2})\frac{p_{1}(\phi_{1,2},\psi_{1},\bsY_{1})}{p_{1}(\phi_{1,2})}
\end{align} for Submodel 1 and \begin{align}\label{eq:meld-posterior-sub3}
    p_{\meld,3}(\phi_{2,3},\psi_{3}|\bsY_{3})\propto p_{\pool,3}(\phi_{2,3})\frac{p_{3}(\phi_{2,3},\psi_{3},\bsY_{3})}{p_{3}(\phi_{2,3})}
\end{align} for Submodel 3, using standard MCMC approaches. Those samples are then used as a collection of proposals for sampling the full melded posterior (which also includes Submodel 2) within a Metropolis-within-Gibbs sampler \citep{tierney1994} in stage two ($s_{2}$). The corresponding proposal distributions are \begin{align*}
    &(\phi_{1,2}^{*},\psi_{1}^{*})\mid (\phi_{2,3},\psi_{2},\psi_{3})\sim p_{\meld,1}(\phi_{1,2}^{*},\psi_{1}^{*}|\bsY_{1})\\
    &(\phi_{2,3}^{*},\psi_{3}^{*}) \mid (\phi_{2,3},\psi_{1},\psi_{2})\sim p_{\meld,3}(\phi_{2,3}^{*},\psi_{3}^{*}|\bsY_{3}).
\end{align*} At each iteration of the sampler, a new proposal for $\phi_{1,2}$ is \textit{uniformly} drawn from the collection; and correspondingly for $\phi_{2,3}$. These proposal distributions are used because the corresponding acceptance probabilities do not depend on the likelihoods in submodels 1 and 3, meaning the approach is `modular' in the sense that the stage 2 sampler requires no information about the stage 1 submodels. Meanwhile, $\psi_{2}$ is also drawn via a generic proposal $q(\psi_{2}^{*}|\psi_{2})$.

\subsection{Divide-and-conquer with SMC}\label{se2.2}
While the two-stage parallel approach, introduced in Section \ref{se2.1}, works well in some settings with $M=3$, as demonstrated by \citet{manderson2023}, it is not robust for the general chain model, in which the number of submodels $M$ could be arbitrarily large. 
To address this challenge, we consider the D\&C-SMC approach proposed by \citet{lindsten2017} that combines the divide-and-conquer technique with SMC.
D\&C-SMC splits the variables of multivariate distributions into disjoint sets that are associated with the node in a tree, with each node associated with an individual \textit{auxiliary target distribution}. Instead of directly sampling from the joint posterior distribution of all parameters of interest, the approach samples from several auxiliary distributions and then merges the results of each set to approximate the distribution of interest using importance sampling. Sampling from the auxiliary distributions can often be implemented in parallel, reducing computation time. 

D\&C-SMC considers the class of  \textit{tree-structured models}, so we first introduce the definition of this class.
Suppose the \textit{target distribution} is $\pi(\bsx)=\gamma(\bsx)/Z$, $\bsx\in X$, with unnormalised density $\gamma(\bsx)$ and normalising constant $Z=\int\gamma(\bsx)d\bsx$.
The \textit{state space} $X$ can be discrete, continuous, or mixed, and we assume the unnormalised density $\gamma(\bsx)$ can be evaluated point-wise.
Let $T = \{1, \dots, n\}$ index the nodes of the tree structure, with $\calC(t)\subset T$ denoting the collection of \textit{child nodes} of node $t \in T$. Conversely, node $t$ is called the \textit{parent node} of its child nodes $c\in\calC(t)$.
Let $r\in T$ denote the \textit{root} of the tree, which is not a child node of any other nodes in the model, and define a \textit{leaf node} as any node with $\calC(t)=\varnothing$. 
If we associate each node $t \in T$ with a corresponding distribution $\pi_{t}$ defined on $X_{t}\subseteq X$, then the following definition provides the conditions under which a tree decomposition is valid. 
\begin{mydef}\label{def:tree-decomp}
    A collection of $\pi_t$, $t \in T$, is a valid tree decomposition for a target distribution $\pi(\bsx)$, $\bsx \in X$, if \begin{enumerate}[label=(\roman*)]
        \item the state spaces and the distribution on the root node and the target are equal i.e. $X_r = X$ and $\pi_{r}=\pi$;
    
        \item the state space at node $t$ is constructed recursively as \begin{align}\label{eq13}
        X_{t}=\left(\otimes_{c\in\calC(t)}X_{c}\right)\times\tilde{X}_{t},
        \end{align} where $\tilde{X}_{t}$ can be chosen arbitrarily.
\end{enumerate}
\end{mydef}

\begin{figure}
    \centering
    \begin{subfigure}{.45\textwidth}
        \begin{tikzpicture}
        \node[circle, draw=black, minimum size=.8cm, inner sep=0pt] (phi) at (0,0) {$\phi$};
        \node[circle, draw=black, minimum size=.8cm, inner sep=0pt] (phi12) at (-1.5,-1) {$\phi_{1,2}$};
        \node[circle, draw=black, minimum size=.8cm, inner sep=0pt] (phi1) at (-2.5,-2) {$\phi_{1}$};
        \node[circle, draw=black, minimum size=.8cm, inner sep=0pt] (phi2) at (-.5,-2) {$\phi_{2}$};
        \node[circle, draw=black, minimum size=.8cm, inner sep=0pt] (phi3) at (1.5,-2) {$\phi_{3}$};

        \node[draw=black, minimum size=.6cm, inner sep=0pt] (y1) at (-2.5,-3.2) {$\bsY_{1}$};
        \node[draw=black, minimum size=.6cm, inner sep=0pt] (y2) at (-.5,-3.2) {$\bsY_{2}$};
        \node[draw=black, minimum size=.6cm, inner sep=0pt] (y3) at (1.5,-3.2) {$\bsY_{3}$};

        \node[circle, draw=black, minimum size=.6cm, inner sep=0pt] (psi1) at (-2.5,-4.4) {$\psi_{1}$};
        \node[circle, draw=black, minimum size=.6cm, inner sep=0pt] (psi2) at (-.5,-4.4) {$\psi_{2}$};
        \node[circle, draw=black, minimum size=.6cm, inner sep=0pt] (psi3) at (1.5,-4.4) {$\psi_{3}$};

        \draw[-stealth] (phi) -- (phi12);
        \draw[-stealth] (phi) -- (phi3);
        \draw[-stealth] (phi12) -- (phi1);
        \draw[-stealth] (phi12) -- (phi2);
        \draw[-stealth] (phi1) -- (y1);
        \draw[-stealth] (phi2) -- (y2);
        \draw[-stealth] (phi3) -- (y3);
        \draw[-stealth] (psi1) -- (y1);
        \draw[-stealth] (psi2) -- (y2);
        \draw[-stealth] (psi3) -- (y3);

        \draw[red, dashed, thick, rounded corners=6pt] ($(phi1.north west) + (-.25, .3)$) -- 
        ($(psi1.south west) + (-.3, -.3)$) -- 
        ($(psi1.south east) + (.3, -.3)$) --
        ($(phi1.north east) + (.25, .3)$) -- cycle;

        \draw[blue, dashed, thick, rounded corners=6pt] ($(phi2.north west) + (-.25, .3)$) -- 
        ($(psi2.south west) + (-.3, -.3)$) -- 
        ($(psi2.south east) + (.3, -.3)$) --
        ($(phi2.north east) + (.25, .3)$) -- cycle;

        \draw[green, dashed, thick, rounded corners=6pt] ($(phi12.north west) + (-1.5, .25)$) -- 
        ($(psi1.south west) + (-.55, -.35)$) -- 
        ($(psi2.south east) + (.55, -.35)$) --
        ($(phi12.north east) + (1.5, .25)$) -- cycle;

        \draw[yellow, dashed, thick, rounded corners=6pt] ($(phi3.north west) + (-.25, .3)$) -- 
        ($(psi3.south west) + (-.3, -.3)$) -- 
        ($(psi3.south east) + (.3, -.3)$) --
        ($(phi3.north east) + (.25, .3)$) -- cycle;
    \end{tikzpicture}
    \end{subfigure}
    \hfill
    \begin{subfigure}{.45\textwidth}
        \begin{tikzpicture}
        \node[circle, fill=black, minimum size=.2cm] (pr) at (0,0){};
        \node[font=\large] () at (0,.5) {$\pi_{r}$};
        \node[circle, fill=green, minimum size=.2cm] (pt1) at (-1.5,-1.5){};
        \node[font=\large] () at (-1.5,-1) {$\pi_{t_{1}}$};
        \node[circle, fill=red, minimum size=.2cm] (pc1) at (-2.5,-3){};
        \node[font=\large] () at (-2.5,-3.5) {$\pi_{c_{1}}$};
        \node[circle, fill=blue, minimum size=.2cm] (pc2) at (-.5,-3){};
        \node[font=\large] () at (-.5,-3.5) {$\pi_{c_{2}}$};
        \node[circle, fill=yellow, minimum size=.2cm] (pt2) at (2,-3){};
        \node[font=\large] () at (2,-3.5) {$\pi_{t_{2}}$};

        \draw[-stealth] (pt1) -- (pc1);
        \draw[-stealth] (pt1) -- (pc2);
        \draw[-stealth] (pr) -- (pt1);
        \draw[-stealth] (pr) -- (pt2);
    \end{tikzpicture}
    \end{subfigure}
    \caption{Left: DAG representation of a hierarchical model with three levels. Right: A tree decomposition of the same hierarchical model. The coloured dashed shapes on the left show the model components incorporated by the distribution at the corresponding node in the tree on the right.
     }
    \label{Fig_hier_dag}
\end{figure}
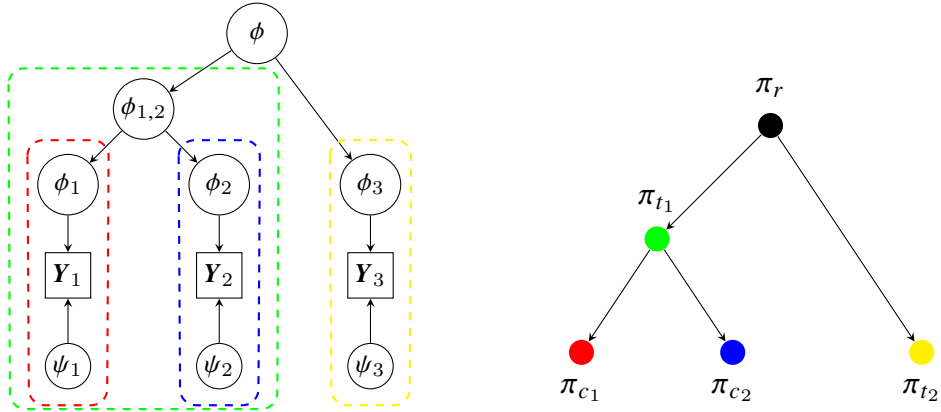

As an example, consider the simple hierarchical model with three levels, shown on the left of Figure~\ref{Fig_hier_dag}.
This can be converted into the tree structure model shown on the right panel of Figure~\ref{Fig_hier_dag}.
Nodes $c_{1},c_{2}$ and $t_{2}$ are the leaves in the tree.
The nodes $c_{1}$ and $c_{2}$ are associated with 
$\pi_{c_{1}}$ and $\pi_{c_{2}}$, which are the posterior distributions of $(\phi_{1},\psi_{1})$ given $\bsY_{1}$ and $(\phi_{2},\psi_{2})$ given $\bsY_{2}$, respectively.
The node $t_2$ is associated with $\pi_{t_{2}}$, which is the posterior distribution of $(\phi_{3},\psi_{3})$ given $\bsY_{3}$.
Nodes $c_{1}$ and $c_{2}$ are the child nodes of node $t_{1}$, and $\pi_{t_{1}}$ represents the posterior distribution of $(\phi_{1},\phi_{2},\phi_{1,2},\psi_{1},\psi_{2})$ given $(\bsY_{1},\bsY_{2})$.
Finally, node $r$ is the root node and $\pi_{r}$ represents the target distribution, which is the overall target posterior distribution of $(\phi_{1},\phi_{2},\phi_{3},\phi_{1,2},\phi,\psi_{1},\psi_{2},\psi_{3})$ given $(\bsY_{1},\bsY_{2},\bsY_{3})$.\\




Note that a tree-structured model does not necessarily correspond to a tree-structured DAG, and that traditional SMC methods consider the special case in which each node in the tree decomponsition has at most one child node, e.g., a state-space model with $n$ state nodes in which the probability function is defined on a product space $X_{t}$ that is described as 
    $X_{t}=\tilde{X}_{1}\times\tilde{X}_{2}\times\cdots\times\tilde{X}_{t},$
where $\tilde{X}_{i}$ is the state at node $i$ for $i=1,...,t$ and $t\in T=\lbrace1,2,...,n\rbrace$. 

\begin{algorithm}[t]
\caption{\texttt{d\&c-smc}(t): }\label{algo1}

    For $c\in\calC(t)$, 
        \nonl\textnormal{(a)} $\lbrace\bsx_{c}^{(i)},w_{c}^{(i)}\rbrace_{i=1}^{N}\leftarrow\text{d\&c-smc}(c)$, where $N$ is the number of particles.
    
        \nonl\textnormal{(b)} Resample $\lbrace\bsx_{c}^{(i)},w_{c}^{(i)}\rbrace_{i=1}^{N}$ to obtain $\lbrace\breve{\bsx}_{c}^{(i)},1\rbrace_{i=1}^{N}$ with equal weights.

    \textnormal{(a)} For $i=1$ to $N$, if $\tilde{X}_{t}\neq\varnothing$, simulate $\tilde{\bsx}_{t}^{(i)}\sim q_{t}(\cdot|\ \breve{\bsx}_{c_{1}}^{(i)},...,\breve{\bsx}_{c_{C}}^{(i)})$; else $\tilde{\bsx}_{t}^{(i)}\leftarrow\varnothing$.

    \nonl\textnormal{(b)} Set $\bsx_{t,0}^{(i)}=(\breve{\bsx}_{c_{1}}^{(i)},...,\breve{\bsx}_{c_{C}}^{(i)},\tilde{\bsx}_{t}^{(i)})$ and $\bsw_{t,0}^{(i)}=1$ for $i=1,...,N$.

    \nonl\textnormal{(c)} \For{SMC sampler iteration $j=1,\ldots, n_{t}$}{
        \nonl\textnormal{i.} Compute $\bsw_{t,j}^{(i)}=\bsw_{t,j-1}^{(i)}\gamma_{t,j}(\bsx_{t,j-1}^{(i)})/\gamma_{t,j-1}(\bsx_{t,j-1}^{(i)}).$

        \nonl\textnormal{ii.} Optionally, resample $\lbrace\bsx_{t,j-1}^{(i)},\bsw_{t,j}^{(i)}\rbrace_{i=1}^{N}$. Override the notation and let $\lbrace\bsx_{t,j-1}^{(i)},\bsw_{t,j}^{(i)}\rbrace_{i=1}^{N}$ refer to the resampled particle system.

        \nonl\textnormal{iii.} Draw $\bsx_{t,j}^{(i)}\sim K_{t,j}(\bsx_{t,j-1}^{(i)},\cdot)$ using a $\pi_{t,j}$-reversible Markov kernel $K_{t,j}$ for $i=1,...,N$.
    } 

    \nonl\textnormal{(d)} Set $\bsx_{t}^{(i)}=\bsx_{n_{t}}^{(i)}$ and $\bsw_{t}^{(i)}=\bsw_{t,n_{t}}^{(i)}.$

    \KwOut{$\lbrace\bsx_{t}^{(i)},\bsw_{t}^{(i)}\rbrace_{i=1}^{N}$.}

\end{algorithm}

Algorithm \ref{algo1} presents the algorithm proposed by \citet{lindsten2017}, where $\bsx^{(i)}_{t}$ denotes the $i^{\thh}$ particle at node $t$ and $\bsw^{(i)}_{t}$ denotes the corresponding weight. The algorithm shown combines the main D\&C-SMC algorithm  with the SMC sampling and tempering extensions \citep[see Section 4.2 of][]{lindsten2017}, which we will adopt for Markov melding in Section~\ref{se3}.
Note that the normalising constant $Z$ can be approximated using D\&C-SMC, but we omit this as it is not required in the settings we consider.

Step 1 of Algorithm \ref{algo1} indicates a recursion of the D\&C-SMC sampler applied for all child nodes of any given node $t$, meaning that the computational flow of the algorithm goes from leaf nodes to the root node (the opposite to the direction of the arrows in the graph). For example, for node $t_{1}$ in Figure \ref{Fig_hier_dag}, that occurs at nodes $c_{1}$ and $c_{2}$; for node $r$, that occurs at nodes $t_{1}$ and $t_{2}$. Step 2 is a tempering SMC that bridges the subposteriors of the child nodes and their parent node by constructing an artificial sequence of distributions via an annealing process \citep{chopin2020}. The sampling proposal $q_{t}$ in Step 2(a) can exploit the information of all the children $c\in\calC(t)$ of node $t$. In practice, a simple choice of $q_{t}$ is to use a prior for $\tilde{\bsx}_{t}$ at node $t$. If a prior has been directly specified for $\tilde{\bsx}_{t}$, then this can be used; otherwise, an artificial prior can be used. Specifically, Step 2(c) indicates that the sampler targets the distribution $\pi_{t}(\bsx_{t})$ with initialization of $\pi_{t,0}(\bsx_{t}):=\prod_{c\in\calC(t)}\pi(\bsx_{c})$ via annealing steps $\pi_{t,j}\propto\pi_{t,0}^{1-\alpha_{j}}\pi_{t,n_{t}}^{\alpha_{j}}$ with $0<\alpha_{1}<\cdots<\alpha_{n_{t}}=1$ and $\pi_{t,n_{t}}(\bsx_{t}):=\pi_{t}(\bsx_{t})$. \\


\section{Markov melding with D\&C-SMC}\label{se3}

In this section, 
we explore the combination of the Markov melding and the D\&C-SMC sampler used for the chain model introduced in Section \ref{se2.1}, and propose a multi-stage sampler that can sample from any chain model of the form in \eqref{eq:chained-meld-joint} formed from $M\geq3$ submodels. We start by constructing a two-stage sampler for the three-submodel case, before considering the general case. For convenience, we use the same notation as in Section \ref{se2.1} to represent probability functions.\\


\subsection{Divide and Conquer Melding with $M=3$ submodels}\label{se3.1}


Let $\Phi_{i,j}$ be the space for the common parameter $\phi_{i,j}$ and $\Psi_{i}$ be the space for the submodel-specific parameter $\psi_{i}$, with $i,j=1,...,M$ and $j>i$. To adopt D\&C-SMC for sampling from a chain model with three submodels, we need to construct a valid tree decomposition for the posterior of \eqref{eq:chained-meld-joint} with $M=3$ \begin{align}\label{eq:meld-posterior-M3}
    &p_{\meld}(\phi_{1,2},\phi_{2,3},\psi_{1},\psi_{2},\psi_{3}\mid\bsY_{1},\bsY_{2},\bsY_{3})\nonumber\\
    &\qquad\propto\ p_{\pool}(\phi_{1,2},\phi_{2,3})\frac{p_{1}(\phi_{1,2},\psi_{1},\bsY_{1})}{p_{1}(\phi_{1,2})}\frac{p_{2}(\phi_{1,2},\phi_{2,3},\psi_{2},\bsY_{2})}{p_{2}(\phi_{1,2},\phi_{2,3})}\frac{p_{3}(\phi_{2,3},\psi_{3},\bsY_{3})}{p_{3}(\phi_{2,3})}
\end{align}
The structure of the two-stage parallel sampler in Section \ref{se2.1} suggests the following decomposition:
\begin{align*}
\pi_{c_{1}} &= p_{\meld,1}(\phi_{1,2},\psi_{1}|\bsY_{1}) & X_{c_{1}} &= \Phi_{1,2} \times \Psi_{1}\\
\pi_{c_{2}} &= p_{\meld,3}(\phi_{2,3},\psi_{3}|\bsY_{3}) & X_{c_{2}} &= \Phi_{2,3} \times \Psi_{2}\\
\pi_{r} &= p_{\meld}(\phi_{1,2},\phi_{2,3},\psi_{1},\psi_{2},\psi_{3}|\bsY_{1},\bsY_{2},\bsY_{3}) & X_{r} &= (\Phi_{1,2} \times \Psi_{1}) \times (\Phi_{2,3} \times \Psi_{2}) \times \Psi_{3}
\end{align*}
with the tree formed by a root note $\pi_r$ that has child notes $\calC(r) = \{\pi_{c_{1}}, \pi_{c_{2}}\}$.
Clearly, according to Definition \ref{def:tree-decomp}, this is a valid tree decomposition, since $\pi_r = p_{\meld}$, with identical state spaces, and 
$X_{r}=(X_{c_{1}}\otimes X_{c_{2}})\times\tilde{X}_{r}$, where $\tilde{X}_{r} = \Psi_{3}$.

\begin{algorithm}[t]
    \caption{\texttt{d\&c-melding}(3): Markov melding with D\&C-SMC for $M=3$}
    \label{algo2}
        \KwIn{data $\bsY$; the choice of $p_{\pool}(\phi_{1,2},\phi_{2,3})$; the subposteriors $p_{\meld,m}(\phi_{m},\psi_{m}\mid\bsY_{m})$ for $m=1,2,3$; the proposal $q_{\meld}(\psi_{2}|\phi_{1,2},\phi_{2,3})$; the number of particles $N$, the number of the annealing steps $n_{t}$.} 
        
        In stage one ($s_{1}$ in Figure \ref{Fig_two_stage_orig}), 
        
            \nonl\textnormal{(a)} Sample $\lbrace\phi_{1,2}^{(i)},w_{1,2}^{(i)}\rbrace_{i=1}^{N}$ and $\lbrace\phi_{2,3}^{(i)},w_{2,3}^{(i)}\rbrace_{i=1}^{N}$ via SMC samplers in parallel.
        
            \nonl\textnormal{(b)} Resample $\lbrace\phi_{1,2}^{(i)},w_{1,2}^{(i)}\rbrace_{i=1}^{N}$ and $\lbrace\phi_{2,3}^{(i)},w_{2,3}^{(i)}\rbrace_{i=1}^{N}$ in parallel to obtain equally weighted particle systems $\lbrace\breve{\phi}_{1,2}^{(i)},1\rbrace_{i=1}^{N}$ and $\lbrace\breve{\phi}_{2,3}^{(i)},1\rbrace_{i=1}^{N}$.
        
        In stage two ($s_{2}$ in Figure \ref{Fig_two_stage_orig}),
        
            \nonl\textnormal{(a)} Initialise $\tilde{\psi}_{2}^{(i)}\sim q_{\meld}(\cdot|\breve{\phi}_{1,2}^{(i)},\breve{\phi}_{2,3}^{(i)})$ for $i=1,...,N$.
        
            \nonl\textnormal{(b)} Set $\bstheta_{0}^{(i)}=(\breve{\phi}_{1,2}^{(i)},\breve{\phi}_{2,3}^{(i)},\tilde{\psi}_{2}^{(i)})$ and $w_{0}^{(i)}=1$ for $i=1,...,N$.
        
            \nonl\textnormal{(c)} \For{SMC sampler iteration $j=1$ to $n_{t}$}{
                \nonl\textnormal{i.} Compute $w_{j}^{(i)}=w_{j-1}^{(i)}\gamma_{\smc,j}(\bstheta_{j-1}^{(i)})/\gamma_{\smc,j-1}(\bstheta_{j-1}^{(i)}).$
        
                \nonl\textnormal{ii.} Optionally, resample $\lbrace\bstheta_{j-1}^{(i)},w_{j}^{(i)}\rbrace_{i=1}^{N}$. Override the notation and let $\lbrace\bstheta_{j-1}^{(i)},w_{j}^{(i)}\rbrace_{i=1}^{N}$ refer to the resampled particle system.
        
                \nonl\textnormal{iii.} Draw $\bstheta_{j}^{(i)}\sim K_{j}(\bstheta_{j-1}^{(i)})$ using a $p_{\smc,j}$-reversible Markov kernel $K_{j}$ for $i=1,...,N$.
            }
        
            \nonl\textnormal{(d)} Set $\bstheta^{(i)}=\bstheta_{n_{t}}^{(i)}$ and $w^{(i)}=w_{n_{t}}$.
            
        \KwOut{$\lbrace\bstheta^{(i)},w^{(i)}\rbrace_{i=1}^{N}$.}

\end{algorithm} 

Our sampler, which we call \textit{D\&C-melding}, for a general $M=3$ setting (Figure \ref{Fig_two_stage_orig}) is presented in Algorithm \ref{algo2}, and is a combination of the two-stage parallel sampler in Section~\ref{se2.1.4} and the D\&C-SMC sampler in Algorithm \ref{algo1}. In stage one (Step 1), we draw particles $\lbrace\phi_{1,2}^{(i)},w_{1,2}^{(i)}\rbrace_{i=1}^{N}$ and $\lbrace\phi_{2,3}^{(i)},w_{2,3}^{(i)}\rbrace_{i=1}^{N}$ from the two submodels by targeting the subposteriors \eqref{eq:meld-posterior-sub1} and \eqref{eq:meld-posterior-sub3} separately.
To do this, we can employ any appropriate SMC sampler
. In stage two (Step 2), the merged particles of $\phi_{1,2}$ and $\phi_{2,3}$ are resampled by combining the auxiliary distribution of Submodel 2. This process occurs at the root node of the tree. In particular, following the idea of D\&C-SMC, we use a tempering SMC to gradually target the joint posterior. Within the tempering sampler, a reversible Markov kernel is applied to draw samples for the submodel-specific parameter $\psi_{2}$ of Submodel 2. 

We construct a synthetic sequence process in the SMC sampler to gradually target the full posterior distribution. The dependence in chained Markov melding of $\psi_{2}$ on $\phi_{1,2}$ and $\phi_{2,3}$ can affect the choice of proposal function for $\psi_{2}$. In the D\&C-melding sampler, the choice of $q_{\meld}$ in Line 2(a) follows this principle: if there is a direct dependence of $\psi_{2}$ on $(\phi_{1,2},\phi_{2,3})$, then $q_{\meld}$ is chosen to be the probability function conditional on the common parameters, $q_{2}(\cdot|\phi_{1,2},\phi_{2,3})$; otherwise, simply, $q_{\meld}$ is chosen to be the (artificial) prior $p_{2}(\psi_{2})$ in Submodel 2.

Let $\bstheta=(\phi_{1,2},\phi_{2,3},\tilde{\psi}_{2})$ be the vector of all variables. Define $p_{\smc,0}(\bstheta)\allowbreak:=\allowbreak p_{\meld,1,3}(\phi_{1,2},\allowbreak\phi_{2,3}|\allowbreak\bsY_{1},\allowbreak\bsY_{3})\allowbreak q_{\meld}(\tilde{\psi}_{2}|\allowbreak\phi_{1,2},\allowbreak\phi_{2,3})$ and $p_{\smc,n_{t}}(\bstheta)\allowbreak:=\allowbreak p_{\meld}(\phi_{1,2},\allowbreak\phi_{2,3},\allowbreak\psi_{1},\allowbreak\psi_{3},\allowbreak\tilde{\psi}_{2}|\allowbreak\bsY_{1},\allowbreak\bsY_{2},\allowbreak\bsY_{3})$. Let $\gamma_{\smc,0}$ and $\gamma_{\smc,n_{t}}$ be unnormalised $p_{\smc,0}$ and $p_{\smc,n_{t}}$, respectively. The sampler then uses the annealing process to target \eqref{eq:meld-posterior-M3} via a geometric path $p_{\meld,j}\propto p_{\meld,0}^{1-\alpha_{j}}p_{\meld,n_{t}}^{\alpha_{j}}$ with $0<\alpha_{1}<\cdots<\alpha_{n_{t}}=1$.

In particular, the update of $\bsw_{j}^{(i)}$ in Line (c)ii can be simplified as \begin{align}\label{eq:weight-update}
    \bsw_{j}^{(i)}=\bsw_{j-1}^{(i)}\ls\frac{\gamma_{\smc,n_{t}}(\bstheta_{j-1}^{(i)})}{\gamma_{\smc,0}(\bstheta_{j-1}^{(i)})}\rs^{\alpha_{j}-\alpha_{j-1}}.
\end{align} Specifically, we can use the following result to implement the update. \begin{mylemma}\label{le:weight-update}
    The increment in \eqref{eq:weight-update} follows \begin{align}\label{eq:weight-update1}
    \frac{\gamma_{\smc,n_{t}}(\bstheta_{j-1}^{(i)})}{\gamma_{\smc,0}(\bstheta_{j-1}^{(i)})}\propto\frac{p_{\pool,2}(\phi_{1,2}^{(i)},\phi_{2,3}^{(i)})p_{2}(\bsY_{2}|\phi_{1,2}^{(i)},\phi_{2,3}^{(i)},\tilde{\psi}_{2}^{(i)})p_{2}(\tilde{\psi}_{2}^{(i)})}{q_{\meld}(\tilde{\psi}_{2}^{(i)}|\phi_{1,2}^{(i)},\phi_{2,3}^{(i)})}.
\end{align} In addition, if we choose $q_{\meld}(\tilde{\psi}_{2}|\phi_{1,2},\phi_{2,3})=p_{2}(\tilde{\psi}_{2})$, then \eqref{eq:weight-update1} simplifies to \begin{align}\label{eq:weight-update2}
    \frac{\gamma_{\smc,n_{t}}(\bstheta_{j-1}^{(i)})}{\gamma_{\smc,0}(\bstheta_{j-1}^{(i)})}\propto p_{\pool,2}(\phi_{1,2}^{(i)},\phi_{2,3}^{(i)})p_{2}(\bsY_{2}|\phi_{1,2}^{(i)},\phi_{2,3}^{(i)},\tilde{\psi}_{2}^{(i)}).
\end{align} 
\end{mylemma} 

\subsection{Generalisation for $M\geq3$}\label{se3.2}
Our extension of the sampler for situations when the number of submodels $M>3$ depends on whether $M$ is odd or even.
%
\begin{figure}[t]
    \centering
    \begin{subfigure}[b]{.6\textheight}
        \centering
        \begin{tikzpicture}[scale=0.75, transform shape]
        \node[dashed, draw=red, minimum width=3.5cm, minimum height=1.2cm, inner sep=0pt] (p2) at (0,0) {\footnotesize $p_{\meld,2}(\textcolor{red}{\phi_{1,2},\phi_{2,3}},\psi_{2}\mid\bsY_{2})$};
        \node[red] () at (0,.4) {$s_{2}$};

        \node[dashed, draw=blue, minimum width=3cm, minimum height=1.2cm, inner sep=0pt] (p1) at (-2.5,-2) {\footnotesize $p_{\meld,1}(\textcolor{blue}{\phi_{1,2}},\psi_{1}\mid\bsY_{1})$};
        \node[blue] () at (-1.2,-1.65) {$s_{1}$};
        \node[dashed, draw=blue, minimum width=3.5cm, minimum height=1.2cm, inner sep=0pt] (p3) at (2.5,-2) {\footnotesize $p_{\meld,3}(\textcolor{blue}{\phi_{2,3},\phi_{3,4}},\psi_{3}\mid\bsY_{3})$};
        \node[blue] () at (2.5,-1.65) {$s_{1}$};
        \node[dashed, draw=blue, minimum width=3cm, minimum height=1.2cm, inner sep=0pt] (p5) at (7.5,-2) {\footnotesize $p_{\meld,5}(\textcolor{blue}{\phi_{4,5}},\psi_{5}\mid\bsY_{5})$};
        \node[blue] () at (6.3,-1.65) {$s_{1}$};

        \node[dashed, draw=safegreen, minimum width=3.5cm, minimum height=1.2cm, inner sep=0pt] (p4) at (5,0) {\footnotesize $p_{\meld,4}(\textcolor{safegreen}{\phi_{3,4},\phi_{4,5}},\psi_{4}\mid\bsY_{4})$};
        \node[safegreen] () at (5,.4) {$s_{3}$};


        \draw[-stealth, red] (p1) -- (p2);
        \draw[-stealth, red] (p3) -- (p2);
        \draw[-stealth, safegreen] (p3) -- (p4);
        \draw[-stealth, safegreen] (p5) -- (p4);
    \end{tikzpicture}
    \caption{}
    \end{subfigure}

    \begin{subfigure}[b]{.6\textheight}
        \centering
        \begin{tikzpicture}[scale=0.75, transform shape]
        \node[dashed, draw=red, minimum width=3.5cm, minimum height=1.2cm, inner sep=0pt] (p2) at (0,0) {\footnotesize $p_{\meld,2}(\textcolor{red}{\phi_{1,2},\phi_{2,3}},\psi_{2}\mid\bsY_{2})$};
        \node[red] () at (0,.35) {$s_{2}$};

        \node[dashed, draw=blue, minimum width=3cm, minimum height=1.2cm, inner sep=0pt] (p1) at (-2,-2) {\footnotesize $p_{\meld,1}(\textcolor{blue}{\phi_{1,2}},\psi_{1}\mid\bsY_{1})$};
        \node[blue] () at (-.8,-1.6) {$s_{1}$};
        \node[dashed, draw=blue, minimum width=3.5cm, minimum height=1.2cm, inner sep=0pt] (p3) at (2,-2) {\footnotesize $p_{\meld,3}(\textcolor{blue}{\phi_{2,3},\phi_{3,4}},\psi_{3}\mid\bsY_{3})$};
        \node[blue] () at (3.5,-1.6) {$s_{1}$};

        \node[dashed, draw=safegreen, minimum width=3.5cm, minimum height=1.2cm, inner sep=0pt] (p4) at (4,0) {\footnotesize $p_{\meld,4}(\textcolor{safegreen}{\phi_{3,4},\phi_{4,5}},\psi_{4}\mid\bsY_{4})$};
        \node[safegreen] () at (4,.35) {$s_{3}$};
        
        \node[dashed, draw=blue, minimum width=3.5cm, minimum height=1.2cm, inner sep=0pt] (p5) at (6,-2) {\footnotesize $p_{\meld,5}(\textcolor{blue}{\phi_{4,5},\phi_{5,6}},\psi_{5}\mid\bsY_{5})$};
        \node[blue] () at (4.5,-1.6) {$s_{1}$};
        \node[dashed, draw=blue, minimum width=3cm, minimum height=1.2cm, inner sep=0pt] (p7) at (10,-2) {\footnotesize $p_{\meld,7}(\textcolor{blue}{\phi_{6,7}},\psi_{7}\mid\bsY_{7})$};
        \node[blue] () at (8.8,-1.6) {$s_{1}$};

        \node[dashed, draw=red, minimum width=3.5cm, minimum height=1.2cm, inner sep=0pt] (p6) at (8,0) {\footnotesize $p_{\meld,6}(\textcolor{red}{\phi_{5,6},\phi_{6,7}},\psi_{6}\mid\bsY_{6})$};
        \node[red] () at (8,.35) {$s_{2}$};


        \draw[-stealth, red] (p1) -- (p2);
        \draw[-stealth, red] (p3) -- (p2);
        \draw[-stealth, red] (p5) -- (p6);
        \draw[-stealth, red] (p7) -- (p6);
        \draw[-stealth, safegreen] (p3) -- (p4);
        \draw[-stealth, safegreen] (p5) -- (p4);
    \end{tikzpicture}
    \caption{}  
    \end{subfigure}
    \caption{Diagram illustrating the sampling process for (a) $M=5$ and (b) $M=7$. The computational stage is denoted by $s_{t}$, with $t=1,2,3$, and the densities shown reflect the additional terms considered at that stage. The arrows represent the merging direction. The common parameters are highlighted.}
    \label{Fig_sampler_odd}
\end{figure}
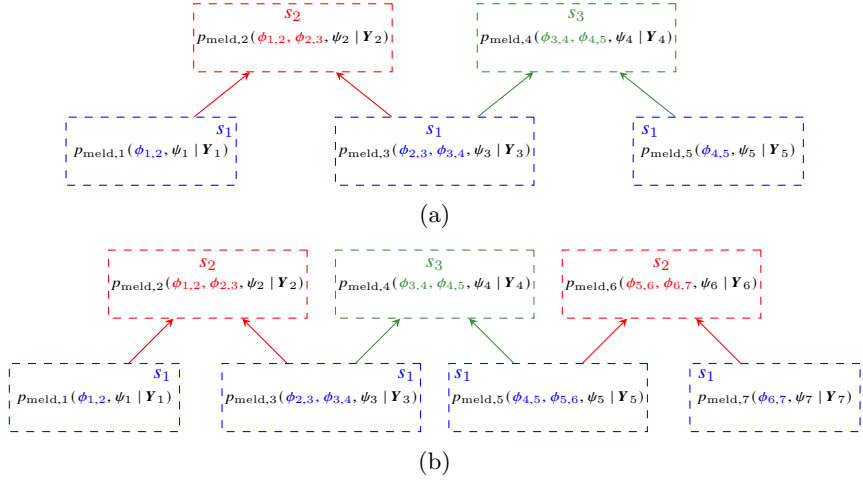 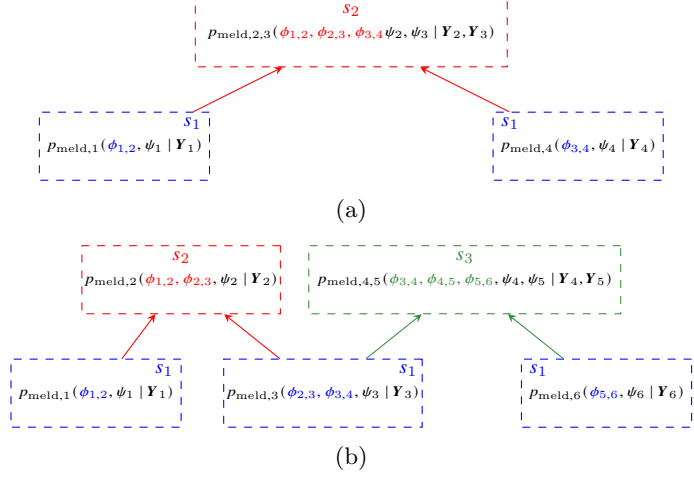
\begin{figure}[t]
    \centering
    \begin{subfigure}[b]{.6\textheight}
        \centering
        \begin{tikzpicture}[scale=0.75, transform shape]
            \node[dashed, draw=red, minimum width=5.5cm, minimum height=1.2cm, inner sep=0pt] (p23) at (2,0) {\footnotesize $p_{\meld,2,3}(\textcolor{red}{\phi_{1,2},\phi_{2,3},\phi_{3,4}}\psi_{2},\psi_{3}\mid\bsY_{2},\bsY_{3})$};
            \node[red] () at (2,.4) {$s_{2}$};
    
            \node[dashed, draw=blue, minimum width=3cm, minimum height=1.2cm, inner sep=0pt] (p1) at (-2,-2) {\footnotesize $p_{\meld,1}(\textcolor{blue}{\phi_{1,2}},\psi_{1}\mid\bsY_{1})$};
            \node[blue] () at (-.8,-1.6) {$s_{1}$};
    
            \node[dashed, draw=blue, minimum width=3cm, minimum height=1.2cm, inner sep=0pt] (p4) at (6,-2) {\footnotesize $p_{\meld,4}(\textcolor{blue}{\phi_{3,4}},\psi_{4}\mid\bsY_{4})$};
            \node[blue] () at (4.8,-1.6) {$s_{1}$};

    
            \draw[-stealth, red] (p1) -- (p23);
            \draw[-stealth, red] (p4) -- (p23);
        \end{tikzpicture}
        \caption{}
    \end{subfigure}

    \begin{subfigure}[b]{.6\textheight}
        \centering
        \begin{tikzpicture}[scale=0.75, transform shape]
            \node[dashed, draw=blue, minimum width=3cm, minimum height=1.2cm, inner sep=0pt] (p1) at (-3.5,-2) {\footnotesize $p_{\meld,1}(\textcolor{blue}{\phi_{1,2}},\psi_{1}\mid\bsY_{1})$};
            \node[blue] () at (-2.3,-1.6) {$s_{1}$};
            \node[dashed, draw=blue, minimum width=3.5cm, minimum height=1.2cm, inner sep=0pt] (p3) at (0.5,-2) {\footnotesize $p_{\meld,3}(\textcolor{blue}{\phi_{2,3},\phi_{3,4}},\psi_{3}\mid\bsY_{3})$};
            \node[blue] () at (2,-1.6) {$s_{1}$};
            \node[dashed, draw=blue, minimum width=3cm, minimum height=1.2cm, inner sep=0pt] (p6) at (5.5,-2) {\footnotesize $p_{\meld,6}(\textcolor{blue}{\phi_{5,6}},\psi_{6}\mid\bsY_{6})$};
            \node[blue] () at (4.3,-1.6) {$s_{1}$};

            \node[dashed, draw=red, minimum width=3.5cm, minimum height=1.2cm, inner sep=0pt] (p2) at (-2,0) {\footnotesize $p_{\meld,2}(\textcolor{red}{\phi_{1,2},\phi_{2,3}},\psi_{2}\mid\bsY_{2})$};
            \node[red] () at (-2,.4) {$s_{2}$};

            \node[dashed, draw=safegreen, minimum width=5.5cm, minimum height=1.2cm, inner sep=0pt] (p45) at (3,0) {\footnotesize $p_{\meld,4,5}(\textcolor{safegreen}{\phi_{3,4},\phi_{4,5},\phi_{5,6}},\psi_{4},\psi_{5}\mid\bsY_{4},\bsY_{5})$};
            \node[safegreen] () at (3,.4) {$s_{3}$};


            \draw[-stealth, red] (p1) -- (p2);
            \draw[-stealth, red] (p3) -- (p2);
            \draw[-stealth, safegreen] (p3) -- (p45);
            \draw[-stealth, safegreen] (p6) -- (p45);
        \end{tikzpicture}
        \caption{}
    \end{subfigure}
    \caption{Diagram illustrating the sampling process for (a) $M=4$ and (b) $M=6$. The computational stage is denoted by $s_{t}$, with $t=1,2,3$, and the densities shown reflect the additional terms considered at that stage. The arrows represent the merging direction. The common parameters are highlighted.}
    \label{Fig_sampler_even}
\end{figure}%
Figure \ref{Fig_sampler_odd} illustrates the sampling process for odd values of $M$ ($M=5$ and $7$); and Figure \ref{Fig_sampler_even} illustrates for even values of $M$ ($M=4$ and $6$). Note that the sampling process for $M=3$ is a special case of the process for $M=7$. In fact, the sampling process for any $M\geq3$ has a pattern that is the same as one of the four cases shown in Figures \ref{Fig_sampler_odd} and \ref{Fig_sampler_even}.

For convenience, define $\phi_{m}:=(\phi_{m-1,m},\,\phi_{m,m+1})$ for $m=2,...,M-1$, with $\phi_{1}=\phi_{1,2}$ and $\phi_{M}=\phi_{M-1,M}$. In terms of this and \eqref{eq:chained-meld-joint}, for any arbitrary $M\geq3$, the target distribution is \begin{align}\label{eq:chained-meld-posterior}
    p_{\meld}(\bsphi,\bspsi\mid\bsY)\propto p_{\pool}(\bsphi)\prod_{m=1}^{M}\frac{p_{m}(\phi_{m},\psi_{m},\bsY_{m})}{p_{m}(\phi_{m})}.
\end{align} Assuming the pooled prior can be decomposed into 
    $p_{\pool}(\bsphi)=\prod_{m=1}^{M}p_{\pool,m}(\phi_{m})$,
 the target distribution follows \begin{align}\label{eq:pooled-prior-posterior-new}
    p_{\meld}(\bsphi,\bspsi\mid\bsY)\propto\prod_{m=1}^{M}p_{\pool,m}(\phi_{m})\frac{p_{m}(\phi_{m},\psi_{m},\bsY_{m})}{p_{m}(\phi_{m})}.
\end{align} Similarly to the sampler for $M=3$, we can choose the $p_{\pool,m}(\phi_{m})$ in the first stage to be the original priors $p_{m}(\phi_{m})$, and then make appropriate adjustments to the remaining $p_{\pool,m}(\phi_{m})$
In addition, the multi-stage sampler requires consideration of the distributions of arbitrary subsets $(i_{1}, \dots, i_{t}) \subseteq M$ of the submodels.  Let $p_{\meld,i_{1},...,i_{t}}$ be the joint distribution of submodels $i_{1},...,i_{t}$: \begin{align*}
    p_{\meld,i_{1},...,i_{t}}\lr\bsphi_{(i_{1},...,i_{t})},\bspsi_{(i_{1},...,i_{t})},\bsY_{(i_{1},...,i_{t})}\rr=\prod_{m=i_{1},...,i_{t}}p_{\pool,m}(\phi_{m})\frac{p_{m}(\phi_{m},\psi_{m},\bsY_{m})}{p_{m}(\phi_{m})},
\end{align*} where $\bsphi_{(i_{1},...,i_{t})},\bspsi_{(i_{1},...,i_{t})},\bsY_{(i_{1},...,i_{t})}$ are the corresponding variables from submodels $i_{1},...,i_{t}$.
The corresponding subposterior is defined analogously. The sampling process structure for the multi-stage sampler is described below in terms of different cases of $M$. \begin{enumerate}[label=\Roman*.]
    \item $M$ is odd. \begin{enumerate}[label=\roman*.]
        \item When $4\mid(M+1)$ (e.g. $M=7$ in Figure \ref{Fig_sampler_odd}), the sampling process follows: \begin{enumerate}[label=a\arabic*)]
            \item In stage 1, target the joint subposterior \begin{align*}
                p_{\meld,1,3,5,...,M}(\bsphi,\bspsi_{(1,3,5,...,M)}\mid\bsY_{(1,3,5,...,M)})
            \end{align*}
                    
        \item In stage $s$ for $s=2,3,..., (M+1)/4$, additionally consider Submodels $m_{L}$ and $m_{R}$, with $m_{L}=2s-2$ and $m_{R}=M+3-2s$, and target the joint subposterior $p_{\meld,I_{s}}(\bsphi,\bspsi_{I_{s}}|\bsY_{I_{s}})$ with \begin{align*}
            I_{s}=\lbrace &1,2,3,...,m_{L},m_{L}+1,m_{L}+3,..., m_{R}-1,m_{R},m_{R}+1,m_{R}+2,...,M\rbrace.
        \end{align*} Note that the subscripts $L$ and $M$ in $m_{L}$ and $m_{R}$ are used to refer to the left and right submodels, such as Submodel 2 and Submodel 6 in Figure \ref{Fig_sampler_odd}(b). The same meaning is used in the following notation.

        \item In stage $(M+5)/4$, additionally consider Submodel $(M+1)/2$, and target the full posterior.
        \end{enumerate}

        \item When $4\mid(M-1)$ (e.g. $M=5$ in Figure \ref{Fig_sampler_odd}), the sampling process follows: \begin{enumerate}[label=b\arabic*)]
            \item In stage 1, target the joint subposterior \begin{align*}
                p_{\meld,1,3,5,...,M}(\bsphi,\bspsi_{(1,3,5,...,M)}\mid\bsY_{(1,3,5,...,M)}).
            \end{align*}

            \item In stage $s$ for $s=2,3,..., (M-1)/4$, additionally consider Submodels $m_{L}$ and $m_{R}$, with $m_{L}=2s-2$ and $m_{R}=M+3-2s$, and target the joint subposterior $p_{\meld,I_{s}}(\bsphi,\bspsi_{I_{s}}\mid\bsY_{I_{s}})$ with \begin{align*}
                I_{s}=\lbrace &1,2,3,...,m_{L},m_{L}+1,m_{L}+3,..., m_{R}-1,m_{R},m_{R}+1,m_{R}+2,...,M \rbrace.
            \end{align*}
    
            \item In stage $(M+3)/4$, additionally consider Submodel $(M-1)/2$, and target the joint subposterior \begin{align*}
                p_{\meld,I_{s}\cup\lbrace(M-1)/2\rbrace}\lr\bsphi,\bspsi_{I_{s}\cup\lbrace(M-1)/2\rbrace}\mid\bsY_{I_{s}\cup\lbrace(M-1)/2\rbrace}\rr,
            \end{align*} 
    
            \item In stage $(M+7)/4$, additionally consider Submodel $(M+1)/2+1$, and target the full posterior. 
        \end{enumerate}
    \end{enumerate}

    \item $M$ is even. \begin{enumerate}[label=\roman*.]
        \item When $4\mid M$ (e.g. $M=4$ in Figure \ref{Fig_sampler_even}), the sampling process follows: \begin{enumerate}[label=a\arabic*)]
            \item In stage 1, target the joint subposterior $p_{\meld,I_{s_{1}}}(\bsphi_{I_{s_{1}}},\bspsi_{I_{s_{1}}}\mid\bsY_{I_{s_{1}}})$ with \begin{align*}
                I_{s_{1}}=\lb 1,3,5,...,\frac{M}{2}-1,\frac{M}{2}+2,\frac{M}{2}+4,...,M-2,M\rb.
            \end{align*}

            \item In stage $s$ for $s=2,3,...,M/4$, additionally consider Submodels $m_{L}$ and $m_{R}$, with $m_{L}=2s-2$ and $m_{R}=M+3-2s$, and target the joint subposterior $p_{\meld,I_{s_{2}}}(\bsphi_{I_{s_{2}}},\bspsi_{I_{s_{2}}}\mid \bsY_{I_{s_{2}}})$ with \begin{align*}
                I_{s_{2}}=\bigg\lbrace &1,2,3,...,m_{L},m_{L}+1,m_{L}+3,...,\nonumber\\
                &\qquad\frac{M}{2}-1,\frac{M}{2}+2,\frac{M}{2}+4,..., m_{R}-1,m_{R},m_{R}+1,m_{R}+2,...,M\bigg\rbrace.
            \end{align*}

            \item In stage $(M+4)/4$, additionally consider Submodels $M/2$ and $(M+2)/2$, and target the full posterior.
        \end{enumerate}

        \item When $4\nmid M$ (e.g. $M=6$ in Figure \ref{Fig_sampler_even}), the sampling process follows: \begin{enumerate}[label=b\arabic*)]
            \item In stage 1, target the joint subposterior $p_{\meld,I_{s_{1}}}(\bsphi_{I_{s_{1}}},\bspsi_{I_{s_{1}}}\mid\bsY_{I_{s_{1}}})$ with \begin{align*}
                I_{s_{1}}=\lb 1,3,5,...,\frac{M}{2},\frac{M}{2}+3,\frac{M}{2}+5,...,M-2,M\rb.
            \end{align*}

            \item In stage $s$ for $s=2,3,...,(M-2)/4$, additionally consider Submodels $m_{L}$ and $m_{R}$, with $m_{L}=2s-2$ and $m_{R}=M+3-2s$, and target the joint subposterior $p_{\meld,I_{s_{2}}}(\bsphi_{I_{s_{2}}},\bspsi_{I_{s_{2}}}\mid\bsY_{I_{s}})$ with \begin{multline*}
                I_{s_{2}}= \bigg\lbrace 1,2,3,...,m_{L},m_{L}+1,m_{L}+3,...,\frac{M}{2},\frac{M}{2}+3,\frac{M}{2}+5,...,\\
                m_{R}-1,m_{R},m_{R}+1,m_{R}+2,...,M\bigg\rbrace.
            \end{multline*}

            \item In stage $(M+2)/4$, additionally consider Submodel $M/2-1$, and target the joint subposterior \begin{align*}
                p_{\meld,I_{s_{2}}\cup\lbrace M/2-1\rbrace}\lr\bsphi_{I_{s_{2}}\cup\lbrace M/2-1\rbrace},\bspsi_{I_{s_{2}}\cup\lbrace M/2-1\rbrace}\mid\bsY_{I_{s_{2}}\cup\lbrace M/2-1\rbrace}\rr.
            \end{align*}

            \item In stage $(M+6)/4$, additionally consider Submodels $M/2+1$ and $M/2+2$, and target the full posterior.
        \end{enumerate}
    \end{enumerate}
\end{enumerate}

\begin{myremark}\label{re:stage_number}
    Compared to methods that simultaneously sample all parameters from the full posterior, such as MCMC applied to the full model, an advantage of the multi-stage sampler is that the sampling process can be split into multiple stages, and sampling across different submodels within a stage can be implemented in parallel. The number of stages into which a multi-stage sampler can be split is summarised below 
\begin{itemize}
\item if $4\mid(M+1)$, then $S = \frac{1}{4}(M+ 5)$
\item if $4\mid(M-1)$, then $S = \frac{1}{4}(M+ 7)$
\item if $4\mid M$, then $S = \frac{1}{4}(M+ 4)$
\item if $4\nmid M$, then $S = \frac{1}{4}(M+ 6)$
\end{itemize}
\end{myremark}

\begin{myremark}\label{re:tree-decomp}
    As with the sampler proposed for $M=3$, to apply the D\&C-SMC technique, we also need to show that the structure of the multi-stage sampler is a valid tree decomposition according to Definition \ref{def:tree-decomp}. We now show this for the case of odd $M$ with $4\mid(M+1)$. The other three cases can be shown similarly. Let $c_{i,j}$ refer to the node in stage $i$ for Submodel $j$. In stage one, which is the leaf node in the tree structure, the distributions and spaces are \begin{align*}
        &\pi_{c_{1,m}}=p_{\meld,m}(\phi_{m},\psi_{m}\mid\bsY_{m})\quad X_{c_{1,m}}=\Phi_{m}\times\Psi_{m},\quad\text{for $m=1,3,5,...,M$}
    \end{align*} In stage $s$ from $2$ to $(M+1)/4$, let $L=2s-2$ and $R=M+3-2s$. The node $c_{s,L}$ has child nodes $\calC(c_{s,L})=\lbrace c_{s-1,L-1},c_{1,L+1}\rbrace$, and the node $c_{s,R}$ has child nodes $\calC(c_{s,R})=\lbrace c_{s-1,R+1},c_{1,R-1}\rbrace$. The corresponding distributions and spaces are \begin{align*}
        &\quad\pi_{c_{s,L}}=p_{\meld,I_{s_{1}}}(\bsphi_{I_{s_{1}}},\bspsi_{I_{s_{1}}}\mid\bsY_{I_{s_{1}}})\quad X_{c_{s,L}}=(\otimes_{c\in\calC(c_{s,L})} X_{c})\times (\Phi_{2s-2}\times\Psi_{2s-2}),\\
        &\text{and}\\
        &\quad\pi_{c_{s,R}}=p_{\meld,I_{s_{2}}}(\bsphi_{I_{s_{2}}},\bspsi_{I_{s_{2}}}\mid\bsY_{I_{s_{2}}})\quad X_{c_{s,R}}=(\otimes_{c\in\calC(c_{s,R}}) X_{c})\times (\Phi_{M+3-2s}\times\Psi_{M+3-2s}),
    \end{align*} where \begin{align*}
        I_{s_{1}}=\lbrace1,2,...,m_{L},m_{L}+1\rbrace\quad \andd\quad I_{s_{2}}=\lbrace m_{R}-1,m_{R},m_{R}+1,...,M\rbrace.
    \end{align*} In the final stage $(M+5)/4$, which is the root in the tree structure, denote the node by $r$. The child nodes of $r$ are $\calC(r)=\lbrace c_{(M+1)/4,(M-3)/2},c_{(M+1)/4,(M+5)/2}\rbrace$. Then, the distribution and spaces on the root are \begin{align*}
        \pi_{r}=p_{\meld}(\bsphi,\bspsi\mid\bsY)\qquad\qquad X_{r}&=(\otimes_{c\in\calC(r)}X_{c})\times(\Phi_{(M+1)/2}\times\Psi_{(M+1)/2})\\
        &=\otimes_{m\in\lbrace1,...,M\rbrace}(\Phi_{m}\times\Psi_{m}).
    \end{align*} Since $\pi_{r}=p_{\meld}$, with identical state spaces, then the structure of the multi-stage sampler in this case is a valid tree decomposition. Since the structures in the other three cases are also a valid tree decomposition, the D\&C-SMC algorithm can be applied in the multi-stage sampler.  
\end{myremark}

Algorithms for the multi-stage sampler for each case and the method for tracking particle trajectories used to obtain posterior samples for all parameters (i.e.\ not just the parameter of the final stage) are provided in the Supplementary. We use the case $M=5$ to demonstrate how the sampler works. \begin{myexample}\label{eg:M5}
    For $M=5$, the target distribution is $p_{\meld}(\phi_{1,2},\allowbreak\phi_{2,3},\allowbreak\phi_{3,4},\allowbreak\phi_{4,5},\allowbreak\psi_{1},...,\allowbreak\psi_{5}|\bsY_{1},...,\allowbreak\bsY_{5})$. \begin{description}
        \item[Stage one ($s_{1}$)] is illustrated by Figure \ref{Fig_sampler_odd}(a). This stage considers Submodels 1, 3 and 5 by targeting the joint subposterior 
        $p_{\meld,1,3,5}(\phi_{1,2},\allowbreak\phi_{2,3},\allowbreak\phi_{3,4},\allowbreak\phi_{4,5},\allowbreak\psi_{1},\allowbreak\psi_{3},\allowbreak\psi_{5}|\bsY_{1},\bsY_{3},\bsY_{5})$.
        Equation \eqref{eq:pooled-prior-posterior-new} indicates that this posterior can factorise into separate submodel-specific terms, implying the three submodels are independent, and thus we can sample each submodel-specific posterior  $p_{\meld,m}(\phi_{m},\psi_{m}|\bsY_{m})$ separately.
        Let $\bstheta_{m}=(\phi_{m},\psi_{m})$. Denote the resulting particles by
        \begin{align*}
        &\lbrace\bstheta_{1}^{(i)},w_{1}^{(i)}\rbrace_{i=1}^{N} & \text{drawn from } p_{\meld,1}(\phi_{1},\psi_{1}|\bsY_{1})\\
        &\lbrace\bstheta_{3}^{(i)},w_{3}^{(i)} \rbrace_{i=1}^{N} & \text{drawn from } p_{\meld,3}(\phi_{3},\psi_{3}|\bsY_{3})\\
        &\lbrace\bstheta_{5}^{(i)},w_{5}^{(i)}\rbrace_{i=1}^{N} &  \text{drawn from } p_{\meld,5}(\phi_{5},\psi_{5}|\bsY_{5}).
        \end{align*}
        Resample the particles above separately to obtain equally weighted particles \begin{align*}
            \lbrace\bstheta_{1}^{(i)},1\rbrace_{i=1}^{N},\quad \lbrace\bstheta_{3}^{(i)},1\rbrace_{i=1}^{N}, \quad \lbrace\bstheta_{5}^{(i)},1\rbrace_{i=1}^{N}.
        \end{align*} 
        
        \item[Stage two ($s_{2}$)] additionally considers Submodel 2, so 
        targets the joint subposterior 
          $p_{\meld,1,2,3,5}(\phi_{1,2},\phi_{2,3},\phi_{3,4},\phi_{4,5},\psi_{1},\psi_{2},\psi_{3},\psi_{5}|\bsY_{1},\bsY_{2},\bsY_{3},\bsY_{5})$.
        To do this, we first merge $\lbrace\phi_{1,2}^{(i)},\allowbreak 1\rbrace_{i=1}^{N}$ and $\lbrace\phi_{2,3}^{(i)},\allowbreak 1\rbrace_{i=1}^{N}$, obtaining particles for the matching $\lbrace (\phi_{1,2}^{(i)}, \phi_{2,3}^{(i)}), 1\rbrace_{i=1}^{N}$.  
        Resample the particles proportional to 
            $p_{\meld,2}(\phi_{1,2},\phi_{2,3},\psi_{2}|\bsY_{2})$
        and then a further resampling to obtain equally weighted particles, $\lbrace\bstheta_{2}^{(i)},1\rbrace_{i=1}^{N}$. Although $\phi_{3,4}$ is not involved in Submodel 2, since its resulting particles in stage one share the same indices with the ones of $\phi_{2,3}$, its resulting particles in stage two can be obtained by tracking the trajectories of $\phi_{2,3}$, obtaining $\lbrace\phi_{3,4}^{(i)},1\rbrace_{i=1}^{N}$. The details of the update process for $\phi_{3,4}$ are provided in the general algorithm in the Supplementary.
        \\
        
        \item[Stage three ($s_{3}$)] additionally considers Submodel 4, and targets the full posterior. We first merge $\lbrace\phi_{4,5}^{(i)},\allowbreak 1\rbrace_{i=1}^{N}$ from stage one and $\lbrace\phi_{3,4}^{(i)},\allowbreak 1\rbrace_{i=1}^{N}$ from stage two. The resampling importance is proportional to \begin{align*}
            p_{\meld,4}(\phi_{3,4},\phi_{4,5},\psi_{4}|\bsY_{4})\propto p_{\pool,4}(\phi_{4})\frac{p_{4}(\phi_{4},\psi_{4},\bsY_{4})}{p_{4}(\phi_{4})}.
        \end{align*} Denote the resulting particles by $\lbrace\bstheta_{4}^{(i)},w_{4}^{(i)}\rbrace_{i=1}^{N}$. By tracking the trajectories for each parameter, we can update the samples for $\phi_{1,2},\phi_{2,3},\psi_{1},\psi_{2}$ and $\psi_{3}$. The details of this update process are provided in the general algorithm in the Supplementary.\\
    \end{description}
\end{myexample}

\subsection{Extension for $\smcsq$}\label{se3.3}
$\smcsq$ algorithm is a special SMC sampler proposed by \citet{chopin2013} to perform sequential Bayesian inference in a state-space model that consists of the observation process, the state process, and static parameters. Specifically, the sampler employs a particle filter to propagate particles for latent variables and a particle MCMC (PMCMC) to rejuvenate particles for static parameters. As a result, $\smcsq$ has the advantage of implementing Bayesian inference for some models with which the standard MCMC kernels struggle, such as the \textit{stochastic volatility} (SV) model, whose latent state variables are strongly autocorrelated and static parameters are strongly correlated with the latent path. To integrate $\smcsq$ into the D\&C-melding sampler, we can simply replace the tempering SMC sampler in Algorithm \ref{algo2} by an $\smcsq$. The algorithm applied to this extension for $M=3$ is given in the Supplementary. 

\section{Simulation study}\label{se4}
In this simulation study, we consider a toy example involving 11 submodels. To test the ability of our D\&C-melding multi-stage sampler, we use several different types of submodels. Specifically, a SV model is assigned as Submodel 6, which is located in the middle of the chain model and involved at the root node in the multi-stage sampler, to demonstrate the extension of the D\&C-melding approach for utilising $\smcsq$. The remaining models are a mix of normal and non-standardised t-distribution $\tstu(\mu,\tau,\nu)$ observation submodels and hidden Markov models, arranged in a symmetric structure.  
The full model is presented below: \begin{align}\label{eq29}
    &\text{\textit{Normal distribution}:}\ \bsY_{1}\sim \normal(\phi_{1,2},\psi_{1}^{2});\quad \bsY_{11}\sim\normal(\psi_{11},\phi_{10,11}^{2});\nonumber\\
    &\text{\textit{Non-standardised t-distribution}:}\nonumber\\
    &\bsY_{2}\sim \tstu(\phi_{1,2},\phi_{2,3},\psi_{2});\quad\bsY_{3}\sim\tstu(\phi_{3,4},\phi_{2,3},\psi_{3});\nonumber\\
    &\bsY_{10}\sim\tstu(\phi_{9,10},\phi_{10,11},\psi_{10});\quad\bsY_{9}\sim\tstu(\phi_{9,10},\phi_{8,9},\psi_{9});\nonumber\\
    &\text{\textit{Hidden Markov model}:}\nonumber\\
    &\bsY_{4,t}=X_{4,t}+\delta_{4,t},\ \with\  X_{4,t+1}=\phi_{3,4}+\psi_{4}(X_{4,t}-\phi_{3,4})+\varepsilon_{4,t+1}\ \andd\ X_{4,1}=\phi_{3,4}+\varepsilon_{4,1},\nonumber\\
    &\qquad\qquad\qquad\quad\where\ \delta_{4,t}\sim\normal(0,1^{2})\ \andd\ \varepsilon_{4,t}\sim\normal(1,\phi_{4,5}^{2})\ \for\ t=1,2,...,T;\nonumber\\
    &\bsY_{5,t}=X_{5,t}+\delta_{5,t},\ \with\ X_{5,t+1}=X_{5,t}+\varepsilon_{5,t+1}\ \andd\ X_{t,1}=1+\varepsilon_{5,1},\nonumber\\
    &\qquad\qquad\qquad\quad\where\ \delta_{5,t}\sim\normal(0,\phi_{4,5}^{2})\ \andd\ \varepsilon_{5,t}\sim\normal(0,\phi_{5,6}^{2})\ \for\ t=1,2,...,T;\nonumber\\
    &\bsY_{8,t}=X_{8,t}+\delta_{8,t},\ \with\ X_{8,t+1}=\phi_{7,8}+\psi_{8}(X_{8,t}-\phi_{7,8})+\varepsilon_{8,t+1}\ \andd\ X_{8,1}=\phi_{7,8}+\varepsilon_{8,1},\nonumber\\
    &\qquad\qquad\qquad\quad\where\ \delta_{8,t}\sim\normal(0,1^{2})\ \andd\ \varepsilon_{8,t}\sim\normal(0,\phi_{8,9}^{2})\ \for\ t=1,2,...,T;\nonumber\\
    &\bsY_{7,t}=\phi_{7,8}+\delta_{7,t},\ \with\ \log(X_{7,t+1})=X_{7,t}+\phi_{6,7}+\varepsilon_{7,t+1}\ \andd\ \log(X_{7,1})=\phi_{6,7}+\varepsilon_{7,1},\nonumber\\
    &\qquad\qquad\qquad\quad\where\ \delta_{7,t}\sim(0,X_{7,t}^{2})\ \andd\ \varepsilon_{7,t}\sim\normal(0,0.1^{2})\ \for\ t=1,2,...,T;\nonumber\\
    &\text{\textit{Stochastic volatility model}:}\nonumber\\
    &\bsY_{6,t}=\exp\lr\frac{X_{6,t}}{2}\rr\delta_{6,t},\ \with\ X_{6,t+1}=\phi_{6,7}+\psi_{6}(X_{6,t}-\phi_{6,7})+\varepsilon_{6,t+1}\ \andd\nonumber\\ &\qquad\qquad\qquad\qquad\qquad\qquad\qquad X_{6,1}=\phi_{6,7}+\varepsilon_{6,1},\nonumber\\
    &\qquad\qquad\qquad\quad\where\ \delta_{6,t}\sim\normal(0,1^{2})\ \andd\ \varepsilon_{6,t}\sim\normal(0,\phi_{5,6}^{2})\ \for\ t=1,2,...,T;\nonumber\\
    &\with\nonumber\\
    &\phi_{1,2}\sim\normal(10,1^{2}),\ \phi_{2,3}\sim\Ga(5,1),\ \phi_{3,4}\sim\normal(1,10^{2}),\ \phi_{4,5}\sim\Ga(5,3),\nonumber\\
    &\psi_{1}\sim\Ga(1,2),\ \psi_{2}=5,\ \psi_{3}=12,\ \psi_{4}=0.87;\nonumber\\
    &\phi_{10,11}\sim\Ga(4,2),\ \phi_{9,10}\sim\normal(7,5^{2}),\ \phi_{8,9}\sim\Ga(12,3),\ \phi_{7,8}\sim\normal(3,1^{2}),\nonumber\\
    &\psi_{11}\sim\normal(4,0.8^{2}),\ \psi_{10}=21,\ \psi_{9}=2,\ \psi_{8}=0.93;\nonumber\\
    &\phi_{5,6}=0.178,\ \phi_{6,7}=-1.024,\ \psi_{6}=0.9702.
\end{align} The true values of $\phi_{5,6}, \phi_{6,7}$ and $\psi_{6}$ match  the stochastic volatility examples in \citet{pitt1999} and \citet{chopin2020}, where these parameters were estimated using an auxiliary particle filter.
We assume $50$ observations are available for the observable quantities in each submodel, except for the state-space submodels (Submodels 4, 5, 6, 7 and 8), for which we assume each of the $T=10$ time points is associated with a single observation.

The priors assigned for $\phi_{1,2},\phi_{3,4},\phi_{6,7},\phi_{7,8},\phi_{9,10}$ and $\psi_{11}$ are $\normal(0,2^{2})$; for $\phi_{2,3},\phi_{4,5},\phi_{5,6}$ and $\phi_{10,11}$ are $\Ga(2,2)$; for $\phi_{8,9}$ is $\Ga(12,2)$; for $\psi_{2},\psi_{3},\psi_{9}$ and $\psi_{10}$ are $\MultN(30,1/30)$; and for $\psi_{4},\psi_{6}$ and $\psi_{8}$ are $\Betatext(9,1)$. We employ the logarithmic pooling for the pooled prior with equal weights $\lambda=(1/2,1/2,...,1/2)$. To increase difficulty, these priors do not align fully with the true values.

We estimate the parameters using three approaches: D\&C-melding approach; D\&C-melding approach with $\smcsq$, in which $\smcsq$ is only applied at the root node; and standard MCMC. In both D\&C-melding-based approaches, we used $N=10,000$ particles. The standard D\&C-melding approach uses an MCMC kernel in the SMC sampler with $5$ iterations at all nodes, and with the annealing process temperature increment $\alpha_{j}-\alpha_{j-1}=0.2$, for all $j$ in \eqref{eq:weight-update}. For D\&C-melding with $\smcsq$, we used 50 particles for the latent variables in $\smcsq$. For the MCMC sampler, we used 10,000 iterations, discarding 5,000 samples as burn-in.
For computational efficiency, in both D\&C-melding-based samplers, some key computations are written in \texttt{C++} \citep{stroustrup2013} and integrated into \texttt{R} via the packages \texttt{Rcpp} \citep{rcpp} and \texttt{RcppArmadillo} \citep{rcpparmadillo}. The MCMC kernel used by the SMC sampler is written in \texttt{BUGS} \citep{lunn2009} and implemented via \texttt{rjags} \citep{rjags}. The PMCMC kernel used in $\smcsq$ is implemented via \texttt{NIMBLE} \citep{devalpine2017} and its \texttt{R} package \texttt{nimbleSMC} \citep{nimblesmc}. The MCMC sampler is written in \texttt{BUGS} and implemented via \texttt{rjags}. We also use these software programs and packages to write the samplers for the case study of an owl example in Section \ref{se5}.

To assess performance of each method, we conducted $500$ replicates. As performance metrics, we consider the average mean squared error (MSE) across all replicates for the pointwise estimate, as well as the empirical coverage and the average width of the $90\%$ credible interval. Since $\phi_{5,6}$ and $\phi_{6,7}$ are the common parameters shared by the middle three submodels in the chain model, they are the priority in this example. Table \ref{tab1} presents the estimates for these two parameters.

\begin{table}[t]
    \scriptsize
    \noindent\makebox[\linewidth][c]{%
      \begin{tabular}{ccccc}
        \hline
        \multicolumn{2}{c}{} & \multicolumn{2}{c}{$\phi_{5,6}$} & \multicolumn{1}{c}{}\\
        \cline{2-5}
        & Stage one & D\&C-melding & MCMC & D\&C-melding $\smcsq$\\
        MSE & 0.535 & 0.486 & 0.190 & 0.092\\
        Coverage & 0.772 & 0.836 & 0.836 & 0.992\\
        CI width & 1.405 & 1.371 & 0.883 & 0.760\\
        \cline{2-5}
        \multicolumn{2}{c}{} & \multicolumn{2}{c}{$\phi_{6,7}$} & \multicolumn{1}{c}{}\\
        \cline{2-5}
        MSE & 0.022 & 0.021 & 0.020 & 0.020\\
        Coverage & 0.888 & 0.892 & 0.886 & 0.910\\
        CI width & 0.301 & 0.301 & 0.297 & 0.295\\
        \cline{2-5}
        \multicolumn{2}{c}{} & \multicolumn{2}{c}{$\phi_{8,9}$} & \multicolumn{1}{c}{}\\
        \cline{2-5}
        MSE & 2.689 & 2.633 & 1.194 & 2.574\\
        Coverage & 0.888 & 0.892 & 0.886 & 0.888\\
        CI width & 3.717 & 3.716 & 2.526 & 3.642\\
        \hline
      \end{tabular}%
    }
    \caption{Average MSE, empirical coverage and average width of $90\%$ credible interval of the estimates for $\phi_{5,6}, \phi_{6,7}$ and $\phi_{8,9}$, respectively, by the D\&C-melding approach, its combination with $\smcsq$ and the full MCMC. In addition, the measurement for the estimates obtained in stage one by the D\&C-melding approach is also provided.}
    \label{tab1}
\end{table}

\begin{figure}
    \centering
    \includegraphics[width=.8\textwidth, height=.32\textheight]{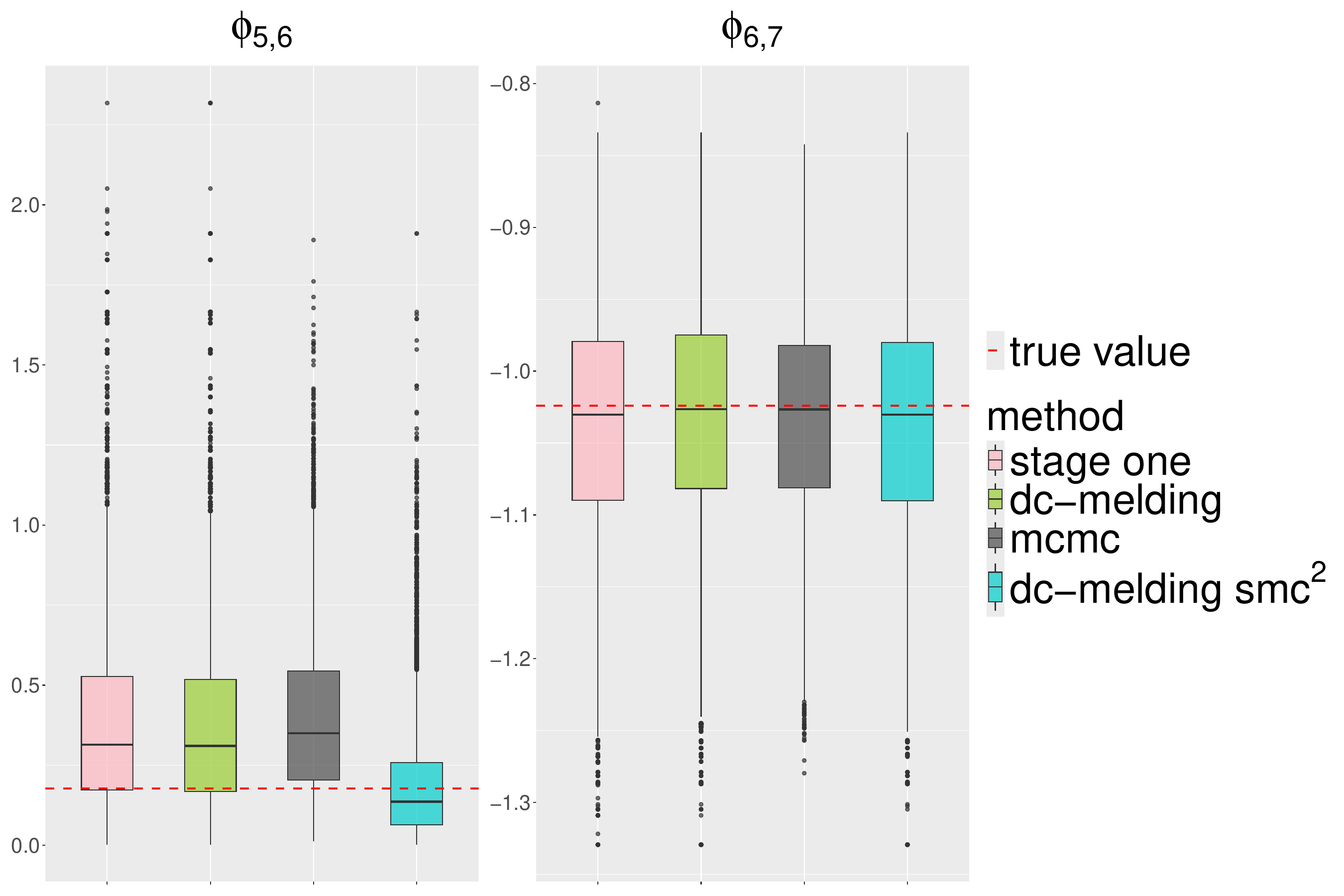}
    \caption{The estimates of $\phi_{5,6}$ and $\phi_{6,7}$ from one randomly-selected replicate.}
    \label{Fig_phi56_67}
\end{figure}

Due to the poor mixing of standard MCMC samplers when state parameters and static parameters are strongly correlated \citep{chopin2020}, the MCMC sampler and the D\&C-melding with a tempering SMC sampler provide poor estimates of $\phi_{5,6}$, the scaling parameter of the latent variables in the model. In fact, using a tempering SMC sampler in the D\&C-melding performs even worse than the MCMC sampler, although there is an obvious improvement in the estimate compared to stage one. This may be due to the weak MCMC rejuvenation in the SMC sampler, which cannot explore the latent path sufficiently. However, performance is improved significantly by replacing the tempering SMC sampler with $\smcsq$, which is well-suited for sampling from SV models. Unsurprisingly, this extension also substantially outperforms standard MCMC in both pointwise and uncertainty estimates. All methods exhibit similarly good performance in estimating the unconditional mean $\phi_{6,7}$, which has a lower correlation with the latent path. These results are illustrated by Figure \ref{Fig_phi56_67}, which presents the estimates for $\phi_{5,6}$ and $\phi_{6,7}$ for a single randomly-selected replicate.  

The estimate for $\phi_{8,9}$ is also provided in Table \ref{tab1}, and the remaining common parameters are presented in the Supplementary.
For most parameters, in both D\&C-melding samplers, the estimation can benefit from including all information of the full model, compared to the results from the first stage. 
The performance of both D\&C-melding samplers is similar to the gold standard MCMC sampler for most parameters. The only exception is $\phi_{8,9}$, for which the stage one estimate is poor. Similar issues may arise in multi-stage samplers when the posterior samples from the first stage are poor.

\section{Little owls example}\label{se5}
An important case study considered by \cite{manderson2023} is an integrated population model (IPM) for little owls. This example drew their attention because of the equivalence of the chained Markov melding model and the IPM for a certain choice of pooling function and pooling weights. Hence, the posterior from the IPM could be a benchmark for their multi-stage sampler. For comparison, we use our method to draw samples from the same IPM as \cite{manderson2023}, which has been considered by \cite{finke2019} and is a variant of the original model.\\

We first introduce some notation. The data and parameters are stratified into two age groups $a\in\lbrace J,F\rbrace$, where $J$ denotes juvenile owls (less than one year old) and $A$ adults; and two sex groups $s\in\lbrace M,F\rbrace$, where $M$ denotes male owls and $F$ female. Data are observed annually at times $t\in\lbrace1,...,T\rbrace$ with $T=25$. The main parameters of the model are the following: \begin{itemize}
    \item $\delta_{a,s,t}$: probability of an owl of sex $s$ from age group $a$ surviving from time $t$ to $t+1$;

    \item $\pi_{s,t+1}$: probability of a previously captured owl of sex $s$ being recaptured at time $t+1$;

    \item $\eta_{t}$: immigration rate of female owls from time $t-1$ to time $t$;

    \item $\rho$: reproductive rate.
\end{itemize}

\subsection{Capture-recapture model $\text{p}_{1}$}\label{se5.1}

The capture-recapture data consist of the number of captured owls released at time $t$ and then recaptured at time $u$ with $t+1\leq u\leq T$, or not recaptured before the study was completed with $u=T+1$. Specifically, let $M_{a,s,t,u}$ be the number of owls of sex $s$ and age $a$ released at time $t$ and recaptured at time $u$. The data is then aggregated into the form of age- and sex-specific matrices $\bsM_{a,s}:=\lbrace M_{a,s,t,u}\rbrace$ with $T$ rows corresponding to the released time and $T+1$ columns corresponding to the recaptured time. Note that $M_{a,s,t,u}=0$ if $u\leq t$. Let $R_{a,s,t}:=\sum_{u=1}^{T+1}M_{a,s,t,u}$ be the number of owls of sex $s$ and age $a$ released at time $t$. The number of subsequent recaptured owls is assumed to follow a multinomial distribution $(M_{a,s,t,1},...,M_{a,s,t,T+1})\sim\text{Multinomial}(R_{a,s,t}, \bsQ_{a,s,t})$, with probabilities $\bsQ_{a,s,t}=(Q_{a,s,t,1},...,Q_{a,s,t,T+1})$ given by \begin{align*}
    Q_{a,s,t,u}=\begin{dcases}
        0,\qquad\qquad&\text{for}\ u=1,...,t,\\
        \delta_{a,s,t}\pi_{s,u}\prod\nolimits_{r=t+1}^{u-1}\delta_{a,s,r}(1-\pi_{s,t}), &\text{for}\ u=t+1,...,T,\\
        1-\sum\nolimits_{r=1}^{T}Q_{a,s,t,r}, &\text{for}\ u=T+1.
    \end{dcases}
\end{align*}

\subsection{Count model $\text{p}_{2}$}
Let $x_{J,t}$ and $x_{A,t}$ be the true population size of juvenile females and adult females at time $t$, respectively. The population abundance is modelled as 
    $y_{t}|x_{t}\sim\Pois(x_{t})$,
where $x_{t}=x_{J,t}+x_{A,t}$ is the true population size for juvenile and adult females at time $t$. Let $\surv_{t}$ be the number of female adults who survive from time $t-1$ to time $t$, and $\imm_{t}$ be the number of female adults who immigrate in this period. The latent variable $x_{t}$ is specified by \begin{align}\label{eq30}
    &x_{J,t}|x_{t-1},\rho,\delta_{J,F,t-1}\sim\Pois\lr x_{t-1}\frac{\rho}{2}\delta_{J,F,t-1}\rr,\nonumber\\
    &\surv_{t}|x_{t-1},\delta_{A,F,t-1}\sim\Bin\lr x_{t-1},\delta_{A,F,t-1}\rr,\nonumber\\
    &\imm_{t}|x_{t-1},\eta_{t}\sim\Pois(x_{t-1},\eta_{t}),\nonumber\\
    &x_{A,t}=\surv_{t}+\imm_{t}.
\end{align} The priors on the initial population sizes $x_{J,1}$ and $x_{J,1}$ are assumed to be discrete uniform distributions on $\lbrace0,1,...,50\rbrace$. Note that the Poisson and binomial distributions have zero point mass if $x_{t-1}=0$.

\subsection{Fecundity model $\text{p}_{3}$}
The data in the fecundity model consist of $N_{t}$, the number of breeding females at time $t$, and $n_{t}$, the number of chicks produced that survive and leave the nest, for which we assume 
    $n_{t}\sim\Pois(N_{t}\rho)$.

\subsection{Specification of parameters and priors}
The parameterisation for the time-dependent quantities considered by \cite{manderson2023} is \begin{align}\label{eq31}
    \logit(\eta_{a,s,t})=\alpha_{0}+\alpha_{1}\bbInd(s=M)+\alpha_{2}\bbInd(a=A),\quad\log(\eta_{t})=\alpha_{6},\nonumber\\
    \logit(\pi_{s,u})=\alpha_{4}\bbInd(s=M)+\alpha_{5,u}, \quad\for\ u=2,,...,T.
\end{align} Based on this parameterisation, with the specification for the chain model, the common parameters are $\phi_{1,2}=(\alpha_{0},\alpha_{2})$ and $\phi_{2,3}=\rho$, and the unique parameters for each of the submodels are $\psi_{1}=(\alpha_{1},\alpha_{4},(\alpha_{5,u})_{u=1}^{T})$, $\psi_{2}=(\alpha_{6}, x_{J,t},\surv_{t},\imm_{t})$ and $\psi_{3}=\varnothing$. It should be noted that (\ref{eq31}) ignores the parameter $\alpha_{3}$ compared to the model variant in \cite{finke2019}.

Let $\bsalpha=(\alpha_{0},\alpha_{1},\alpha_{2},\alpha_{4},\alpha_{6})$. The priors specified for $\bsalpha$ in $\text{p}_{1}$ and $\text{p}_{2}$ are independent Normal $N(0,2^{2})$ truncated to $[-10,10]$. The time-dependent parameter $\alpha_{5,u}$ has the same truncated Normal prior. A $\Unif(0,10)$ prior is specified for $\rho$ in $\text{p}_{1}$ and $\text{p}_{2}$. For comparability, we employ a logarithmic pooled prior with $\lambda=(1/2,1/2,1/2)$ for $(\phi_{1,2},\phi_{2,3})$, which ensures that the melded posterior is the same as the original IPM.

\subsection{Results}
The advantages of the sampler described in Algorithm \ref{algo2} allow us to draw particles from $\text{p}_{1}$ and $\text{p}_{3}$ in parallel, and then use the merged particles targeting the posterior $p(\bsphi,\bspsi|\bsY)$ in $\text{p}_{2}$.
We use $N=16,000$ particles for each parameter. In stage one, the SMC sampler employs $\alpha=0.2$ as the temperature parameter and implements $10$ iterations in the MCMC kernel for both submodels. In stage two, the SMC sampler employs $\alpha=0.5$ and $20$ iterations in the MCMC kernel. 

We compare the performance of our sampler to two main samplers: the two-stage sampler with the standard Markov melding in Section \ref{se2.1.4}, which employs the same pooled prior as our sampler for comparison; and the standard MCMC sampler, which directly samples from the original IPM. In addition, we also consider a two-stage approach commonly used for model integration, which uses the pointwise estimates, such as mean, median or mode, of $\phi_{1,2}$ and $\phi_{2,3}$ obtained from the subposteriors to target the full distribution in stage two.

\begin{figure}[tb]
    \centering
    \begin{subfigure}[c]{1\textwidth}
        \centering
        \includegraphics[width=.65\textwidth, height=0.25\textheight]{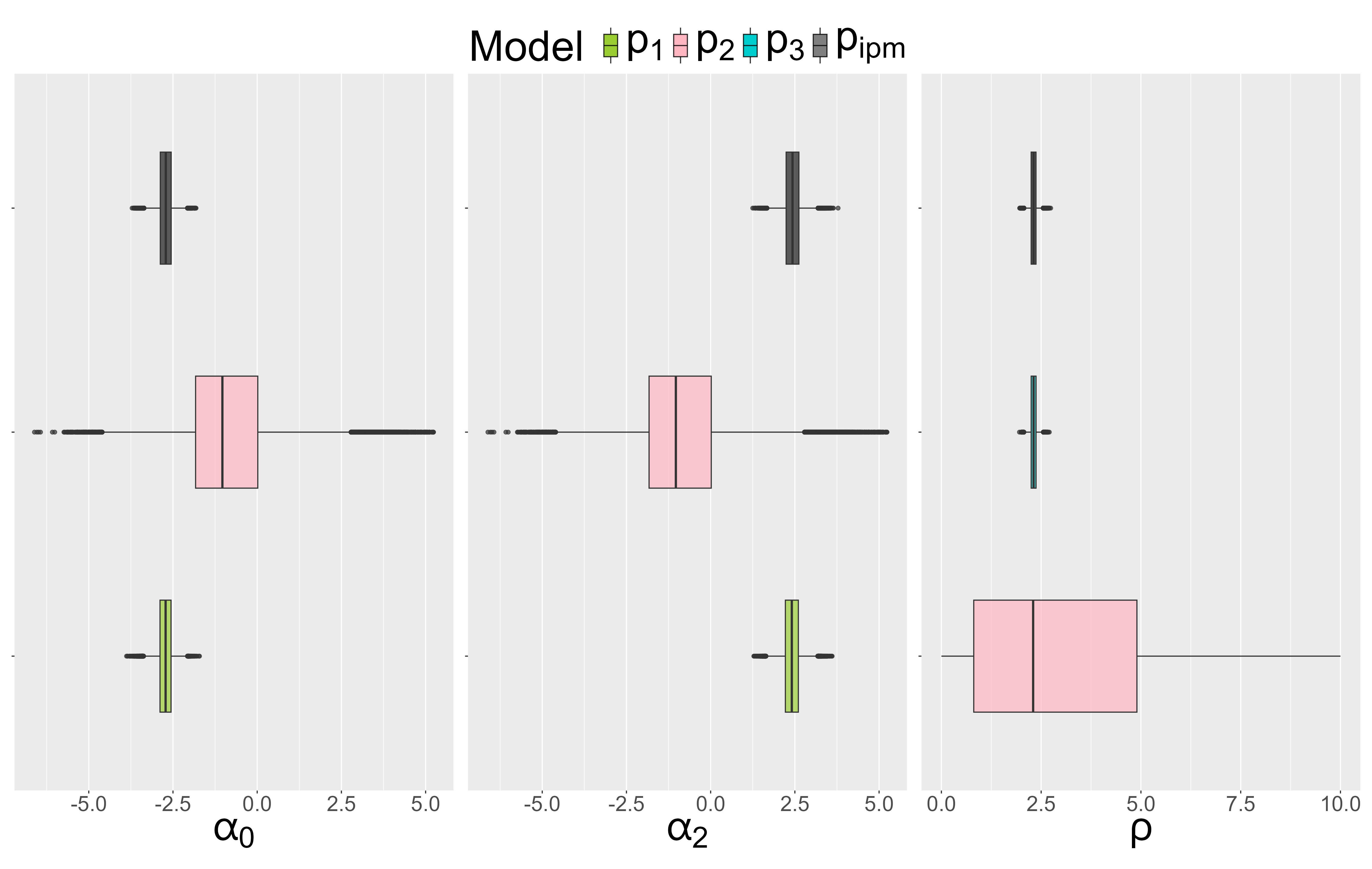}
    \end{subfigure}

    \begin{subfigure}[c]{1\textwidth}
        \centering
        \includegraphics[width=.65\textwidth, height=0.25\textheight]{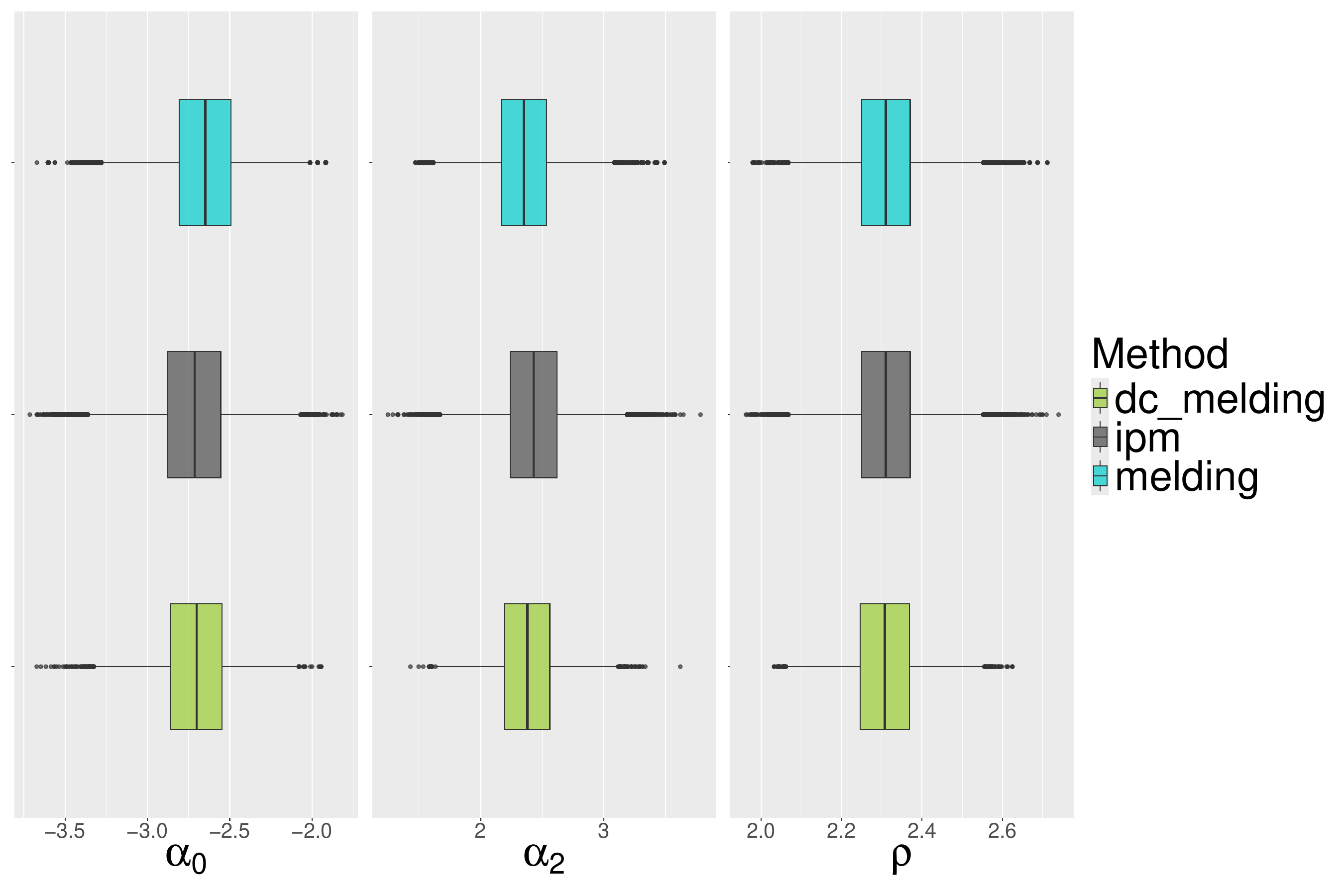}
        \vspace{0\baselineskip}
    \end{subfigure}
    \caption{Top row: the subposteriors for the common parameters $\alpha_{0},\alpha_{2}$ and $\rho$ from Submodels $p_{1}, p_{2}$ and $p_{3}$, respectively, and the posteriors for the same parameters from the original integrated population model (IPM). Bottom row: the posteriors for the common parameters from the original IPM (repeated from (a)), and the melded models with the original Markov melding and the D\&C-melding, respectively.}
    \label{fig_owls_common}
\end{figure}

Figure \ref{fig_owls_common} shows the posterior credible interval for the common parameters of the individual submodels, the melded models estimated by the original Markov melding and the D\&C-melding, respectively, and the original IPM. The wide intervals in the top row in Figure \ref{fig_owls_common} for the count model $p_{2}$ indicate that this model provides little information about $\alpha_{0},\alpha_{2}$ and $\rho$. The results for the full model (Figure \ref{fig_owls_common}; bottom row) show that the D\&C-melding sampler produces similar posterior estimates to the original IPM and the original Markov melding sampler for the common parameters.

\begin{figure}[tb]
    \centering
    \begin{subfigure}[t]{0.48\textwidth}
        \centering
        \includegraphics[width=\linewidth]{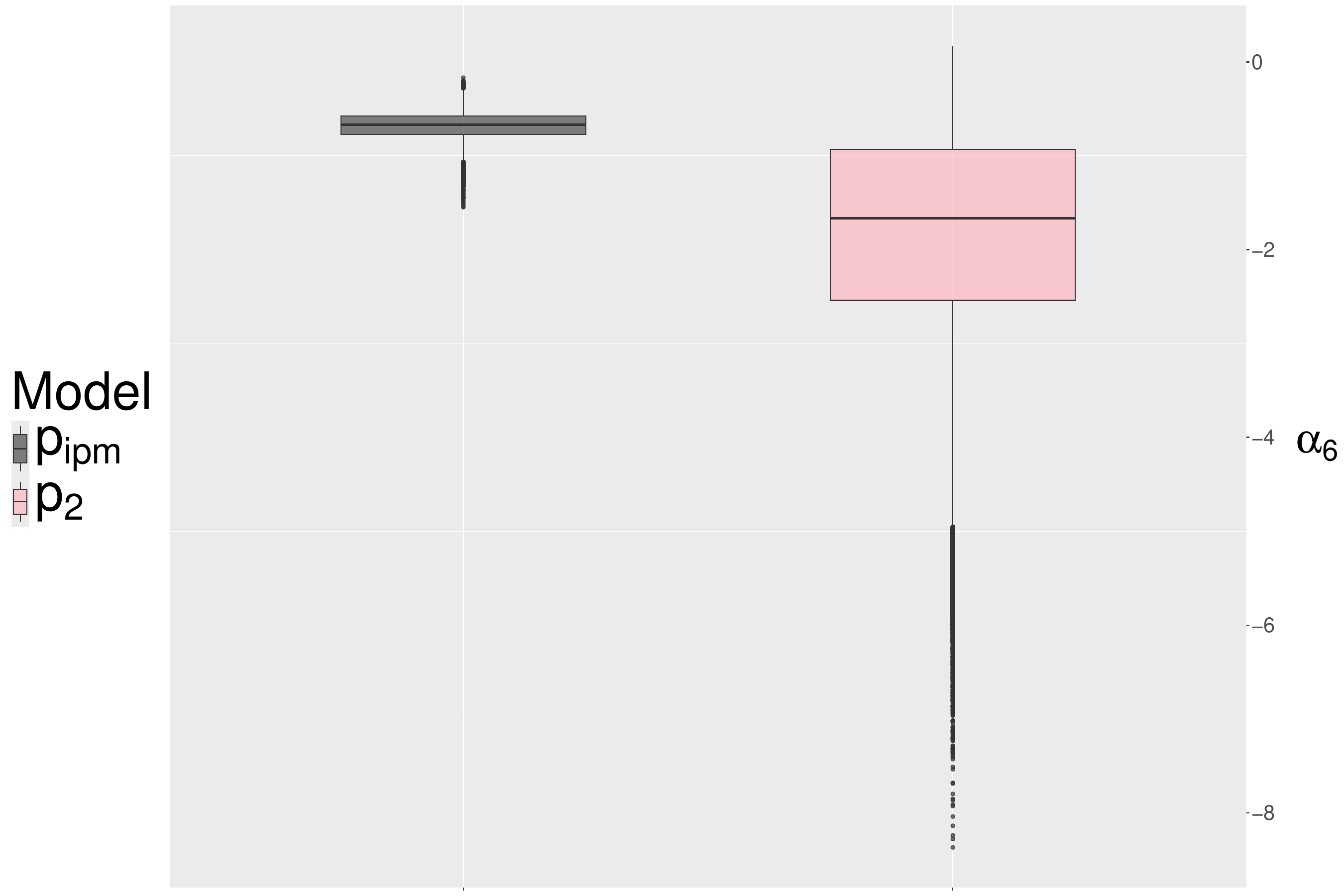}
    \end{subfigure}
    \hfill
    \begin{subfigure}[t]{0.48\textwidth}
        \centering
        \includegraphics[width=\linewidth]{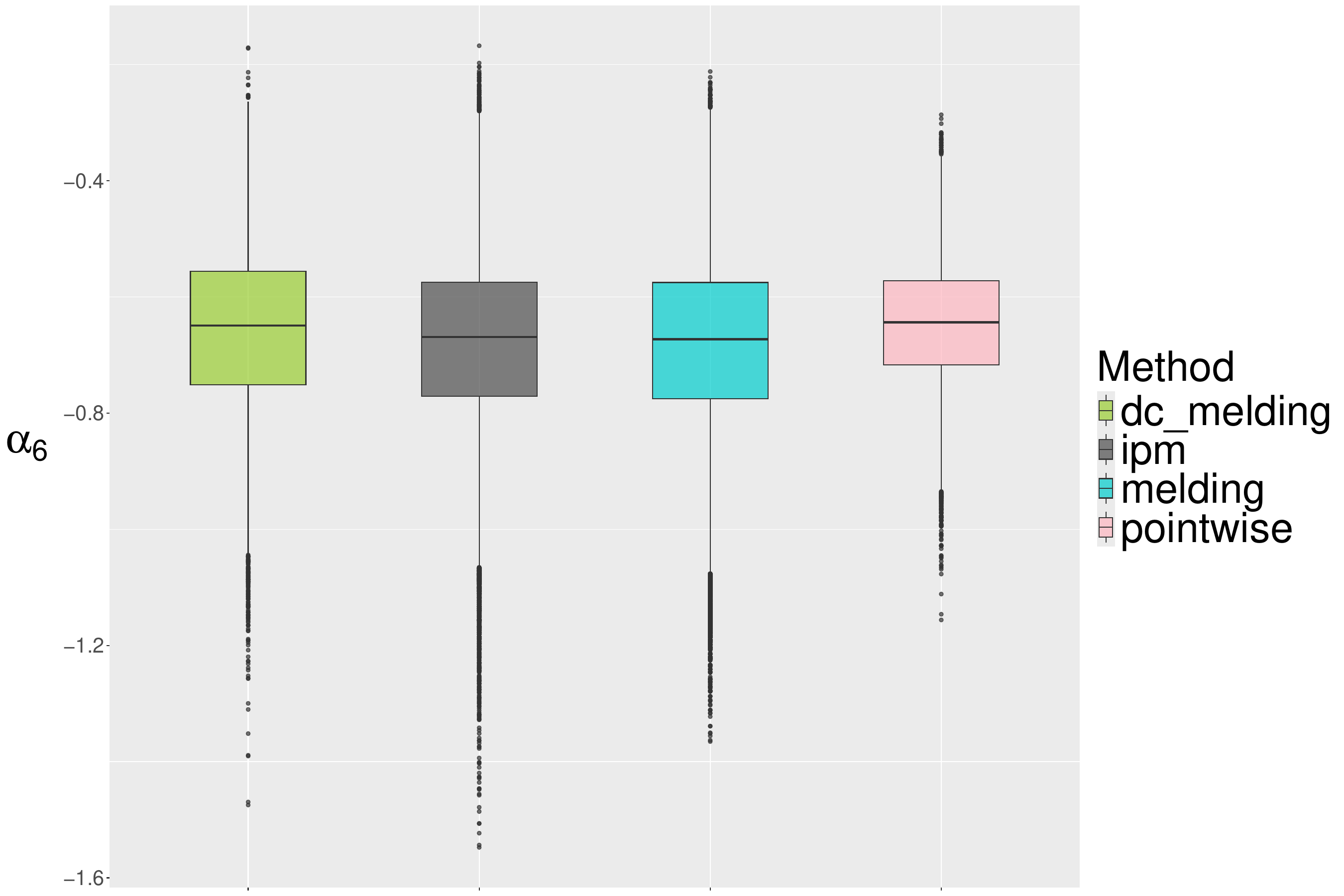}
        \vspace{0\baselineskip}
    \end{subfigure}
    \caption{Left: the subposterior for $\alpha_{6}$ from Submodel $p_{2}$ and the corresponding posterior from the original IPM. Right: posteriors for $\alpha_{6}$ from the original IPM (repeated from (a)), the melded models with the original Markov melding and the D\&C-melding, respectively, and the model fitted by the pointwise estimate.}
    \label{Fig_owls_alpha6}
\end{figure}

Similar results are also obtained for $\alpha_{6}$, which appears only in model $p_{2}$. As for the common parameters, the count model $p_{2}$ provides little information about $\alpha_{6}$ (Figure~\ref{Fig_owls_alpha6}; top row), and additional information from other models is required. The bottom row shows that the estimates obtained from those the three sampling approaches are similar. In addition, as expected, using the fixed pointwise estimates of $\alpha_{0},\alpha_{2}$ and $\rho$ in $p_{2}$ results a (small) underestimation  the uncertainty of $\alpha_{6}$.

\section{Conclusion}\label{se6}

This paper proposes a novel approach for Bayesian inference in the chained Markov melded model. Specifically, we introduce a multi-stage sampler, D\&C-melding, that targets the full posterior distribution by progressively merging samples obtained from submodels. This approach enables separate sampling from individual submodels rather than requiring simultaneous sampling from the full joint model, and allows for fully parallelised sampling at each stage. 

The proposed multi-stage sampler builds on the two-stage sampler for the chained model with three submodels introduced by \citet{manderson2023} and incorporates the divide-and-conquer sequential Monte Carlo (D\&C-SMC) framework of \citet{lindsten2017} by demonstrating that the sampler structure induces a tree decomposition of the chained model. We first present the sampler for the case of three submodels, which may be regarded as a combination of the original two-stage sampler \citep{manderson2023} and the D\&C-SMC method. We then generalise the sampler to accommodate an arbitrary number of submodels. It is shown that the structure of the sampler depends on the number of submodels and differs between odd and even cases. We also derive the number of stages required by the sampler under the different settings. Furthermore, owing to the SMC framework employed, we present an extension that incorporates the $\smcsq$ approach.  

We demonstrate the application of the sampler through both a toy example and a real case study. The toy example consists of 11 submodels of various types, including a stochastic volatility (SV) model. The results indicate that the multi-stage sampler performs comparably to standard full-MCMC for most parameters. In particular, combining the sampler with $\smcsq$ substantially improves performance for the SV model and can outperform full-MCMC. The example also highlights the importance of incorporating full information when combining models, with parameter estimates generally improving as additional information is introduced. Similar findings are observed in the owls example. Although the multi-stage sampler does not outperform full-MCMC in these examples overall, it offers a more flexible alternative for sampling from complex joint models, and we expect this flexibility to become increasingly advantageous as the number of submodels grows. 

However, the poor performance for $\phi_{8,9}$ in the toy example suggests that the multi-stage sampler may be sensitive to the quality of samples obtained in the first stage. The sampler fixes the first-stage samples and only resamples the corresponding particles, without propagating new ones. Consequently, poor samples generated in the first stage cannot be substantially corrected in later stages. One intuitive solution is to introduce a rejuvenation step within the SMC procedure for particles generated in the first stage, with optional resampling of the newly propagated particles. Nevertheless, this modification requires careful consideration, as rejuvenation may also affect parameters that are not included in the current submodel. For instance, in the toy example, updating $\phi_{8,9}$ in stage three using Submodel 8 also affects the update of $\phi_{6,7}$, despite the latter not being involved in that submodel. 

Another possible approach is to modify the selection of submodels sampled in stage one. In one of the four cases for the number of submodels discussed in Section \ref{se3.2}, the proposed multi-stage sampler samples from a fixed selection of submodels in the first stage. Future work may relax this restriction and allow the sampler to select the stage-one submodels more flexibly. This improvement can enable the sampler to begin with submodels from which sampling is relatively straightforward, potentially improving the quality of the initial samples obtained in stage one.

Datasets from different hospitals are often subject to privacy constraints and therefore cannot be jointly analysed, rendering Bayesian inference under the full model infeasible. In such settings, our multi-stage sampler provides an effective approach for conducting inference while preserving data privacy. However, practical applications often involve more complex model structures than the chained model. For instance, three submodels may share a common parameter, denoted by $\phi_{1,2,3}$, while two of these submodels additionally share another parameter, $\phi_{i,j}$, specific to Submodels $i$ and $j$. Consequently, there is a need to generalise the Markov melding to accommodate arbitrary structures of DAGs in which submodels are connected through multiple shared parameters.
To achieve this, it is necessary to develop a generalised \textit{marginal replacement} that extends the original one proposed by \citet{goudie2019}, potentially incorporating ideas from the \textit{Markov combination} \citep{dawid1993, massa2010, byrne2015}. In addition, an appropriate pooling function is required for constructing the pooled prior distribution. Logarithmic pooling is an initial choice because it can account for dependence structures among submodels. Alternative approaches, including linear pooling and dictatorial pooling, may also be considered \citep{goudie2019}. Furthermore, an efficient sampling algorithm is required, either in the form of a multi-stage sampler that allows parallel computation or a sequential sampler that performs inference sequentially across submodels.

\backmatter

\bmhead{Acknowledgements}


YL and RJBG are funded by UKRI Medical Research Council (MRC) (programme codes MC\_UU\_00002/20 and MC\_UU\_00040/04). 
For the purpose of open access, the author has applied a Creative Commons Attribution (CC BY) licence to any Author Accepted Manuscript version arising.

\FloatBarrier
\bibliography{sn-bibliography}

@article{kuntz2024,
  title = {The Divide-and-Conquer Sequential {{Monte Carlo}} Algorithm: {{Theoretical}} Properties and Limit Theorems},
  shorttitle = {The Divide-and-Conquer Sequential {{Monte Carlo}} Algorithm},
  author = {Kuntz, Juan and Crucinio, Francesca R. and Johansen, Adam M.},
  year = 2024,
  month = feb,
  journal = {The Annals of Applied Probability},
  volume = {34},
  number = {1B},
  pages = {1469--1523},
  publisher = {Institute of Mathematical Statistics},
  issn = {1050-5164, 2168-8737},
  doi = {10.1214/23-AAP1996}
}

@article{manderson2022,
  title = {A Numerically Stable Algorithm for Integrating {{Bayesian}} Models Using {{Markov}} Melding},
  author = {Manderson, Andrew A. and Goudie, Robert J. B.},
  year = 2022,
  month = feb,
  journal = {Statistics and Computing},
  volume = {32},
  number = {2},
  pages = {24},
  issn = {1573-1375},
  doi = {10.1007/s11222-022-10086-2}
}

@article{birrell2025,
  title = {Real-Time Modelling of the {{SARS-CoV-2}} Pandemic in {E}ngland 2020--2023: A Challenging Data Integration},
  author = {Birrell, Paul J and Blake, Joshua and Kandiah, Joel and Alexopoulos, Angelos and {van Leeuwen}, Edwin and Pouwels, Koen B and Ghosh, Sanmitra and Starr, Colin and Walker, Ann Sarah and House, Thomas A and Gay, Nigel and Finnie, Thomas and Gent, Nick and Charlett, Andr{\'e} and De Angelis, Daniela},
  year = {2025},
  journal = {Journal of the Royal Statistical Society Series A: Statistics in Society},
  doi = {10.1093/jrsssa/qnaf030}
}

@article{vaneeMeldedIntegratedPopulation2025,
  title = {Melded {{Integrated Population Models}}},
  author = {Van Ee, Justin J. and Hagen, Christian A. and Pavlacky, David C. and Haukos, David A. and Lawrence, Andrew J. and Tanner, Ashley M. and Grisham, Blake A. and Fricke, Kent A. and Rossi, Liza G. and Beauprez, Grant M. and Kuklinski, Kurt E. and Martin, Russell L. and Koslovsky, Matthew D. and Rintz, Troy B. and Hooten, Mevin B.},
  year = 2025,
  month = sep,
  journal = {Journal of Agricultural, Biological and Environmental Statistics},
  volume = {30},
  number = {3},
  pages = {769--799},
  issn = {1537-2693},
  doi = {10.1007/s13253-024-00620-2}
}

@article{vaneeMeldingWildlifeSurveys2023,
  title = {Melding Wildlife Surveys to Improve Conservation Inference},
  author = {Van Ee, Justin J. and Hagen, Christian A. and Jr., David C. Pavlacky and Fricke, Kent A. and Koslovsky, Matthew D. and Hooten, Mevin B.},
  year = 2023,
  journal = {Biometrics},
  volume = {79},
  number = {4},
  pages = {3941--3953},
  issn = {1541-0420},
  doi = {10.1111/biom.13903}
}

@article{nicholsonInteroperabilityStatisticalModels2022,
  title = {Interoperability of Statistical Models in Pandemic Preparedness: Principles and Reality},
  shorttitle = {Interoperability of {{Statistical Models}} in {{Pandemic Preparedness}}},
  author = {Nicholson, George and Blangiardo, Marta and Briers, Mark and Diggle, Peter J. and Fjelde, Tor Erlend and Ge, Hong and Goudie, Robert J. B. and Jersakova, Radka and King, Ruairidh E. and Lehmann, Brieuc C. L. and Mallon, Ann-Marie and Padellini, Tullia and Teh, Yee Whye and Holmes, Chris and Richardson, Sylvia},
  year = 2022,
  month = may,
  journal = {Statistical Science},
  volume = {37},
  number = {2},
  pages = {183--206},
  publisher = {Institute of Mathematical Statistics},
  issn = {0883-4237, 2168-8745},
  doi = {10.1214/22-STS854}
}

@article{goudie2019,
	author = {R. J. B. Goudie and A. M. Presanis and D. Lunn and D. {De Angelis} and L. Wernisch},
	title = {Joining and splitting models with {M}arkov melding},
	journal = {Bayesian Analysis},
	year = {2019},
	volume = {14},
        issue = {1},
	pages = {81--109},
	note = {doi: \url{10.1214/18-BA1104}}
}

@article{dawid1993,
	author = {A. P. Dawid and S. L. Lauritzen},
	title = {Hyper {M}arkov laws in the statistical analysis of decomposable graphical models},
	journal = {Annals of Statistics},
	year = {1993},
	volume = {21},
        issue = {3},
	pages = {1272--1317},
	doi = {10.1214/aos/1176349260}
}

@inproceedings{massa2010,
	author = {M. S. Massa and S. L. Lauritzen},
	title = {Combining statistical models},
        editor = {M. A. G. Viana and H. P. Wynn},
	booktitle = {Algebraic Methods in Statistics and Probability II},
	year = {2010},
	volume = {516},
	pages = {239--260},
        publisher = {American Mathematical Society},
        address = {Providence, Rhode Island},
	doi = {10.1090/conm/516/10180}
}

@article{byrne2015,
	author = {S. Byrne and A. P. Dawid},
	title = {Structural {M}arkov graph laws for {B}ayesian model uncertainty},
	journal = {Annals of Statistics},
	year = {2015},
	volume = {43},
        issue = {4},
	pages = {1647--1681},
	doi = {10.1214/15-AOS1319}
}

@article{manderson2023,
	author = {A. Manderson and R. J. B. Goudie},
	title = {Combining chains of {B}ayesian models with
{M}arkov melding},
	journal = {Bayesian Analysis},
	year = {2023},
	volume = {18},
    issue = {3},
	pages = {807--840},
	doi = {10.1214/22-BA1327}
}

@article{tierney1994,
	author = {L. Tierney},
	title = {Markov chains for exploring posterior distributions},
	journal = {Ann. Stat.},
	year = {1994},
	volume = {22(4)},
	pages = {1701--1762},
	doi = {10.1214/aos/1176325750}
}

@article{lindsten2017,
	author = {F. Lindsten and A. M. Johansen and C. A. Naesseth and B. Kirkpatrick and T. B. Sch\"{o}n and J. A. D. Aston and A. Bouchard-C\^{o}t\'{e}},
	title = {Divide-and-conquer with sequential {M}onte {C}arlo},
	journal = {Journal of Computational and Graphical Statistics},
	year = {2017},
	volume = {26},
	pages = {445--458},
	doi = {0.1080/10618600.2016.1237363}
}

@article{lin2005,
	author = {M. T. Lin and J. L. Zhang and Q. Cheng and R. Chen},
	title = {Independent particle filters},
	journal = {Journal of American Statistical Association},
	year = {2005},
	volume = {100},
	pages = {1412--1421},
	doi = {10.1198/016214505000000349}
}

@article{pitt1999,
	author = {M. K. Pitt and N. Shephard},
	title = {Filtering via simulation: auxiliary particle filters},
	journal = {Journal of American Statistical Association},
	year = {1999},
	volume = {94},
	pages = {590--599},
	doi = {10.2307/2670179}
}

@article{finke2019,
	author = {A. Finke and R. King and A. Beskos and P. Dellaportas},
	title = {Effcient sequential {M}onte {C}arlo algorithms for integrated population models},
	journal = {Journal of Agricultural, Biological and Environmental Statistics},
	year = {2019},
	volume = {24},
	pages = {204--224},
	doi = {10.1007/s13253-018-00349-9}
}

@article{lunn2009,
	author = {D. Lunn and D. Spiegelhalter and A. Thomas and N. Best},
	title = {The {BUG} project: evolution, critique and future direction},
	journal = {Statistics in Medicine},
	year = {2009},
	volume = {28(25)},
	pages = {3049--3067},
	doi = {10.1002/sim.3680}
}

@book{stroustrup2013,
    author = {B. Stroustrup},
    title = {The C++ Programming Language},
    publisher = {Addison-Wesley},
    edition = {4th},
    year = {2013},
    address = {Reading, Mass}
}

@Manual{rcpp,
    title = {Rcpp: Seamless R and C++ Integration},
    author = {D. Eddelbuettel and R. Francois and J. Allaire and K. Ushey and Q. Kou and N. Russell and I. Ucar and D. Bates and J. Chambers},
    note = {R package version 1.1.0},
    year = {2025},
    doi = {10.32614/CRAN.package.Rcpp}
}

@Manual{rcpparmadillo,
    title = {RcppArmadillo: 'Rcpp' Integration for the 'Armadillo' Templated Linear Algebra},
    author = {D. Eddelbuettel and R. Francois and  D. Bates and B. Ni and C. Sanderson},
    note = {R package version 15.0.2-2},
    year = {2025},
    doi = {10.32614/CRAN.package.RcppArmadillo}
}

@article{hinton2002,
	author = {G. E. Hinton},
	title = {Training products of experts by minimizing contrastive divergence},
	journal = {Neural Computation},
	year = {2002},
	volume = {14(8)},
	pages = {1771--1800},
	doi = {10.1162/089976602760128018}
}

@book{chopin2020,
    author = {N. Chopin and O. Papaspiliopoulos},
    title = {An Introduction to Sequential Monte Carlo},
    publisher = {Springer Cham},
    address = {Switzerland},
    edition = {1st},
    series = {Springer Series in Statistics},
    year = {2020},
    doi = {10.1007/978-3-030-47845-2}
}

@article{chopin2013,
	author = {N. Chopin and P. E. Jacob and O. Papaspiliopoulos},
	title = {$\smcsq$: an efficient algorithm for sequential analysis of state sapce models},
	journal = {Journal of the Royal Statistical Society Series B: Statistical Methodology},
	year = {2013},
	volume = {75(3)},
	pages = {397--426},
	doi = {10.1111/j.1467-9868.2012.01046.x}
}

@inbook{ohagan2006,
    author = {A. O{'}Hagan and C. Buck and A. Daneshkhah and J. Eiser and P. Garthwaite and D. Jenkinson and J. Oakley and T. Rakow},
    chapter = {Statistics in Practice},
    title = {Uncertain Judgements: Eliciting Experts{'} Probabilities},
    publisher = {John Wiley \& Sons, Ltd},
    address = {Chichester, UK},
    year = {2006},
    pages = {322--323},
    doi = {10.1002/0470033312.scard}
}

@article{genest1986,
  author  = {Genest, C. and McConway, K. J. and Schervish, M. J.},
  title   = {Characterization of externally {B}ayesian pooling operators},
  journal = {The Annals of Statistics},
  year    = {1986},
  volume  = {14},
  number  = {2},
  pages   = {487--501},
  doi     = {10.1214/aos/1176349934},
  mrnumber = {MR0840510}
}

@Manual{rjags,
    title = {{r}jags: Bayesian Graphical Models using MCMC},
    author = {M. Plummer and A. Stukalov and M. Denwood},
    note = {R package version 4-17},
    year = {2025},
    doi = {10.32614/CRAN.package.rjags}
}

@article{devalpine2017,
  author  = {de Valpine, P. and Turek, D. and Paciorek, C. J. and Anderson-Bergman, C. and Lang, D. T. and Bodik, R.},
  title   = {Programming with models: writing statistical algorithms for general model structures with {NIMBLE}},
  journal = {Journal of Computational and Graphical Statistics},
  year    = {2017},
  volume  = {26},
  number  = {2},
  pages   = {403--413},
  doi     = {10.1080/10618600.2016.1172487}
}

@Manual{nimblesmc,
  title = {{nimbleSMC}: Sequential Monte Carlo Methods for {NIMBLE}},
  author = {Michaud, Nick and de Valpine, Perry and Paciorek, Christopher and Turek, Daniel and Goldstein, Benjamin R. and Nguyen, Dao},
  year = {2025},
  note = {R package version 0.11.1},
  organization = {The Regents of the University of California},
  doi = {10.32614/CRAN.package.nimbleSMC}
}

@article{alvares2023,
  author    = {Alvares, Diego and Leiva-Yamaguchi, Victor},
  title     = {A two-stage approach for Bayesian joint models: reducing complexity while maintaining accuracy},
  journal   = {Statistics and Computing},
  volume    = {33},
  number    = {115},
  year      = {2023},
  doi       = {10.1007/s11222-023-10281-9}
}

@article{alvares2025,
    author = {Alvares, Danilo and Barrett, Jessica K. and Mercier, François and Roumpanis, Spyros and Yiu, Sean and Castro, Felipe and Schulze, Jochen and Zhu, Yajing},
    title = {A {B}ayesian Joint Model of Multiple Nonlinear Longitudinal and Competing Risks Outcomes for Dynamic Prediction in Multiple Myeloma: Joint Estimation and Corrected Two-Stage Approaches},
    journal = {Statistics in Medicine},
    volume = {44},
    number = {3-4},
    pages = {e10322},
    doi = {10.1002/sim.10322},
    year = {2025}
}

@article{leiva-yamaguchi2021,
  author    = {Leiva-Yamaguchi, Victor and Alvares, Diego},
  title     = {A Two-Stage Approach for {B}ayesian Joint Models of Longitudinal and Survival Data: Correcting Bias with Informative Prior},
  journal   = {Entropy},
  year      = {2021},
  volume    = {23},
  number    = {1},
  pages     = {50},
  doi       = {10.3390/e23010050}
}

@article{murawska2012,
  author    = {Murawska, Magdalena and Rizopoulos, Dimitris and Lesaffre, Emmanuel},
  title     = {A Two-Stage Joint Model for Nonlinear Longitudinal Response and a Time-to-Event with Application in Transplantation Studies},
  journal   = {Journal of Probability and Statistics},
  year      = {2012},
  volume    = {2012},
  doi       = {10.1155/2012/194194}
}

@article{ades2003,
  author    = {Ades, A. E.},
  title     = {A chain of evidence with mixed comparisons: models for multi-parameter synthesis and consistency of evidence},
  journal   = {Statistics in Medicine},
  year      = {2003},
  volume    = {22},
  number    = {19},
  pages     = {2995--3016},
  doi       = {10.1002/sim.1566}
}

@article{ades2006,
  author    = {Ades, A. E. and Sutton, A. J.},
  title     = {Multiparameter Evidence Synthesis in Epidemiology and Medical Decision-Making: Current Approaches},
  journal   = {Journal of the Royal Statistical Society: Series A (Statistics in Society)},
  year      = {2006},
  volume    = {169},
  number    = {1},
  pages     = {5--35},
  doi       = {10.1111/j.1467-985X.2005.00377.x}
}

@article{presanis2014,
  author    = {Presanis, Anne M. and Pebody, Richard G. and Birrell, Paul J. and Tom, Brian D. M. and Green, Helen K. and Durnall, Hannah and Fleming, Douglas and De Angelis, Daniela},
  title     = {Synthesising Evidence to Estimate Pandemic (2009) A/H1N1 Influenza Severity in 2009--2011},
  journal   = {Annals of Applied Statistics},
  year      = {2014},
  volume    = {8},
  number    = {4},
  pages     = {2378--2403},
  doi       = {10.1214/14-AOAS775}
}

@article{lunn2013,
  author    = {Lunn, David and Barrett, Jessica and Sweeting, Michael and Thompson, Simon},
  title     = {Fully {B}ayesian Hierarchical Modelling in Two Stages, with Application to Meta-Analysis},
  journal   = {Journal of the Royal Statistical Society: Series C (Applied Statistics)},
  year      = {2013},
  volume    = {62},
  number    = {4},
  pages     = {551--572},
  doi       = {10.1111/rssc.12007}
}

@book{ibrahim2001,
  author    = {Ibrahim, Joseph G. and Chen, Ming-Hui and Sinha, Debajyoti},
  title     = {Bayesian Survival Analysis},
  series    = {Springer Series in Statistics},
  publisher = {Springer},
  address   = {New York, NY},
  year      = {2001},
  edition   = {1},
  doi       = {10.1007/978-1-4757-3447-8}
}

@article{lawrence-gould2015,
    author = {Lawrence Gould, A. and Boye, Mark Ernest and Crowther, Michael J. and Ibrahim, Joseph G. and Quartey, George and Micallef, Sandrine and Bois, Frederic Y.},
    title = {Joint modeling of survival and longitudinal non-survival data: current methods and issues. {R}eport of the {DIA} {B}ayesian joint modeling working group},
    journal = {Statistics in Medicine},
    volume = {34},
    number = {14},
    pages = {2181-2195},
    doi = {https://doi.org/10.1002/sim.6141},
    year = {2015}
}

@article{guo2004,
  author    = {Guo, Xu and Carlin, Bradley P.},
  title     = {Separate and Joint Modeling of Longitudinal and Event Time Data Using Standard Computer Packages},
  journal   = {The American Statistician},
  year      = {2004},
  volume    = {58},
  number    = {1},
  pages     = {16--24},
  doi       = {10.1198/0003130042854}
}

@article{chen2025,
  author    = {Chen, Sida and Alvares, Danilo and Palma, Marco and Barrett, Jessica K.},
  title     = {Bayesian shared parameter joint models for heterogeneous populations},
  journal   = {Statistics and Computing},
  year      = {2025},
  volume    = {35},
  number    = {5},
  pages     = {125},
  doi       = {10.1007/s11222-025-10647-1}
}

@article{mauff2020,
  author    = {Mauff, K. and Steyerberg, E. W. and Kardys, I. and Boersma, E. and Rizopoulos, D.},
  title     = {Joint models with multiple longitudinal outcomes and a time-to-event outcome: a corrected two-stage approach},
  journal   = {Statistics and Computing},
  year      = {2020},
  volume    = {30},
  pages     = {999--1014},
  doi       = {10.1007/s11222-020-09927-9}
}

@article{martins2016,
    author = {Martins, Rui and Silva, Giovani L. and Andreozzi, Valeska},
    title = {Bayesian joint modeling of longitudinal and spatial survival {AIDS} data},
    journal = {Statistics in Medicine},
    volume = {35},
    number = {19},
    pages = {3368-3384},
    doi = {https://doi.org/10.1002/sim.6937},
    year = {2016}
}

@InProceedings{donnat2020,
  title = 	 {A {B}ayesian Hierarchical Network for Combining Heterogeneous Data Sources in Medical Diagnoses},
  author =       {Donnat, Claire and Miolane, Nina and Bunbury, Freddy and Kreindler, Jack},
  booktitle = 	 {Proceedings of the Machine Learning for Health NeurIPS Workshop},
  pages = 	 {53--84},
  year = 	 {2020},
  volume = 	 {136},
  series = 	 {Proceedings of Machine Learning Research},
  publisher =    {PMLR},
  address = {New York, NY}
}

@article{rendall2009,
  author    = {Rendall, Michael S. and Handcock, Mark S. and Jonsson, Stefan H.},
  title     = {Bayesian estimation of {H}ispanic fertility hazards from survey and population data},
  journal   = {Demography},
  year      = {2009},
  volume    = {46},
  number    = {1},
  pages     = {65--83},
  doi       = {10.1353/dem.0.0041}
}

@article{abadi2010a,
    author = {Abadi, Fitsum and Gimenez, Olivier and Arlettaz, Rapha\"{e}l and Schaub, Michael},
    title = {An assessment of integrated population models: bias, accuracy, and violation of the assumption of independence},
    journal = {Ecology},
    volume = {91},
    number = {1},
    pages = {7-14},
    doi = {doi.org/10.1890/08-2235.1},
    year = {2010a}
}

@article{abadi2010b,
    author = {Abadi, Fitsum and Gimenez, Olivier and Ullrich, Bruno and Arlettaz, Raphaël and Schaub, Michael},
    title = {Estimation of immigration rate using integrated population models},
    journal = {Journal of Applied Ecology},
    volume = {47},
    number = {2},
    pages = {393-400},
    doi = {10.1111/j.1365-2664.2010.01789.x},
    year = {2010b}
}

@article{king2008,
    author = {King, Ruth and Brooks, Stephen P. and Mazzetta, Chiara and Freeman, Stephen N. and Morgan, Byron J. T.},
    title = {Identifying and Diagnosing Population Declines: A Bayesian Assessment of Lapwings in the UK},
    journal = {Journal of the Royal Statistical Society Series C: Applied Statistics},
    volume = {57},
    number = {5},
    pages = {609-632},
    year = {2008},
    doi = {10.1111/j.1467-9876.2008.00633.x}
}

@article{rhodes2011,
    author = {Jonathan R. Rhodes and Chooi Fei Ng and Deidré L. {de Villiers} and Harriet J. Preece and Clive A. McAlpine and Hugh P. Possingham},
    title = {Using integrated population modelling to quantify the implications of multiple threatening processes for a rapidly declining population},
    journal = {Biological Conservation},
    volume = {144},
    number = {3},
    pages = {1081-1088},
    year = {2011},
    doi = {10.1016/j.biocon.2010.12.027}
}

@article{woodworth2017,
  author    = {Woodworth, Bradley K. and Wheelwright, N. T. and Newman, A. E. and Schaub, M. and Norris, D. Ryan},
  title     = {Winter temperatures limit population growth rate of a migratory songbird},
  journal   = {Nature Communications},
  year      = {2017},
  volume    = {8},
  pages     = {14812},
  doi       = {10.1038/ncomms14812}
}

@article{hooten2021,
    author = {Mevin B. Hooten and Devin S. Johnson and Brian M. Brost},
    title = {Making Recursive {B}ayesian Inference Accessible},
    journal = {The American Statistician},
    volume = {75},
    number = {2},
    pages = {185--194},
    year = {2021},
    doi = {10.1080/00031305.2019.1665584}
}

@article{carpenter1999,
    author = {J. Carpenter  and P. Clifford  and P. Fearnhead },
    title = {Improved particle filter for nonlinear
    problems},
    journal = {IEE Proceedings - Radar, Sonar and Navigation},
    volume = {146},
    issue = {1},
    pages = {2-7},
    year = {1999},
    doi = {10.1049/ip-rsn:19990255}
}

\clearpage

\appendix

\section*{Appendix}

\setcounter{section}{0}
\renewcommand{\thesection}{\arabic{section}}

\setcounter{equation}{0}
\setcounter{figure}{0}
\setcounter{table}{0}
\setcounter{section}{0}

\renewcommand{\theequation}{S\arabic{equation}}
\renewcommand{\thefigure}{S\arabic{figure}}
\renewcommand{\thetable}{S\arabic{table}}
\renewcommand{\thesection}{S\arabic{section}}

\section{More on $M=3$}
\subsection{Details on the SMC sampler}

\begin{mypf}{Lemma}{\ref{le:weight-update}}
    Let \begin{align}\label{eq:proof-lemma1-eq1}
        p_{c}(\phi_{1,2},\phi_{2,3}):=\int\frac{p_{1}(\phi_{1,2},\psi_{1},\bsY_{1})p_{3}(\phi_{2,3},\psi_{3},\bsY_{3})}{p_{1}(\phi_{1,2})p_{3}(\phi_{2,3})}d(\psi_{1},\psi_{3}).
    \end{align} Then, \eqref{eq:meld-subposterior-1-3} indicates that \begin{align}\label{eq:proof-lemma1-eq2}
        &\ p_{\meld,1,3}(\phi_{1,2},\phi_{2,3}\mid\bsY_{1},\bsY_{3})\nonumber\\
        =&\int p_{\meld,1,3}(\phi_{1,2},\phi_{2,3},\psi_{1},\psi_{3}|\bsY_{1},\bsY_{3})d(\psi_{1},\psi_{3})\nonumber\\
        \propto&\int p_{\pool,1}(\phi_{1,2})\frac{p_{1}(\phi_{1,2},\psi_{1},\bsY_{1})}{p_{1}(\phi_{1,2})}p_{\pool,3}(\phi_{2,3})\frac{p_{3}(\phi_{2,3},\psi_{3},\bsY_{3})}{p_{3}(\phi_{2,3})}d(\psi_{1},\psi_{3})\nonumber\\
        =&\ p_{\pool,1}(\phi_{1,2})p_{\pool,3}(\phi_{2,3})\int\frac{p_{1}(\phi_{1,2},\psi_{1},\bsY_{1})p_{3}(\phi_{2,3},\psi_{3},\bsY_{3})}{p_{1}(\phi_{1,2})p_{3}(\phi_{2,3})}d(\psi_{1},\psi_{3})\nonumber\\
        =&\  p_{\pool,1}(\phi_{1,2})p_{\pool,3}(\phi_{2,3})p_{c}(\phi_{1,2},\phi_{2,3}).
    \end{align} In addition, \eqref{eq:meld-posterior-M3} implies that the marginal posterior of $\bstheta$ follows \begin{align}\label{eq:proof-lemma1-eq3}
        p_{\meld}(\bstheta\mid\bsY_{1},\bsY_{2},\bsY_{3})&=\int p_{\meld}(\phi_{1,2},\phi_{2,3},\psi_{1},\tilde{\psi}_{2},\psi_{3}|\bsY_{1},\bsY_{2},\bsY_{3})d(\psi_{1},\psi_{3})\nonumber\\
        &\propto p_{\pool}(\phi_{1,2},\phi_{2,3})\frac{p_{2}(\phi_{1,2},\phi_{2,3},\tilde{\psi}_{2},\bsY_{2})}{p_{2}(\phi_{1,2},\phi_{2,3})}p_{c}(\phi_{1,2},\phi_{2,3})\nonumber\\
        &=p_{\pool}(\phi_{1,2},\phi_{2,3})p_{2}(\tilde{\psi}_{2})p_{2}(\bsY_{2}|\phi_{1,2},\phi_{2,3},\tilde{\psi}_{2})p_{c}(\phi_{1,2},\phi_{2,3}).
    \end{align} With abuse of notation, let us assume that the particles obtained from the $j-1$ step are $(\phi_{1,2}^{(i)},\phi_{2,3}^{(i)},\tilde{\psi}_{2}^{(i)})_{i=1}^{N}$. Based on \eqref{eq:proof-lemma1-eq2} and \eqref{eq:proof-lemma1-eq3}, the increment in \eqref{eq:weight-update} follows \begin{align}\label{eq:proof-lemma1-eq4}
        \frac{\gamma_{\smc,n_{t}}(\bstheta_{j-1}^{(i)})}{\gamma_{\smc,0}(\bstheta_{j-1}^{(i)})}&\propto\frac{p_{\pool}(\phi_{1,2}^{(i)},\phi_{2,3}^{(i)})p_{2}(y_{2}|\phi_{1,2}^{(i)},\phi_{2,3}^{(i)},\tilde{\psi}_{2}^{(i)})p_{2}(\tilde{\psi}_{2}^{(i)})}{p_{\pool,1}(\phi_{1,2}^{(i)})p_{\pool,3}(\phi_{2,3}^{(i)})q_{\meld}(\tilde{\psi}_{2}^{(i)}|\phi_{1,2}^{(i)},\phi_{2,3}^{(i)})}\nonumber\\
        &=\frac{p_{\pool,2}(\phi_{1,2}^{(i)},\phi_{2,3}^{(i)})p_{2}(y_{2}|\phi_{1,2}^{(i)},\phi_{2,3}^{(i)},\tilde{\psi}_{2}^{(i)})p_{2}(\tilde{\psi}_{2}^{(i)})}{q_{\meld}(\tilde{\psi}_{2}^{(i)}|\phi_{1,2}^{(i)},\phi_{2,3}^{(i)})}.
    \end{align} \eqref{eq:weight-update2} is a direct result of \eqref{eq:proof-lemma1-eq4}.
\end{mypf}

\subsection{Extension for merging subpopulations}

In Algorithm \ref{algo1}, the naive merging process, which uses $\pi_{t,0}(\bsx_{t}):=\prod_{c\in\calC(t)}\pi(\bsx_{c})$, has a low computational cost. However, it struggles when $\prod_{c\in\mathcal {C} (t)}\pi(\bsx_{c})$ significantly differs from the corresponding marginal of $\pi_{t}$, which can result in large variance. To amend this, \citet{lindsten2017} adopts the idea of the auxiliary particle filter \citep{pitt1999} or the mixture proposal approach \citep{carpenter1999} in the merging step. Specifically, they proposed two versions of this extension. The first version is closely related to the original auxiliary particle filter technique, while the second one is a computationally efficient version that employs the independent particle filter proposed by \citet{lin2005}. For details of this extension, we refer to Section 4.1 in the original article. Here, we propose an extension of the second version for use in our samplers, which is demonstrated in Algorithm \ref{algo:dc-melding-extened-merging}.\\


{\renewcommand{\thealgocf}{3}
\begin{algorithm}
    \caption{\texttt{d\&c-melding}(3): D\&C-melding with the extended merging step for $M=3$}
    \label{algo:dc-melding-extened-merging}
        \KwIn{data $\bsY$; the choice of $p_{\pool}(\phi_{1,2},\phi_{2,3})$; the subposteriors $p_{\meld,m}(\phi_{m},\psi_{m}\mid\bsY_{m})$ for $m=1,2,3$; the distribution $p_{\meld}(\phi_{1,2},\phi_{2,3}|\bsY_{1},\bsY_{2},\bsY_{3})$; the proposal $q_{\meld}(\psi_{2}|\phi_{1,2},\phi_{2,3})$; the number of particles $N$, the number of the annealing steps $n_{t}$.} 

        In stage one,

            \nonl\textnormal{(a)} In parallel, sample $\lbrace\phi_{1,2}^{(i)},w_{1,2}^{(i)}\rbrace_{i=1}^{N}$ and $\lbrace\phi_{2,3}^{(i)},w_{2,3}^{(i)}\rbrace_{i=1}^{N}$ via SMC samplers.
        
            \nonl\textnormal{(b)} Sample $\tilde{n}N$ times with replacement from $\lbrace\phi_{1,2}^{(i)},w_{1,2}^{(i)}\rbrace_{i=1}^{N}$ and $\lbrace\phi_{2,3}^{(i)},w_{2,3}^{(i)}\rbrace_{i=1}^{N}$. Override the notation and let $\lbrace\phi_{1,2}^{(i)}\rbrace_{i=1}^{\tilde{n}N}$ and $\lbrace\phi_{2,3}^{(i)}\rbrace_{i=1}^{\tilde{n}N}$ refer to the resampled particles.

            \nonl\textnormal{(c)} Matching the particles with the same index $i$ from each of the children nodes to form tuples $\lbrace(\phi_{1,2}^{(i)},\phi_{2,3}^{(i)})\rbrace_{i=1}^{mN}$ and compute the corresponding weights {\nonl\begin{align*}
                &v_{t}^{(i)}=\check{p}_{\meld}(\phi_{1,2}^{(i)},\phi_{2,3}^{(i)}\mid\bsY_{1},\bsY_{2},\bsY_{3})\bigg/\lr p_{\meld,1,3}(\phi_{1,2}^{(i)},\phi_{2,3}^{(i)}\mid\bsY_{1},\bsY_{3})\rr,\\
                &\qquad\qquad\qquad\qquad\qquad\qquad\qquad\qquad\qquad\qquad\qquad\qquad\qquad i=1,...,mN.
            \end{align*}}

            \nonl\textnormal{(d)} Sample with replacement $N$ tuples $\lbrace(\breve{\phi}_{1,2}^{(i)},\breve{\phi}_{2,3}^{(i)})\rbrace_{i=1}^{N}$ from the $mN$ tuples proportional to $v_{t}^{(j)},j=1,...,mN$.

        In stage two,

            \nonl\textnormal{(a)} Initialise $\tilde{\psi}_{2}^{(i)}\sim q_{\meld}(\cdot|\breve{\phi}_{1,2}^{(i)},\breve{\phi}_{2,3}^{(i)})$ for $i=1,...,N$.
            
            \nonl\textnormal{(b)} Set $\bstheta_{0}^{(i)}=(\breve{\phi}_{1,2}^{(i)},\breve{\phi}_{2,3}^{(i)},\tilde{\psi}_{2}^{(i)})$ and $w_{0}^{(i)}=1$ for $i=1,...,N$.

            \nonl\textnormal{(c)} \For{SMC sampler iteration $j=1$ to $n_{t}$}{
                \nonl\textnormal{i.} Compute $w_{j}^{(i)}=w_{j-1}^{(i)}\gamma_{\smc,j}(\bstheta_{j-1}^{(i)})/\gamma_{\smc,j-1}(\bstheta_{j-1}^{(i)}).$
        
                \nonl\textnormal{ii.} Optionally, resample $\lbrace\bstheta_{j-1}^{(i)},w_{j}^{(i)}\rbrace_{i=1}^{N}$. Override the notation and let $\lbrace\bstheta_{j-1}^{(i)},w_{j}^{(i)}\rbrace_{i=1}^{N}$ refer to the resampled particle system.
        
                \nonl\textnormal{iii.} Draw $\bstheta_{j}^{(i)}\sim K_{j}(\bstheta_{j-1}^{(i)})$ using a $p_{\smc,j}$-reversible Markov kernel $K_{j}$ for $i=1,...,N$.
            }
        
            \nonl\textnormal{(d)} Set $\bstheta^{(i)}=\bstheta_{n_{t}}^{(i)}$ and $w^{(i)}=w_{n_{t}}$.

        \KwOut{$\lbrace\bstheta^{(i)},w^{(i)}\rbrace_{i=1}^{N}$.}
    
\end{algorithm}
}


A critical part of the merging step is the computation of the weights $v_{t}^{(i)}, i=1,...,mN$, which can aid the process. We define $\check{p}_{\meld}$ to be \begin{align}\label{eq:p-check1}
    &\check{p}_{\meld}(\phi_{1,2}^{(i)},\phi_{2,3}^{(i)}\mid\bsY_{1},\bsY_{2},\bsY_{3})\nonumber\\
    \propto&\ls p_{\meld,1,3}(\phi_{1,2}^{(i)},\phi_{2,3}^{(i)}\mid\bsY_{1},\bsY_{3})\rs^{1-\alpha^{*}}\ls\int p_{\meld}(\phi_{1,2}^{(i)},\phi_{2,3}^{(i)},\tilde{\psi}_{2}^{(i)}\mid\bsY_{1},\bsY_{2},\bsY_{3})d\tilde{\psi}_{2}^{(i)}\rs^{\alpha^{*}},
\end{align}  where $\alpha^{*}\in[0,1]$. The second term on the right-hand side of (\ref{eq:p-check1}) follows the idea of the auxiliary particle filter technique. 
Specifically, based on \eqref{eq:meld-posterior-M3}, we have \begin{align}\label{eq:p-int-eq1}
        &\int p_{\meld}(\phi_{1,2}^{(i)},\phi_{2,3}^{(i)},\tilde{\psi}_{2}^{(i)}\mid\bsY_{1},\bsY_{2},\bsY_{3})d\tilde{\psi}_{2}\nonumber\\
        =&\iint p_{\meld}(\phi_{1,2}^{(i)},\phi_{2,3}^{(i)},\psi_{1}^{(i)},\psi_{3}^{(i)},\tilde{\psi}_{2}^{(i)}\mid\bsY_{1},\bsY_{2},\bsY_{3})d(\psi_{1}^{(i)},\psi_{3}^{(i)})d(\tilde{\psi}_{2}^{(i)})\nonumber\\
        \propto&\iint p_{\pool}(\phi_{1,2}^{(i)},\phi_{2,3}^{(i)})\frac{p_{1}(\phi_{1,2}^{(i)},\psi_{1}^{(i)},\bsY_{1})}{p_{1}(\phi_{1,2}^{(i)})}\frac{p_{2}(\phi_{1,2}^{(i)},\phi_{2,3}^{(i)},\tilde{\psi}_{2}^{(i)},\bsY_{2})}{p_{2}(\phi_{1,2}^{(i)},\phi_{2,3}^{(i)})}\nonumber\\
        &\qquad\qquad\qquad\qquad\qquad\qquad \cdot \frac{p_{3}(\phi_{2,3}^{(i)},\psi_{3}^{(i)},\bsY_{3})}{p_{3}(\phi_{2,3}^{(i)})}d(\psi_{1}^{(i)},\psi_{3}^{(i)})d\tilde{\psi}_{2}^{(i)}\nonumber\\
        =&\ p_{\pool}(\phi_{1,2}^{(i)},\phi_{2,3}^{(i)})p_{c}(\phi_{1,2}^{(i)},\phi_{2,3}^{(i)})\int\frac{p_{2}(\phi_{1,2}^{(i)},\phi_{2,3}^{(i)},\tilde{\psi}_{2}^{(i)},\bsY_{2})}{p_{2}(\phi_{1,2}^{(i)},\phi_{2,3}^{(i)})}d\tilde{\psi}_{2}^{(i)}\nonumber\\
        =&\ p_{\pool}(\phi_{1,2}^{(i)},\phi_{2,3}^{(i)})p_{c}(\phi_{1,2}^{(i)},\phi_{2,3}^{(i)})\int p_{2}(\bsY_{2}|\phi_{1,2}^{(i)},\phi_{2,3}^{(i)},\tilde{\psi}_{2}^{(i)})p_{2}(\tilde{\psi}_{2}^{(i)})d\tilde{\psi}_{2}^{(i)}.
    \end{align} We use $p_{2}(\bsY_{2}|\phi_{1,2}^{(i)},\phi_{2,3}^{(i)},\tilde{\mu}_{2}^{(i)})$ to approximate $p_{2}(\bsY_{2}|\phi_{1,2}^{(i)},\phi_{2,3}^{(i)},\tilde{\psi}_{2}^{(i)})$ , where $\tilde{\mu}_{2}^{(i)}$ can be the mean, the mode, or a draw of $p_{2}(\psi_{2}^{(i)})$. Then, \eqref{eq:p-int-eq1} follows \begin{align}\label{eq:p-int-eq2}
        &\int p_{\meld}(\phi_{1,2}^{(i)},\phi_{2,3}^{(i)},\tilde{\psi}_{2}^{(i)}\mid\bsY_{1},\bsY_{2},\bsY_{3})d\tilde{\psi}_{2}\nonumber\\
        \approx&\ p_{\pool}(\phi_{1,2}^{(i)},\phi_{2,3}^{(i)})p_{c}(\phi_{1,2}^{(i)},\phi_{2,3}^{(i)})p_{2}(\bsY_{2}|\phi_{1,2}^{(i)},\phi_{2,3}^{(i)},\tilde{\mu}_{2}^{(i)})\int p_{2}(\tilde{\psi}_{2}^{(i)})d\tilde{\psi}_{2}^{(i)}\nonumber\\
        =&\ p_{\pool}(\phi_{1,2}^{(i)},\phi_{2,3}^{(i)})p_{c}(\phi_{1,2}^{(i)},\phi_{2,3}^{(i)})p_{2}(\bsY_{2}|\phi_{1,2}^{(i)},\phi_{2,3}^{(i)},\tilde{\mu}_{2}^{(i)}).
    \end{align}
\noindent Hence, based on \eqref{eq:proof-lemma1-eq2} and \eqref{eq:p-int-eq2}, we have \begin{align}\label{eq:p-check2}
    &\check{p}_{\meld}(\phi_{1,2}^{(i)},\phi_{2,3}^{(i)}\mid\bsY_{1},\bsY_{2},\bsY_{3})\nonumber\\
    \propto&\ls p_{\pool,1}(\phi_{1,2}^{(i)})p_{\pool,3}(\phi_{2,3}^{(i)})p_{c}(\phi_{1,2}^{(i)},\phi_{2,3}^{(i)})\rs^{1-\alpha^{*}}\nonumber\\
    &\qquad\cdot\ls p_{\pool}(\phi_{1,2}^{(i)},\phi_{2,3}^{(i)})p_{2}(\bsY_{2}|\phi_{1,2}^{(i)},\phi_{2,3}^{(i)},\tilde{\mu}_{2}^{(i)})p_{c}(\phi_{1,2}^{(i)},\phi_{2,3}^{(i)})\rs^{\alpha^{*}}\nonumber\\
    =&\ls p_{\pool,1}(\phi_{1,2}^{(i)})p_{\pool,3}(\phi_{2,3}^{(i)})\rs^{1-\alpha^{*}}\ls p_{\pool}(\phi_{1,2}^{(i)},\phi_{2,3}^{(i)})p_{2}(\bsY_{2}|\phi_{1,2}^{(i)},\phi_{2,3}^{(i)},\tilde{\mu}_{2}^{(i)})\rs^{\alpha^{*}}\nonumber\\
    &\qquad\qquad\qquad\qquad\qquad\qquad\qquad\qquad\qquad\qquad\qquad\cdot p_{c}(\phi_{1,2}^{(i)},\phi_{2,3}^{(i)}).
\end{align} Therefore, by \eqref{eq:proof-lemma1-eq2} and (\ref{eq:p-check2}), the weights $v$ on Line 1(c) can be computed by \begin{align}\label{eq3.13}
    v^{(i)}&=\frac{\check{p}_{\meld}(\phi_{1,2}^{(i)},\phi_{2,3}^{(i)}\mid\bsY_{1},\bsY_{2},\bsY_{3})}{p_{\meld,1,3}(\phi_{1,2}^{(i)},\phi_{2,3}^{(i)}\mid\bsY_{1},\bsY_{3})}\nonumber\\
    &\propto\lr\frac{p_{\pool}(\phi_{1,2}^{(i)},\phi_{2,3}^{(i)})}{p_{\pool,1}(\phi_{1,2}^{(i)})p_{\pool,3}(\phi_{2,3}^{(i)})}p_{2}(\bsY_{2}|\phi_{1,2}^{(i)},\phi_{2,3}^{(i)},\tilde{\mu}_{2}^{(i)})\rr^{\alpha^{*}}\nonumber\\
    &=\lr p_{\pool,2}(\phi_{1,2}^{(i)},\phi_{2,3}^{(i)})p_{2}(\bsY_{2}|\phi_{1,2}^{(i)},\phi_{2,3}^{(i)},\tilde{\mu}_{2}^{(i)})\rr^{\alpha^{*}}.
\end{align} In addition, in the SMC sampler in Line 2(c), $p_{\smc,0}(\bstheta)\allowbreak:=\allowbreak\check{p}_{\meld}\allowbreak(\phi_{1,2},\allowbreak\phi_{2,3}\allowbreak|\allowbreak\bsY_{1},\allowbreak\bsY_{2},\allowbreak\bsY_{3})\allowbreak q_{\meld}(\tilde{\psi}_{2}|\phi_{1,2},\phi_{2,3})$ and $p_{\smc,n_{t}}(\bstheta):=p_{\meld}(\phi_{1,2},\phi_{2,3},\psi_{1},\psi_{3},\tilde{\psi}_{2}|\bsY_{1},\bsY_{2},\bsY_{3})$

\begin{manremark}{S1}\label{re:mu-choice}
    The strategy of choosing $\tilde{\mu}_{2}$ in practice is classified into the following three cases: \begin{enumerate}[label=\alph*)]
        \item if $\psi_{2}$ is directly dependent on $(\phi_{1,2},\phi_{2,3})$, we can use any appropriate descriptive statistics associated with $p_{2}(\psi_{2}^{(i)}|\phi_{1,2}^{(i)},\phi_{2,3}^{(i)})$ for $i=1,...,mN$;

        \item if there is no direct dependence of $\psi_{2}$ on $(\phi_{1,2},\phi_{2,3})$, we can use a Markov kernel to draw posterior samples for $\psi_{2}$ in submodel 2 only based on any descriptive statistics of the particles $(\phi_{1,2}^{(i)})_{i=1}^{mN}$ and $(\phi_{2,3}^{(i)})_{i=1}^{mN}$, and then we use a descriptive statistic of the posterior samples as a fixed value for all $\tilde{\mu}_{2}^{(i)},i=1,...,mN;$ or simply, using the artificial prior of $\psi_{2}$.
    \end{enumerate} 
\end{manremark} 

For Case b) in Remark \ref{re:mu-choice}, using both a Markov kernel and the artificial prior can potentially result in bias for the merged particles, and the former choice also brings a further computational burden due to the large number, $mN$, of tuples of particles involved. Consequently, in practice, if there is no direct dependence between $(\phi_{1,2},\phi_{2,3})$, we suggest using the naive merging step rather than the extended version.

\section{Multi-stage sampler for any $M\geq3$}
\subsection{Particle update}

In the multi-stage sampler, tracking particle trajectories is critical for updating particles. This includes forward updates, e.g., using the trajectories of $\phi_{2,3}$ to update $\phi_{3,4}$ in stage two in Example \ref{eg:M5} for $M=5$, and backward updates, e.g., the updates for $\phi_{1,2},\phi_{2,3},\psi_{1},\psi_{2},\psi_{5}$ in stage three in Example \ref{eg:M5}. We provide the methods for updating particles in Algorithms \ref{algo:index_forward}-\ref{algo:index_rightward}, in which we use the \textit{multiset} to collect index. Recall that a multiset is a generalised concept of a set that allows duplicate elements. We use $X^{\dagger}$ to represent a multiset.\\

{\renewcommand{\thealgocf}{3}
\begin{algorithm}[H]
    \caption{\texttt{forward\_update()}: forward particle update}
    \label{algo:index_forward}
        \KwIn{Index collections $A_{m}^{\dagger}=\lbrace a_{m}^{(1)},\ldots,a_{m}^{(N)}\rbrace$ and $B_{r}^{\dagger}=\lbrace b_{r}^{(1)},\ldots,b_{r}^{(N)}\rbrace$.} 
    
        Use $A_{m}^{\dagger}$ to update $B_{r}^{\dagger}$, obtaining \begin{align*}
            B_{r}^{*\dagger}=\lb b_{r}^{(a_{m}^{(1)})},\ldots,b_{r}^{(a_{m}^{(N)})}\rb.
        \end{align*}
    
        \KwOut{$B_{r}^{*\dagger}$.} 
\end{algorithm}}

{\renewcommand{\thealgocf}{4a}
\begin{algorithm}[H]
    \caption{\texttt{back\_left\_update()}: backward particle update in a leftward order of submodels}
    \label{algo:index_leftward}
        \KwIn{Index collections $A_{m}^{\dagger}=\lbrace a_{m}^{(1)},\ldots,a_{m}^{(N)}\rbrace$ for $m=m_{1},\ldots,m_{T}$ in an increasing order; samples $\lbrace\bstheta_{m}^{(i)}\rbrace_{i=1}^{N}$ for $m=m_{1},\ldots,m_{T-1}$.}
        \For{$s=0,\ldots,T-2$}{
            \nonl \textnormal{(i)} Use $A_{m_{T-s}}$ to override $A_{m_{T-s-1}}$, obtaining \[
                A_{m_{T-s-1}}^{\dagger}=\lb a_{m_{T-s-1}}^{(a_{m_{T-s}}^{(1)})},\ldots,a_{m_{T-s-1}}^{(a_{m_{T-s}}^{(N)})}\rb.
            \]

            \nonl \textnormal{(ii)} Update samples $\lbrace \bstheta_{m_{T-s-1}}^{\dagger}\rbrace_{i=1}^{N}$ by picking the indices in $A_{m_{T-s-1}}^{\dagger}$, obtaining $\lbrace\bstheta_{m_{T-s-1}}^{(i)}\rbrace_{i\in A_{m_{T-s-1}}^{\dagger}}$.
            } 
        \KwOut{Updated $\lbrace\bstheta_{m}^{(i)}\rbrace_{i=1}^{N}$ and $A_{m}^{\dagger}$ for $m=m_{1},\ldots,m_{T-1}$.}
\end{algorithm}}

{\renewcommand{\thealgocf}{4b}
\begin{algorithm}
    \caption{\texttt{back\_right\_update()}: backward particle update in a rightward order of submodels}
    \label{algo:index_rightward}
        \KwIn{Index collections $A_{m}^{\dagger}=\lbrace a_{m}^{(1)},\ldots,a_{m}^{(N)}\rbrace$ for $m=m_{1},\ldots,m_{T}$ in a decreasing order; samples $\lbrace\bstheta_{m}^{(i)}\rbrace_{i=1}^{N}$ for $m=m_{1},\ldots,m_{T+1}$.}
        \For{$s=0,\ldots,T-2$}{
            \nonl \textnormal{(i)} Use $A_{m_{T+s}}$ to override $A_{m_{T+s+1}}$, obtaining \begin{align*}
                A_{m_{T+s+1}}^{\dagger}=\lb a_{m_{T+s+1}}^{(a_{m_{T+s}}^{(N)})},\ldots, a_{m_{T+s+1}}^{(a_{m_{T+s}}^{(N)})}\rb.
            \end{align*}

            \nonl \textnormal{(ii)} Update samples $\lbrace \bstheta_{m_{T+s+1}}^{\dagger}\rbrace_{i=1}^{N}$ by picking the indices in $A_{s+1}^{\dagger}$, obtaining $\lbrace\bstheta_{m_{T+s+1}}^{(i)}\rbrace_{i\in A_{m_{T+s+1}}^{\dagger}}$.
        }
        \KwOut{Updated $\lbrace\phi_{m,m+1}^{(i)}\rbrace_{i=1}^{N}$ and $A_{m}^{\dagger}$ for $m=m_{1},\ldots,m_{T-1}$.} 
\end{algorithm}}

\subsection{Multi-stage sampler}

To clarify the sampler, Algorithm \ref{algo:smc} first presents the SMC sampler employed within the main multi-stage sampler. The algorithm’s output also includes the collection of indices used to reconstruct particle trajectories. The samplers corresponding to the four cases described in Section 3.2 are then given in Algorithms \ref{algo:odd_M_1}-\ref{algo:even_M_2}.

{\renewcommand{\thealgocf}{5}
\begin{algorithm}[H]
    \caption{\texttt{SMC\_sampler()}: a generic SMC sampler}
    \label{algo:smc}
        \KwIn{data $\bsY$; the equally weighted particles $\lbrace\phi_{m}^{(i)},1\rbrace_{i=1}^{N}$; the number of the annealing steps $n_{t}$; the posterior $p_{\meld}(\cdot\mid\bsY)$; the proposal distributions $q_{\meld}()$ for $\psi_{m}$.}

         \textnormal{(a)} Initialise $\tilde{\psi}_{m}^{(i)}\sim q_{\meld}(\cdot|\phi_{m}^{(i)})$ for $i=1,\ldots,N$, and index collection $A_{m}^{\dagger}=\lbrace1,\ldots,N\rbrace$.

         \nonl\textnormal{(b)} Set $\bstheta_{0}^{(i)}=(\phi_{m}^{(i)},\tilde{\psi}_{m}^{(i)})$ and $\tilde{w}_{0}^{(i)}=1$ for $i=1,\ldots,N$, and $A_{0}^{\dagger}=A^{\dagger}$.

        \nonl\textnormal{(c)} \For{SMC sampler iteration $j=1,\ldots,n_{t}$}{
            \nonl\textnormal{i.} Compute $\tilde{w}_{j}^{(i)}=\tilde{w}_{j-1}^{(i)}\gamma_{\meld,j}(\bstheta_{j-1}^{(i)})/\gamma_{\meld,j-1}(\bstheta_{j-1}^{(i)})$ based on \eqref{eq:weight-update1}. Let $A_{j}^{\dagger}=A_{j-1}^{\dagger}$.

             \nonl\textnormal{ii.} Optionally, resample $\lbrace\bstheta_{j-1}^{(i)},\tilde{w}_{j}^{(i)}\rbrace_{i=1}^{N}$ and $A_{j}^{\dagger}$. Override the notation and let $\lbrace\bstheta_{j-1}^{(i)},\tilde{w}_{j}^{(i)}\rbrace$ and $A_{j}^{\dagger}$ refer to the resampled particle system and index collection, respectively.

             \nonl\textnormal{iii.} Draw $\bstheta_{j}^{(i)}\sim K_{j}(\bstheta_{j-1}^{(i)})$ using a $\pi_{\meld,j}$-reversible Markov kernel $K_{j}$ for $i=1,\ldots,N$.
        }

         \nonl\textnormal{(d)} Set $\bstheta^{(i)}=\bstheta_{n_{t}}^{(i)}, w_{m}^{(i)}=\tilde{w}_{n_{t}}^{(i)}$ and $A^{\dagger}=A_{n_{t}}^{\dagger}$.

        \KwOut{$\lbrace\bstheta^{(i)},w_{m}^{(i)}\rbrace_{i=1}^{N}$ and $A^{\dagger}$.}
\end{algorithm}}

{\renewcommand{\thealgocf}{6a}
\SetAlFnt{\footnotesize}
\begin{algorithm}[H]
    \caption{\texttt{d\&c-melding($M$)}: for odd $M$ satisfying $4|(M+1)$}
    \label{algo:odd_M_1}
        \KwIn{data $\bsY=(\bsY_{1},\ldots,\bsY_{M})$; the choice of $p_{\pool}(\bsphi)$; the subposteriors $p_{\meld,m}(\phi_{m},\psi_{m}\mid\bsY_{m})$ for $m=1,\ldots,M$; the proposals $q_{\meld,m}(\psi_{m}\mid\phi_{m})$ for $m=1,\ldots,M$; the number of particles $N$; the number of annealing steps $n_{t}$.}

        In stage one, in parallel for each  $m=1,3,5,\ldots,M$, \textbf{do}
        
            \nonl\textnormal{(a)} Initialise $\phi_{m}^{(i,0)}$ for $\phi_{m}^{(i)}$ to obtain equally weighted particle systems $\lbrace\phi_{m}^{(i,0)},1\rbrace_{i=1}^{N}$.
    
            \nonl\textnormal{(b)} Call the SMC sampler \texttt{SMC\_sampler()} in Algorithm \ref{algo:smc}, obtaining \begin{align*}
                    (\lbrace\bstheta_{m}^{(i)},w_{m}^{(i)}\rbrace_{i=1}^{N},A_{m}^{\dagger})=\text{\texttt{SMC\_sampler$\lr\bsY_{m},\lbrace\phi_{m}^{(i,0)},1\rbrace_{i=1}^{N},p_{\meld,m},q_{\meld,m}\rr$}}.
                \end{align*}
    
             \nonl\textnormal{(c)} Resample $\lbrace\bstheta_{m}^{(i)},w_{m}^{(i)}\rbrace_{i=1}^{N}$ to obtain equally weighted particle systems $\lbrace\bstheta_{m}^{(i)},1\rbrace_{i=1}^{N}$.
    
             \nonl\textnormal{(d)} Record the indices for $\lbrace\bstheta_{m}^{(i)},1\rbrace_{i=1}^{N}$ in a collection $B_{m}^{\dagger}=\lbrace1,\ldots,N\rbrace$.

        In stage $s$, let $m_{L}=2s-2$ and $m_{R}=M+3-2s$. In parallel for each $m=m_{L}$ and $m_{R}$, \For{$s=2,3,\ldots,\frac{M+1}{4}$}{ 

            \nonl\textnormal{(a)} Merge $\lbrace\phi_{m_{L}-1,m_{L}}^{(i)},1\rbrace_{i=1}^{N}$ and $\lbrace\phi_{m_{L},m_{L}+1}^{(i)},1\rbrace_{i=1}^{N}$ to get $\lbrace\phi_{m_{L}}^{(i)},1\rbrace_{i=1}^{N}$. Call the SMC sampler \texttt{SMC\_sampler()} in Algorithm \ref{algo:smc}, obtaining \begin{align*}
                    (\lbrace\bstheta_{m_{L}}^{(i)},w_{m_{L}}^{(i)}\rbrace_{i=1}^{N},A_{m_{L}}^{\dagger})=\text{\texttt{SMC\_sampler$\lr\bsY_{m_{L}},\lbrace\phi_{m_{L}}^{(i)},1\rbrace_{i=1}^{N},p_{\meld,m_{L}},q_{\meld,m_{L}}\rr$}}
                \end{align*} and \begin{align*}
                    (\lbrace\bstheta_{m_{R}}^{(i)},w_{m_{R}}^{(i)}\rbrace_{i=1}^{N},A_{m_{R}}^{\dagger})=\text{\texttt{SMC\_sampler$\lr\bsY_{m_{R}},\lbrace\phi_{m_{R}}^{(i)},1\rbrace_{i=1}^{N},p_{\meld,m_{R}},q_{\meld,m_{R}}\rr$}}.
                \end{align*}

            \nonl\textnormal{(b)} Resample $(\lbrace\bstheta_{m_{L}}^{(i)},w_{m_{L}}^{(i)}\rbrace_{i=1}^{N},A_{m_{L}}^{\dagger})$ and $(\lbrace\bstheta_{m_{R}}^{(i)},w_{m_{R}}^{(i)}\rbrace_{i=1}^{N},A_{m_{R}}^{\dagger})$ to obtain equally weighted particle systems $\lbrace\bstheta_{m_{L}}^{(i)},1\rbrace_{i=1}^{N}$ and $\lbrace\bstheta_{m_{R}}^{(i)},1\rbrace_{i=1}^{N}$, and corresponding index collections $A_{m_{L}}^{*\dagger}$ and $A_{m_{R}}^{*\dagger}$.

            \nonl\textnormal{(c)} Update the particle index for $\phi_{m_{L}}$ and $\phi_{m_{R}}$ by calling \texttt{forward\_update()} in Algorithm \ref{algo:index_forward}, resulting in \nonl\begin{align*}
                &B_{m,L-1}^{\dagger}=\text{\texttt{forward\_update($A_{m_{L}}^{*\dagger},B_{m_{L-1}}^{\dagger}$)}},\ &B_{m,L+1}^{\dagger}=\text{\texttt{forward\_update($A_{m_{L}}^{*\dagger},B_{m_{L+1}}^{\dagger}$)}},\\
                &B_{m,R-1}^{\dagger}=\text{\texttt{forward\_update($A_{m_{L}}^{*\dagger},B_{m_{R-1}}^{\dagger}$)}},\ &B_{m,R+1}^{\dagger}=\text{\texttt{forward\_update($A_{m_{L}}^{*\dagger},B_{m_{R+1}}^{\dagger}$)}}.
            \end{align*}

            \nonl\textnormal{(d)} Update $\phi_{m_{L}+1,m_{L}+2}$ and $\phi_{m_{R}-2,m_{R}-1}$, by picking the indices in $B_{m_{L}+1}^{\dagger}$ and $B_{m_{R}-1}^{\dagger}$, respectively.
            }

        In stage $s=\frac{M+5}{4}$, \textbf{do}

            \nonl\textnormal{(a)} Merge $\lbrace\phi_{(M+1)/2-1,(M+1)/2}^{(i)},1\rbrace_{i=1}^{N}$ and $\lbrace\phi_{(M+1)/2,(M+1)/2+1}^{(i)},1\rbrace_{i=1}^{N}$ to get $\lbrace\phi_{(M+1)/2}^{(i)},1\rbrace_{i=1}^{N}$. Call the SMC sampler \texttt{SMC\_sampler()} in Algorithm \ref{algo:smc}, obtaining \nonl\begin{align*}
                &(\lbrace\bstheta_{(M+1)/2}^{(i)},w_{(M+1)/2}^{(i)}\rbrace_{i=1}^{N}, A_{(M+1)/2}^{\dagger})\\
                =&\ \text{\texttt{SMC\_sampler$\lr\bsY_{(M+1)/2},\lbrace\phi_{(M+1)/2}^{(i)},1\rbrace_{i=1}^{N},p_{\meld,(M+1)/2},q_{\meld,(M+1)/2}\rr$}}.
            \end{align*}
    
            \nonl\textnormal{(b)} Update particles for $\bstheta_{2},\bstheta_{4}\ldots,\allowbreak\bstheta_{(M+1)/2-2}$ by calling \texttt{back\_left\_update()} in Algorithm \ref{algo:index_leftward} as \begin{align*}
                &\lr\bstheta_{2},\bstheta_{4}\ldots,\bstheta_{(M+1)/2-2},A_{2}^{**\dagger},A_{4}^{**\dagger},\ldots,A_{(M+1)/2-2}^{**\dagger}\rr\\
                =&\ \text{\texttt{back\_left\_update$\lr A_{2}^{*\dagger},A_{4}^{*\dagger},\ldots, A_{(M+1)/2-2}^{*\dagger}, A_{(M+1)/2}^{\dagger},\bstheta_{1},\ldots,\bstheta_{(M+1)/2-2}\rr$}};
            \end{align*} and update $\psi_{1}$ by picking the indices in $A_{2}^{**\dagger}$ and update $\psi_{m}$ for $m=3,5,\ldots,(M+1)/2-1$ by picking the indices in $A_{m-1}^{**\dagger}$.
                
            \nonl\textnormal{(c)} Update particles for $\bstheta_{(M+1)/2+2},\bstheta_{(M+1)/2+4},\ldots,\allowbreak\bstheta_{M-1}$ by calling \texttt{back\_right\_update()} in Algorithm \ref{algo:index_rightward} as \begin{align*}
                &\lr\bstheta_{(M+1)/2+2},\bstheta_{(M+1)/2+4},\ldots,\bstheta_{M-1},A_{2}^{**\dagger},A_{(M+1)/2+2}^{**\dagger},A_{(M+1)/2+4}^{**\dagger},\ldots,A_{M-1}^{**\dagger}\rr\\
                =&\ \text{\texttt{back\_right\_update$\lr A_{M-1}^{*\dagger},A_{M-3}^{*\dagger},\ldots,A_{(M+1)/2+2}^{*\dagger},A_{(M+1)/2}^{\dagger},\bstheta_{M-1},\bstheta_{M-3},\ldots,\bstheta_{(M+1)/2+2}\rr$}};
            \end{align*} and update $\psi_{M}$ by picking the indices in $A_{M-1}^{**\dagger}$ and update $\psi_{m}$ for $m=(M+1)/2+1,(M+1)/2+3,\ldots,M-2$ by picking the indices in $A_{m+1}^{**\dagger}$.

        \KwOut{$\lbrace\bstheta_{m}^{(i)}\rbrace_{i=1}^{N}$ for $m=1,\ldots,M$.} 
\end{algorithm}}

{\renewcommand{\thealgocf}{6b}
\SetAlFnt{\footnotesize}
\begin{algorithm}[H]
    \caption{\texttt{d\&c-melding($M$)}: for odd $M$ satisfying $4|(M-1)$}
    \label{algo:odd_M_2}
        \KwIn{data $\bsY=(\bsY_{1},\ldots,\bsY_{M})$; the choice of $p_{\pool}(\bsphi)$; the subposteriors $p_{\meld,m}(\phi_{m},\psi_{m}\mid\bsY_{m})$ for $m=1,\ldots,M$; the proposals $q_{\meld,m}(\psi_{m}\mid\phi_{m})$ for $m=1,\ldots,M$; the number of particles $N$; the number of annealing steps $n_{t}$.}

    In stage one, in parallel for each $m=1,3,5,\ldots,M$, \textbf{do}

        \nonl\textnormal{(a)} Initialise $\phi_{m}^{(i,0)}$ for $\phi_{m}^{(i)}$ to obtain equally weighted particle systems $\lbrace\phi_{m}^{(i,0)},1\rbrace_{i=1}^{N}$ for $m=1,3,5,\ldots,M$.

        \nonl\textnormal{(b)} Call the SMC sampler \texttt{SMC\_sampler()} in Algorithm \ref{algo:smc}, obtaining {\nonl\begin{align*}
            (\lbrace\bstheta_{m}^{(i)},w_{m}^{(i)}\rbrace_{i=1}^{N},A_{m}^{\dagger})=\text{\texttt{SMC\_sampler$\lr\bsY_{m},\lbrace\phi_{m}^{(i,0)},1\rbrace_{i=1}^{N},p_{\meld,m},q_{\meld,m}\rr$}},
        \end{align*}} for $m=1,3,5,\ldots,M$.

        \nonl\textnormal{(c)} Resample $\lbrace\bstheta_{m}^{(i)},w_{m}^{(i)}\rbrace_{i=1}^{N}$ to obtain equally weighted particle systems $\lbrace\bstheta_{m}^{(i)},1\rbrace_{i=1}^{N}$ for $m=1,3,5,\ldots,N$.

        \nonl\textnormal{(d)} Record the indices for $\lbrace\bstheta_{m}^{(i)},1\rbrace_{i=1}^{N}$ in a collection $B_{m}^{\dagger}=\lbrace1,\ldots,N\rbrace$ for $m=1,3,5,\ldots,M$.

    In stage $s$, let $m_{L}=2s-2$ and $m_{R}=M+3-2s$. In parallel, for each $m=m_{L}$ and $m_{R}$, \For{$s=2,3,\ldots,\frac{M-1}{4}$}{

        \nonl\textnormal{(a)} Merge $\lbrace\phi_{m_{L}-1,m_{L}}^{(i)},1\rbrace_{i=1}^{N}$ and $\lbrace\phi_{m_{L},m_{L}+1}^{(i)},1\rbrace_{i=1}^{N}$ to get $\lbrace\phi_{m_{L}}^{(i)},1\rbrace_{i=1}^{N}$. Call the SMC sampler \texttt{SMC\_sampler()} in Algorithm \ref{algo:smc}, obtaining {\nonl\begin{align*}
            (\lbrace\bstheta_{m_{L}}^{(i)},w_{m_{L}}^{(i)}\rbrace_{i=1}^{N},A_{m_{L}}^{\dagger})=\text{\texttt{SMC\_sampler$\lr\bsY_{m_{L}},\lbrace\phi_{m_{L}}^{(i)},1\rbrace_{i=1}^{N},p_{\meld,m_{L}},q_{\meld,m_{L}}\rr$}}
        \end{align*}} and {\nonl\begin{align*}
            (\lbrace\bstheta_{m_{R}}^{(i)},w_{m_{R}}^{(i)}\rbrace_{i=1}^{N},A_{m_{R}}^{\dagger})=\text{\texttt{SMC\_sampler$\lr\bsY_{m_{R}},\lbrace\phi_{m_{R}}^{(i)},1\rbrace_{i=1}^{N},p_{\meld,m_{R}},q_{\meld,m_{R}}\rr$}}.
        \end{align*}}

        \nonl\textnormal{(b)} Resample $(\lbrace\bstheta_{m_{L}}^{(i)},w_{m_{L}}^{(i)}\rbrace_{i=1}^{N},A_{m_{L}}^{\dagger})$ and $(\lbrace\bstheta_{m_{R}}^{(i)},w_{m_{R}}^{(i)}\rbrace_{i=1}^{N},A_{m_{R}}^{\dagger})$ to obtain equally weighted particle systems $\lbrace\bstheta_{m_{L}}^{(i)},1\rbrace_{i=1}^{N}$ and $\lbrace\bstheta_{m_{R}}^{(i)},1\rbrace_{i=1}^{N}$, and corresponding index collections $A_{m_{L}}^{*\dagger}$ and $A_{m_{R}}^{*\dagger}$.

        \nonl\textnormal{(c)} Update the particle index for $\phi_{m_{L}}$ and $\phi_{m_{R}}$ by calling \texttt{forward\_update()} in Algorithm \ref{algo:index_forward}, resulting in {\nonl\begin{align*}
            &B_{m,L-1}^{\dagger}=\text{\texttt{forward\_update($A_{m_{L}}^{*\dagger},B_{m_{L-1}}^{\dagger}$)}},\ &B_{m,L+1}^{\dagger}=\text{\texttt{forward\_update($A_{m_{L}}^{*\dagger},B_{m_{L+1}}^{\dagger}$)}},\\
            &B_{m,R-1}^{\dagger}=\text{\texttt{forward\_update($A_{m_{L}}^{*\dagger},B_{m_{R-1}}^{\dagger}$)}},\ &B_{m,R+1}^{\dagger}=\text{\texttt{forward\_update($A_{m_{L}}^{*\dagger},B_{m_{R+1}}^{\dagger}$)}}.
        \end{align*}}

        \nonl\textnormal{(d)} Update $\phi_{m_{L}+1,m_{L}+2}$ and $\phi_{m_{R}-2,m_{R}-1}$, by picking the indices in $B_{m_{L}+1}^{\dagger}$ and $B_{m_{R}-1}^{\dagger}$, respectively.
    }

    In stage $s=\frac{M+3}{4}$, \textbf{do}

        \nonl\textnormal{(a)} Merge $\lbrace\phi_{(M-1)/2-1,(M-1)/2}^{(i)},1\rbrace_{i=1}^{N}$ and $\lbrace\phi_{(M-1)/2,(M-1)/2+1}^{(i)},1\rbrace_{i=1}^{N}$ to get $\lbrace\phi_{(M-1)/2}^{(i)},1\rbrace_{i=1}^{N}$. Call the SMC sampler \texttt{SMC\_sampler()} in Algorithm \ref{algo:smc}, obtaining {\nonl\begin{align*}
                &(\lbrace\bstheta_{(M-1)/2}^{(i)},w_{(M-1)/2}^{(i)}\rbrace_{i=1}^{N}, A_{(M-1)/2}^{\dagger})\\
                =&\ \text{\texttt{SMC\_sampler$\lr\bsY_{(M-1)/2},\lbrace\phi_{(M-1)/2}^{(i)},1\rbrace_{i=1}^{N},p_{\meld,(M-1)/2},q_{\meld,(M-1)/2}\rr$}}.
        \end{align*}}

        \nonl\textnormal{(b)} Resample $(\lbrace\bstheta_{(M-1)/2}^{(i)},w_{(M-1)/2}^{(i)}\rbrace_{i=1}^{N}, A_{(M-1)/2}^{\dagger})$ to obtain equally weighted particle system $\lbrace\bstheta_{(M-1)/2}^{(i)},1\rbrace_{i=1}^{N}$, and the corresponding index collection $A_{(M-1)/2}^{*\dagger}$.

        \nonl\textnormal{(c)} Update the particle index for $\phi_{(M-1)/2}$ by calling \texttt{forward\_update()} in Algorithm \ref{algo:index_forward}, resulting in {\nonl\begin{align*}
                B_{(M-1)/2-1}^{\dagger}=\text{\texttt{forward\_update($A_{(M-1)/2}^{*\dagger},B_{(M-1)/2-1}^{\dagger}$)}}
        \end{align*}} and {\nonl\begin{align*}
                B_{(M-1)/2+1}^{\dagger}=\text{\texttt{forward\_update($A_{(M-1)/2}^{*\dagger},B_{(M-1)/2+1}^{\dagger}$)}}.
        \end{align*}}

        \nonl\textnormal{(d)} Update $\phi_{(M+3)/4+1,(M+3)/4+2}$, by picking the indices in $B_{(M-1)/2+1}^{\dagger}$.

\end{algorithm}}

{\renewcommand{\thealgocf}{6b}
\SetAlFnt{\footnotesize}
\begin{algorithm}[H]
    \caption{\texttt{d\&c-melding($M$)}: for odd $M$ satisfying $4|(M-1)$ (continue)}

    \nonl\nlset{4} In stage $s=\frac{M+7}{4}$, \textbf{do}

        \nonl\textnormal{(a)} Merge $\lbrace\phi_{(M+1)/2,(M+1)/2+1}^{(i)},1\rbrace_{i=1}^{N}$ and $\lbrace\phi_{(M+1)/2+1,(M+1)/2+2}^{(i)},1\rbrace_{i=1}^{N}$ to get $\lbrace\phi_{(M+1)/2+1}^{(i)},1\rbrace_{i=1}^{N}$. Call the SMC sampler \texttt{SMC\_sampler()} in Algorithm \ref{algo:smc}, obtaining {\nonl\begin{align*}
            &(\lbrace\bstheta_{(M+1)/2+1}^{(i)},w_{(M+1)/2+1}^{(i)}\rbrace_{i=1}^{N}, A_{(M+1)/2+1})\\
            =&\ \text{\texttt{SMC\_sampler$\lr\bsY_{(M+1)/2+1},\lbrace\phi_{(M+1)/2+1}^{(i)},1\rbrace_{i=1}^{N},p_{\meld,(M+1)/2+1},q_{\meld,(M+1)/2+1}\rr$}}.
        \end{align*}}

        \nonl\textnormal{(b)} Update particles for $\bstheta_{2},\bstheta_{4}\ldots,\allowbreak\bstheta_{(M-1)/2}$ by calling \texttt{back\_left\_update()} in Algorithm \ref{algo:index_leftward} as \begin{align*}
            &\lr\bstheta_{2},\bstheta_{4}\ldots,\bstheta_{(M-1)/2},A_{2}^{**\dagger},A_{4}^{**\dagger},\ldots,A_{(M-1)/2}^{**\dagger}\rr\\
            =&\ \text{\texttt{back\_left\_update$\lr A_{2}^{*\dagger},A_{4}^{*\dagger},\ldots, A_{(M-1)/2}^{*\dagger}, A_{(M+1)/2+1}^{\dagger},\bstheta_{2},\bstheta_{4}\ldots,\bstheta_{(M-1)/2}\rr$}};
        \end{align*} and update $\psi_{1}$ by picking the indices in $A_{2}^{**\dagger}$ and update $\psi_{m}$ for $m=3,5,\ldots,(M+1)/2$ by picking the indices in $A_{m-1}^{**\dagger}$.
                
        \nonl\textnormal{(c)} Update particles for $\bstheta_{(M+3)/2+2},\bstheta_{(M+3)/2+4},\ldots,\allowbreak\bstheta_{M-1}$ by calling \texttt{back\_right\_update()} in Algorithm \ref{algo:index_rightward} as {\nonl\begin{align*}
            &\lr\bstheta_{(M+3)/2+2},\bstheta_{(M+3)/2+4},\ldots,\bstheta_{M-1},A_{(M+3)/2+2}^{**\dagger},A_{(M+3)/2+4}^{**\dagger},\ldots,A_{M-1}^{**\dagger}\rr\\
            =&\ \text{\texttt{back\_right\_update$\lr A_{(M+1)/2+1}^{\dagger},A_{(M+3)/2+2}^{*\dagger},A_{(M+3)/2+4}^{*\dagger},\ldots,A_{M-1}^{*\dagger},\bstheta_{(M+3)/2+2},\bstheta_{(M+3)/2+4},\ldots,\bstheta_{M-1}\rr$}};
        \end{align*}} and update $\psi_{M}$ by picking the indices in $A_{M-1}^{**\dagger}$ and update $\psi_{m}$ for $m\allowbreak=\allowbreak(M+1)/2+2,\allowbreak(M+1)/2+4,\allowbreak\ldots,\allowbreak M-2$ by picking the indices in $A_{m+1}^{**\dagger}$.

    \KwOut{$\lbrace\bstheta_{m}^{(i)}\rbrace_{i=1}^{N}$ for $m=1,\ldots,M$.}
\end{algorithm}}

{\renewcommand{\thealgocf}{7a}
\SetAlFnt{\footnotesize}
\begin{algorithm}[H]
    \caption{\texttt{d\&c-melding($M$)}: for even $M$ satisfying $4\mid M$}
    \label{algo:even_M_1}

    \KwIn{data $\bsY=(\bsY_{1},\ldots,\bsY_{M})$; the choice of $p_{\pool}(\bsphi)$; the subposteriors $p_{\meld,m}(\phi_{m},\psi_{m}\mid\bsY_{m})$ for $m=1,\ldots,M$; the proposals $q_{\meld,m}(\psi_{m}\mid\phi_{m})$ for $m=1,\ldots,M$; the number of particles $N$; the number of annealing steps $n_{t}$.}

    In stage one, in parallel for each $m=1,3,5,\ldots,M/2-1,M/2+2,M/2+4,\ldots,M-2,M$, \textbf{do}

        \nonl\textnormal{(a)} Initialise $\phi_{m}^{(i,0)}$ for $\phi_{m}^{(i)}$ to obtain equally weighted particle systems $\lbrace\phi_{m}^{(i,0)},1\rbrace_{i=1}^{N}$.

        \nonl\textnormal{(b)} Call the SMC sampler \texttt{SMC\_sampler()} in Algorithm \ref{algo:smc}, obtaining {\tiny\begin{align*}
            (\lbrace\bstheta_{m}^{(i)},w_{m}^{(i)}\rbrace_{i=1}^{N},A_{m}^{\dagger})=\text{\texttt{SMC\_sampler$\lr\bsY_{m},\lbrace\phi_{m}^{(i,0)},1\rbrace_{i=1}^{N},p_{\meld,m},q_{\meld,m}\rr$}}.
        \end{align*}}

        \nonl\textnormal{(c)} Resample $\lbrace\bstheta_{m}^{(i)},w_{m}^{(i)}\rbrace_{i=1}^{N}$ to obtain equally weighted particle systems $\lbrace\bstheta_{m}^{(i)},1\rbrace_{i=1}^{N}$.

        \nonl\textnormal{(d)} Record the indices for $\lbrace\bstheta_{m}^{(i)},1\rbrace_{i=1}^{N}$ in a collection.

    In stage $s$, let $m_{L}=2s-2$ and $m_{R}=M+3-2s$. In parallel for each $m=m_{L}$ and $m_{R}$, \For{$s=2,3,\ldots,\frac{M}{4}$}{ 

        \nonl\textnormal{(a)} Merge $\lbrace\phi_{m_{L}-1,m_{L}}^{(i)},1\rbrace_{i=1}^{N}$ and $\lbrace\phi_{m_{L},m_{L}+1}^{(i)},1\rbrace_{i=1}^{N}$ to get $\lbrace\phi_{m_{L}}^{(i)},1\rbrace_{i=1}^{N}$. Call the SMC sampler \texttt{SMC\_sampler()} in Algorithm \ref{algo:smc}, obtaining {\tiny\begin{align*}
            (\lbrace\bstheta_{m_{L}}^{(i)},w_{m_{L}}^{(i)}\rbrace_{i=1}^{N},A_{m_{L}}^{\dagger})=\text{\texttt{SMC\_sampler$\lr\bsY_{m_{L}},\lbrace\phi_{m_{L}}^{(i)},1\rbrace_{i=1}^{N},p_{\meld,m_{L}},q_{\meld,m_{L}}\rr$}}
        \end{align*}} and {\tiny\begin{align*}
            (\lbrace\bstheta_{m_{R}}^{(i)},w_{m_{R}}^{(i)}\rbrace_{i=1}^{N},A_{m_{R}}^{\dagger})=\text{\texttt{SMC\_sampler$\lr\bsY_{m_{R}},\lbrace\phi_{m_{R}}^{(i)},1\rbrace_{i=1}^{N},p_{\meld,m_{R}},q_{\meld,m_{R}}\rr$}}.
        \end{align*}}

        \nonl\textnormal{(b)} Resample $(\lbrace\bstheta_{m_{L}}^{(i)},w_{m_{L}}^{(i)}\rbrace_{i=1}^{N},A_{m_{L}}^{\dagger})$ and $(\lbrace\bstheta_{m_{R}}^{(i)},w_{m_{R}}^{(i)}\rbrace_{i=1}^{N},A_{m_{R}}^{\dagger})$ to obtain equally weighted particle systems $\lbrace\bstheta_{m_{L}}^{(i)},1\rbrace_{i=1}^{N}$ and $\lbrace\bstheta_{m_{R}}^{(i)},1\rbrace_{i=1}^{N}$, and corresponding index collections $A_{m_{L}}^{*\dagger}$ and $A_{m_{R}}^{*\dagger}$.

        \nonl\textnormal{(c)} Update the particle index for $\phi_{m_{L}}$ and $\phi_{m_{R}}$ by calling \texttt{forward\_update()} in Algorithm \ref{algo:index_forward}, resulting in {\nonl\tiny\begin{align*}
            &B_{m,L-1}^{\dagger}=\text{\texttt{forward\_update($A_{m_{L}}^{*\dagger},B_{m_{L-1}}^{\dagger}$)}},\ &B_{m,L+1}^{\dagger}=\text{\texttt{forward\_update($A_{m_{L}}^{*\dagger},B_{m_{L+1}}^{\dagger}$)}},\\
            &B_{m,R-1}^{\dagger}=\text{\texttt{forward\_update($A_{m_{L}}^{*\dagger},B_{m_{R-1}}^{\dagger}$)}},\ &B_{m,R+1}^{\dagger}=\text{\texttt{forward\_update($A_{m_{L}}^{*\dagger},B_{m_{R+1}}^{\dagger}$)}}.
        \end{align*}}

        \nonl\textnormal{(d)} Update $\phi_{m_{L}+1,m_{L}+2}$ and $\phi_{m_{R}-2,m_{R}-1}$, by picking the indices in $B_{m_{L}+1}^{\dagger}$ and $B_{m_{R}-1}^{\dagger}$, respectively.
    }

    In stage $s=\frac{M+4}{4}$, \textbf{do}

        \nonl\textnormal{(a)} Merge $\lbrace\phi_{M/2-1,M/2}^{(i)},1\rbrace_{i=1}^{N}$ and $\lbrace\phi_{(M+2)/2,(M+2)/2+1}^{(i)},1\rbrace_{i=1}^{N}$ to get $\lbrace(\phi_{M/2-1,M/2}^{(i)},\allowbreak\phi_{(M+2)/2,(M+2)/2+1}^{(i)}),\allowbreak 1\rbrace_{i=1}^{N}$. Call the SMC sampler \texttt{SMC\_sampler()} in Algorithm \ref{algo:smc}, obtaining {\nonl\tiny\begin{align*}
            &(\lbrace\bstheta_{M/2,(M+2)/2}^{(i)}, w_{M/2,(M+2)/2}^{(i)}\rbrace_{i=1}^{N}, A_{M/2,(M+2)/2}^{\dagger})\\
            &\qquad =\text{\texttt{SMC\_sampler$\bigg((\bsY_{M/2},\bsY_{(M+2)/2}),\lbrace(\phi_{M/2-1,M/2}^{(i)},\phi_{(M+2)/2,(M+2)/2+1}^{(i)}),1\rbrace_{i=1}^{N},$}}\\
            &\qquad\qquad\qquad\qquad\qquad\qquad\qquad\qquad\text{$p_{\meld,M/2,(M+2)/2},q_{\meld,M/2},q_{\meld,(M+2)/2}\bigg)$}.
        \end{align*}}
    
        \nonl\textnormal{(b)} Update $\bstheta_{2},\bstheta_{4},\ldots,\allowbreak\bstheta_{M/2-2}$ by calling \texttt{back\_left\_update()} in Algorithm \ref{algo:index_leftward} as {\nonl\tiny\begin{align*}
            &\lr\bstheta_{2},\bstheta_{4}\ldots,\bstheta_{M/2-2},A_{2}^{**\dagger},A_{4}^{**\dagger},\ldots,A_{M/2-2}^{**\dagger}\rr\\
            &\quad=\text{\texttt{back\_left\_update$\lr A_{2}^{*\dagger},A_{4}^{*\dagger},\ldots, A_{M/2-2}^{*\dagger}, A_{M/2,(M+2)/2}^{\dagger},\bstheta_{2},\bstheta_{4},\ldots,\bstheta_{M/2-2}\rr$}};
        \end{align*}} and update $\psi_{1}$ by picking the indices in $A_{2}^{**\dagger}$ and update $\psi_{m}$ for $m=3,5,\ldots,M/2-1$ by picking the indices in $A_{m-1}^{**\dagger}$. 
        
        \nonl\textnormal{(c)} Update $\bstheta_{(M+2)/2+2},\bstheta_{(M+2)/2+4},\ldots,\allowbreak\bstheta_{M-1}$ by calling \texttt{back\_right\_update()} in Algorithm \ref{algo:index_rightward} as {\nonl\tiny\begin{align*}
            &\lr \bstheta_{(M+2)/2+2},\bstheta_{(M+2)/2+4},\ldots,\bstheta_{M-1},A_{(M+2)/2+2}^{**\dagger},A_{(M+2)/2+4}^{**\dagger},\ldots,A_{M-1}^{**\dagger}\rr\\
            =&\ \text{\texttt{back\_right\_update$\lr A_{M/2,(M+2)/2}^{\dagger}, A_{(M+2)/2+2}^{*\dagger},A_{(M+2)/2+4}^{*\dagger},\ldots, A_{M-1}^{*\dagger},\bstheta_{(M+2)/2+2},\bstheta_{(M+2)/2+4},\ldots,\bstheta_{M-1}\rr$}};
        \end{align*}} and update $\psi_{M}$ by picking the indices in $A_{M-1}^{**\dagger}$ and update $\psi_{m}$ for $m=(M+2)/2+1,\allowbreak (M+2)/2+3,\allowbreak\ldots,\allowbreak M-1$ by picking the indices in $A_{m+1}^{**\dagger}$.

    \KwOut{$\lbrace\bstheta_{m}^{(i)}\rbrace_{i=1}^{N}$ for $m=1,\ldots,M$.}

\end{algorithm}}

{\renewcommand{\thealgocf}{7b}
\SetAlFnt{\footnotesize}
\begin{algorithm}[H]
    \caption{\texttt{d\&c-melding($M$)}: for even $M$ satisfying $4\nmid M$}
    \label{algo:even_M_2}

    \KwIn{data $\bsY=(\bsY_{1},\ldots,\bsY_{M})$; the choice of $p_{\pool}(\bsphi)$; the subposteriors $p_{\meld,m}(\phi_{m},\psi_{m}\mid\bsY_{m})$ for $m=1,\ldots,M$; the proposals $q_{\meld,m}(\psi_{m}\mid\phi_{m})$ for $m=1,\ldots,M$; the number of particles $N$; the number of annealing steps $n_{t}$.}

    In stage one, in parallel for each $m=1,3,5,\ldots,M/2,M/2+3,M/2+5,\ldots,M-2,M$, \textbf{do}

        \nonl\textnormal{(a)} Initialise $\phi_{m}^{(i,0)}$ for $\phi_{m}^{(i)}$ to obtain equally weighted particle systems $\lbrace\phi_{m}^{(i,0)},1\rbrace_{i=1}^{N}$.

        \nonl\textnormal{(b)} Call the SMC sampler \texttt{SMC\_sampler()} in Algorithm \ref{algo:smc}, obtaining {\nonl\begin{align*}
            (\lbrace\bstheta_{m}^{(i)},w_{m}^{(i)}\rbrace_{i=1}^{N},A_{m}^{\dagger})=\text{\texttt{SMC\_sampler$\lr\bsY_{m},\lbrace\phi_{m}^{(i,0)},1\rbrace_{i=1}^{N},p_{\meld,m},q_{\meld,m}\rr$}}.
        \end{align*}}

        \nonl\textnormal{(c)} Resample $\lbrace\bstheta_{m}^{(i)},w_{m}^{(i)}\rbrace_{i=1}^{N}$ to obtain equally weighted particle systems $\lbrace\bstheta_{m}^{(i)},1\rbrace_{i=1}^{N}$.

        \nonl\textnormal{(d)} Record the indices for $\lbrace\bstheta_{m}^{(i)},1\rbrace_{i=1}^{N}$ in a collection.

    In stage $s$, let $m_{L}=2s-2$ and $m_{R}=M+3-2s$. In parallel for each $m=m_{L}$ and $m_{R}$, \For{$s=2,3,\ldots,\frac{M-2}{4}$}{ 

        \nonl\textnormal{(a)} Merge $\lbrace\phi_{m_{L}-1,m_{L}}^{(i)},1\rbrace_{i=1}^{N}$ and $\lbrace\phi_{m_{L},m_{L}+1}^{(i)},1\rbrace_{i=1}^{N}$ to get $\lbrace\phi_{m_{L}}^{(i)},1\rbrace_{i=1}^{N}$. Call the SMC sampler \texttt{SMC\_sampler()} in Algorithm \ref{algo:smc}, obtaining {\nonl\begin{align*}
            (\lbrace\bstheta_{m_{L}}^{(i)},w_{m_{L}}^{(i)}\rbrace_{i=1}^{N},A_{m_{L}}^{\dagger})=\text{\texttt{SMC\_sampler$\lr\bsY_{m_{L}},\lbrace\phi_{m_{L}}^{(i)},1\rbrace_{i=1}^{N},p_{\meld,m_{L}},q_{\meld,m_{L}}\rr$}}
        \end{align*}} and {\nonl\begin{align*}
            (\lbrace\bstheta_{m_{R}}^{(i)},w_{m_{R}}^{(i)}\rbrace_{i=1}^{N},A_{m_{R}}^{\dagger})=\text{\texttt{SMC\_sampler$\lr\bsY_{m_{R}},\lbrace\phi_{m_{R}}^{(i)},1\rbrace_{i=1}^{N},p_{\meld,m_{R}},q_{\meld,m_{R}}\rr$}}.
        \end{align*}}

        \nonl\textnormal{(b)} Resample $(\lbrace\bstheta_{m_{L}}^{(i)},w_{m_{L}}^{(i)}\rbrace_{i=1}^{N},A_{m_{L}}^{\dagger})$ and $(\lbrace\bstheta_{m_{R}}^{(i)},w_{m_{R}}^{(i)}\rbrace_{i=1}^{N},A_{m_{R}}^{\dagger})$ to obtain equally weighted particle systems $\lbrace\bstheta_{m_{L}}^{(i)},1\rbrace_{i=1}^{N}$ and $\lbrace\bstheta_{m_{R}}^{(i)},1\rbrace_{i=1}^{N}$, and corresponding index collections $A_{m_{L}}^{*\dagger}$ and $A_{m_{R}}^{*\dagger}$.

        \nonl\textnormal{(c)} Update the particle index for $\phi_{m_{L}}$ and $\phi_{m_{R}}$ by calling \texttt{forward\_update()} in Algorithm \ref{algo:index_forward}, resulting in {\nonl\begin{align*}
            &B_{m,L-1}^{\dagger}=\text{\texttt{forward\_update($A_{m_{L}}^{*\dagger},B_{m_{L-1}}^{\dagger}$)}},\ &B_{m,L+1}^{\dagger}=\text{\texttt{forward\_update($A_{m_{L}}^{*\dagger},B_{m_{L+1}}^{\dagger}$)}},\\
            &B_{m,R-1}^{\dagger}=\text{\texttt{forward\_update($A_{m_{L}}^{*\dagger},B_{m_{R-1}}^{\dagger}$)}},\ &B_{m,R+1}^{\dagger}=\text{\texttt{forward\_update($A_{m_{L}}^{*\dagger},B_{m_{R+1}}^{\dagger}$)}}.
            \end{align*}}

        \nonl\textnormal{(d)} Update $\phi_{m_{L}+1,m_{L}+2}$ and $\phi_{m_{R}-2,m_{R}-1}$, by picking the indices in $B_{m_{L}+1}^{\dagger}$ and $B_{m_{R}-1}^{\dagger}$, respectively.
    }

    In stage $(M+2)/4$, \textbf{do}
    
        \nonl\textnormal{(a)} Merge $\lbrace\phi_{M/2-2,M/2-1}^{(i)},1\rbrace_{i=1}^{N}$ and $\lbrace\phi_{M/2-1,M/2}^{(i)},1\rbrace_{i=1}^{N}$ to get $\lbrace\phi_{M/2-1}^{(i)},1\rbrace_{i=1}^{N}$. Call the SMC sampler \texttt{SMC\_sampler()} in Algorithm \ref{algo:smc}, obtaining {\nonl\begin{align*}
            &(\lbrace\bstheta_{M/2-1}^{(i)},w_{M/2-1}^{(i)}\rbrace_{i=1}^{N}, A_{M/2-1})\\
            &\qquad=\text{\texttt{SMC\_sampler$\lr\bsY_{M/2-1},\lbrace\phi_{M/2-1}^{(i)},1\rbrace_{i=1}^{N},p_{\meld,M/2-1},q_{\meld,M/2-1}\rr$}}.
        \end{align*}}

        \nonl\textnormal{(b)} Resample $(\lbrace\bstheta_{M/2-1}^{(i)},w_{M/2-1}^{(i)}\rbrace_{i=1}^{N}, A_{M/2-1}^{\dagger})$ to obtain equally weighted particle system $\lbrace\bstheta_{M/2-1}^{(i)},1\rbrace_{i=1}^{N}$, and the corresponding index collection $A_{M/2-1}^{*\dagger}$.

        \nonl\textnormal{(c)} Update the particle index for $\phi_{M/2-1}$ by calling \texttt{forward\_update()} in Algorithm \ref{algo:index_forward}, resulting in {\nonl\begin{align*}
                B_{M/2-2}^{\dagger}=\text{\texttt{forward\_update($A_{M/2-1}^{*\dagger},B_{M/2-2}^{\dagger}$)}}
        \end{align*}} and {\nonl\begin{align*}
                B_{M/2}^{\dagger}=\text{\texttt{forward\_update($A_{M/2-1}^{*\dagger},B_{M/2}^{\dagger}$)}}.
        \end{align*}}

        \nonl\textnormal{(d)} Update $\phi_{M/2,M/2+1}$, by picking the indices in $B_{M/2}^{\dagger}$.

\end{algorithm}}

{\renewcommand{\thealgocf}{7b}
\SetAlFnt{\footnotesize}
\begin{algorithm}[H]
    \caption{\texttt{d\&c-melding($M$)}: for even $M$ satisfying $4\nmid M$ (continue)}

    \nonl\nlset{4} In stage $s=\frac{M+6}{4}$, \textbf{do}

        \nonl\textnormal{(a)} Merge $\lbrace\phi_{M/2,M/2+1}^{(i)},1\rbrace_{i=1}^{N}$ and $\lbrace\phi_{(M+2)/2+1,(M+2)/2+2}^{(i)},1\rbrace_{i=1}^{N}$ to get $\lbrace(\phi_{M/2,M/2+1}^{(i)},\allowbreak\phi_{(M+2)/2+1,(M+2)/2+2}^{(i)}),\allowbreak 1\rbrace_{i=1}^{N}$. Call the SMC sampler \texttt{SMC\_sampler()} in Algorithm \ref{algo:smc}, obtaining {\nonl\begin{align*}
            &(\lbrace\bstheta_{M/2+1,M/2+2}^{(i)}, w_{M/2+1,M/2+2}^{(i)}\rbrace_{i=1}^{N}, A_{M/2+1,M/2+2}^{\dagger})\\
            &\qquad =\text{\texttt{SMC\_sampler$\bigg((\bsY_{M/2+1},\bsY_{M/2+2}),\lbrace(\phi_{M/2,M/2+1}^{(i)},\phi_{(M+2)/2+1,(M+2)/2+2}^{(i)}),1\rbrace_{i=1}^{N},$}}\\
            &\qquad\qquad\qquad\qquad\qquad\qquad\qquad\qquad\text{$p_{\meld,M/2+1,M/2+2},q_{\meld,M/2+1},q_{\meld,M/2+2}\bigg)$}.
        \end{align*}}
    
        \nonl\textnormal{(b)} Update $\bstheta_{2},\bstheta_{4},\ldots,\allowbreak\bstheta_{M/2-1}$ by calling \texttt{back\_left\_update()} in Algorithm \ref{algo:index_leftward} as {\nonl\begin{align*}
            &\lr\bstheta_{2},\bstheta_{4},\ldots,\bstheta_{M/2-1},A_{2}^{**\dagger},A_{4}^{**\dagger},\ldots,A_{M/2-1}^{**\dagger}\rr\\         &\quad=\text{\texttt{back\_left\_update$\lr A_{2}^{*\dagger},A_{4}^{*\dagger},\ldots, A_{M/2-1}^{*\dagger}, A_{M/2+1,M/2+2}^{\dagger},\bstheta_{2},\bstheta_{4},\ldots,\bstheta_{M/2-1}\rr$}};
        \end{align*}} and update $\psi_{1}$ by picking the indices in $A_{2}^{**\dagger}$ and update $\psi_{m}$ for $m=3,5,\ldots,M/2$ by picking the indices in $A_{m-1}^{**\dagger}$.  
        
        \nonl\textnormal{(c)} Update $\bstheta_{(M+2)/2+3},\bstheta_{(M+2)/2+5},\ldots,\allowbreak\bstheta_{M-1}$ by calling \texttt{back\_right\_update()} in Algorithm \ref{algo:index_rightward} as {\nonl\begin{align*}
            &\lr\bstheta_{(M+2)/2+3},\bstheta_{(M+2)/2+5},\ldots,\bstheta_{M-1},A_{(M+2)/2+3}^{**\dagger},A_{(M+2)/2+5}^{**\dagger},\ldots,A_{M-1}^{**\dagger}\rr\\
            =&\ \text{\texttt{back\_right\_update$\lr A_{M/2+1,M/2+2}^{\dagger},A_{(M+2)/2+3}^{\dagger},A_{(M+2)/2+5}^{\dagger},\ldots, A_{M-1}^{*\dagger},\bstheta_{(M+2)/2+3},\bstheta_{(M+2)/2+5},\ldots,\allowbreak\bstheta_{M-1}\rr$}};
        \end{align*}} and update $\psi_{M}$ by picking the indices in $A_{M-1}^{**\dagger}$ and update $\psi_{m}$ for $m=(M+2)/2+2,(M+2)/2+4,\ldots,M-2$ by picking the indices in $A_{m+1}^{**\dagger}$.

    \KwOut{$\lbrace\bstheta_{m}^{(i)}\rbrace_{i=1}^{N}$ for $m=1,\ldots,M$.} 
    
\end{algorithm}}

\section{More simulation results}

\begin{table}[!htbp]
    \scriptsize
    \noindent\makebox[\linewidth][c]{%
    \begin{tabular}{ccccc}
    \hline
    
    \multicolumn{2}{c}{} & \multicolumn{2}{c}{$\phi_{4,5}$} & \multicolumn{1}{c}{}\\
    \cline{2-5}
    & Stage one & D\&C-melding & MCMC & D\&C-melding $\smcsq$\\
    MSE & 0.411 & 0.376 & 0.202 & 0.270\\
    Coverage & 0.858 & 0.896 & 0.896 & 0.914\\
    CI width & 1.315 & 1.300 & 0.983 & 1.145\\
    \cline{2-5}
    
    \multicolumn{2}{c}{} & \multicolumn{2}{c}{$\phi_{7,8}$} & \multicolumn{1}{c}{}\\
    \cline{2-5}
    MSE  & 0.110 & 0.110 & 0.094 & 0.111\\
    Coverage & 0.870 & 0.874 & 0.866 & 0.878\\
    CI width & 0.750 & 0.752 & 0.692 & 0.759\\
    \cline{2-5}
    
    \multicolumn{2}{c}{} & \multicolumn{2}{c}{$\phi_{3,4}$} & \multicolumn{1}{c}{}\\
    \cline{2-5}
    MSE & 0.011 & 0.011 & 0.011 & 0.011\\
    Coverage & 0.894 & 0.876 & 0.892 & 0.880\\
    CI width & 0.237 & 0.238 & 0.241 & 0.237\\
    \cline{2-5}
    
    \multicolumn{2}{c}{} & \multicolumn{2}{c}{$\phi_{2,3}$} & \multicolumn{1}{c}{}\\
    \cline{2-5}
    MSE & 2.741 & 2.211 & 1.962 & 2.206\\
    Coverage & 0.850 & 0.878 & 0.868 & 0.872\\
    CI width & 3.191 & 2.837 & 2.526 & 2.839\\
    \cline{2-5}
    
    \multicolumn{2}{c}{} & \multicolumn{2}{c}{$\phi_{9,10}$} & \multicolumn{1}{c}{}\\
    \cline{2-5}
    MSE & 0.020 & 0.015 & 0.011 & 0.015\\
    Coverage & 0.868 & 0.896 & 0.868 & 0.888\\
    CI width & 0.309 & 0.287 & 0.235 & 0.286\\
    \cline{2-5}
    
    \multicolumn{2}{c}{} & \multicolumn{2}{c}{$\phi_{1,2}$} & \multicolumn{1}{c}{}\\
    \cline{2-5}
    MSE & 0.027 & 0.006 & 0.005 & 0.006\\
    Coverage & 0.890 & 0.916 & 0.884 & 0.902\\
    CI width & 0.238 & 0.154 & 0.129 & 0.153\\
    \cline{2-5}
    
    \multicolumn{2}{c}{} & \multicolumn{2}{c}{$\phi_{10,11}$} & \multicolumn{1}{c}{}\\
    \cline{2-5}
    MSE & 0.085 & 0.080 & 0.075 & 0.080\\
    Coverage & 0.908 & 0.928 & 0.902 & 0.924\\
    CI width & 0.611 & 0.603 & 0.573 & 0.602\\
    
    \hline
    \end{tabular}}
    \caption{Average MSE, empirical coverage and average width of $90\%$ credible intervals for $\phi_{4,5}$, $\phi_{7,8}$, $\phi_{3,4}$, $\phi_{8,9}$, $\phi_{2,3}$, $\phi_{9,10}$, $\phi_{1,2}$ and $\phi_{10,11}$, obtained by the D\&C-melding approach, its combination with $\smcsq$ and the full MCMC. Measurements from stage one are also provided.}
    \label{tab:S1}
\end{table}

\begin{figure}[tb]
    \centering
    \includegraphics[width=1\textwidth, height=.42\textheight]{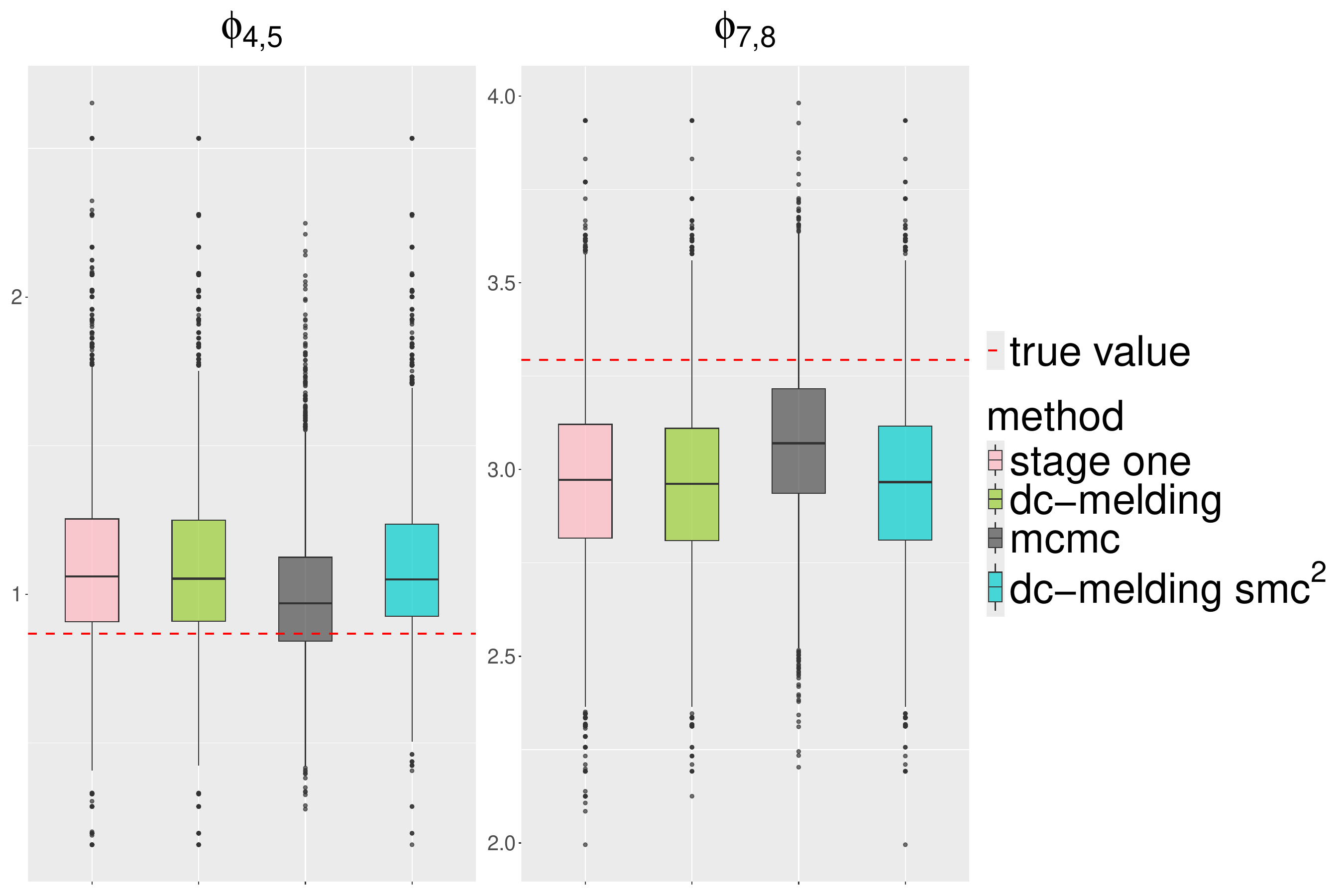}
    \caption{The estimates of $\phi_{3,4}$ and $\phi_{8,9}$ from same replicate as the one in Figure \ref{Fig_phi45_78}.}
    \label{Fig_phi45_78}
\end{figure}

\begin{figure}[tb]
    \centering
    \includegraphics[width=1\textwidth, height=.42\textheight]{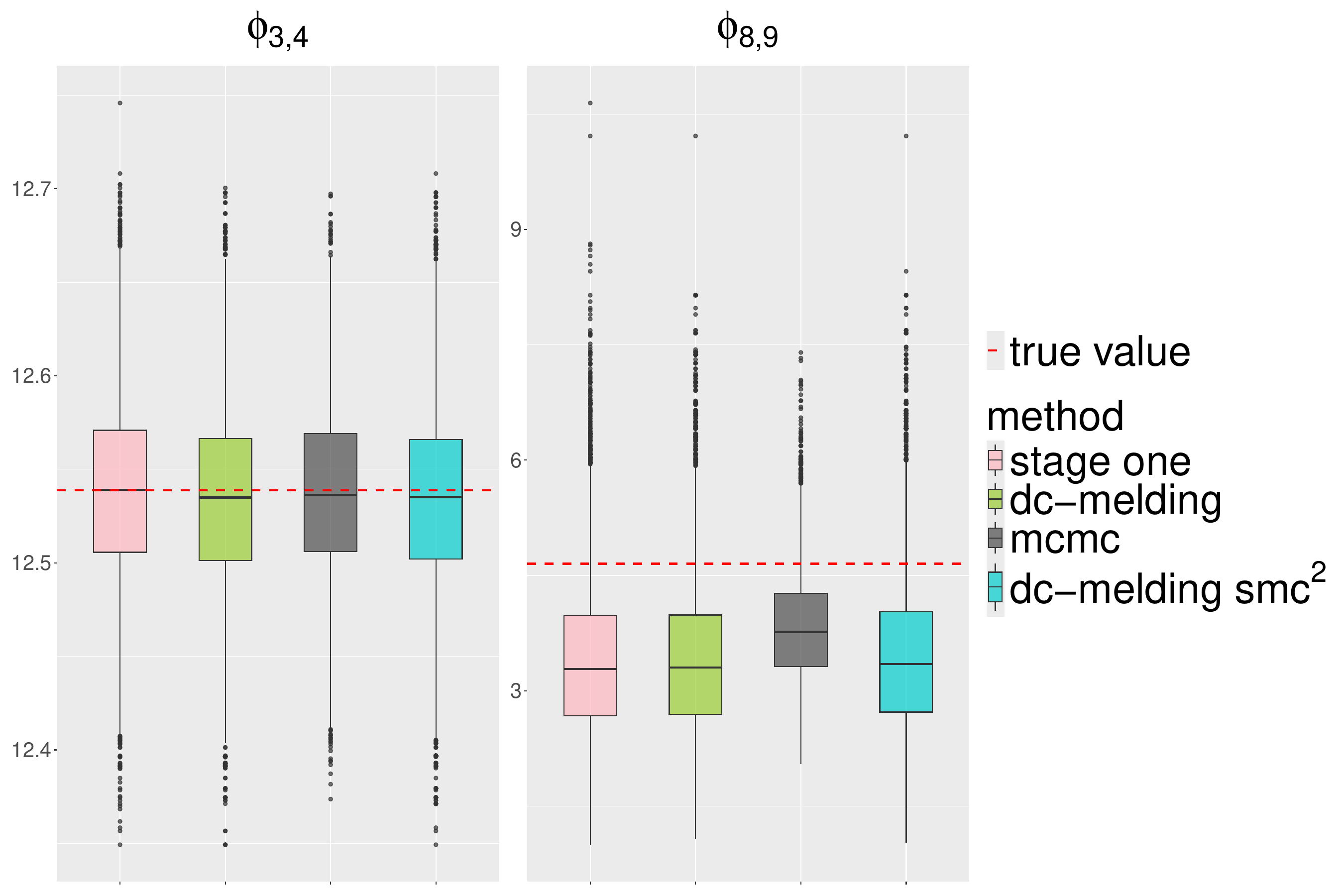}
    \caption{The estimates of $\phi_{3,4}$ and $\phi_{8,9}$ from same replicate as the one in Figure \ref{Fig_phi45_78}.}
    \label{Fig_phi34_89}
\end{figure}

\begin{figure}[tb]
    \centering
    \includegraphics[width=1\textwidth, height=.42\textheight]{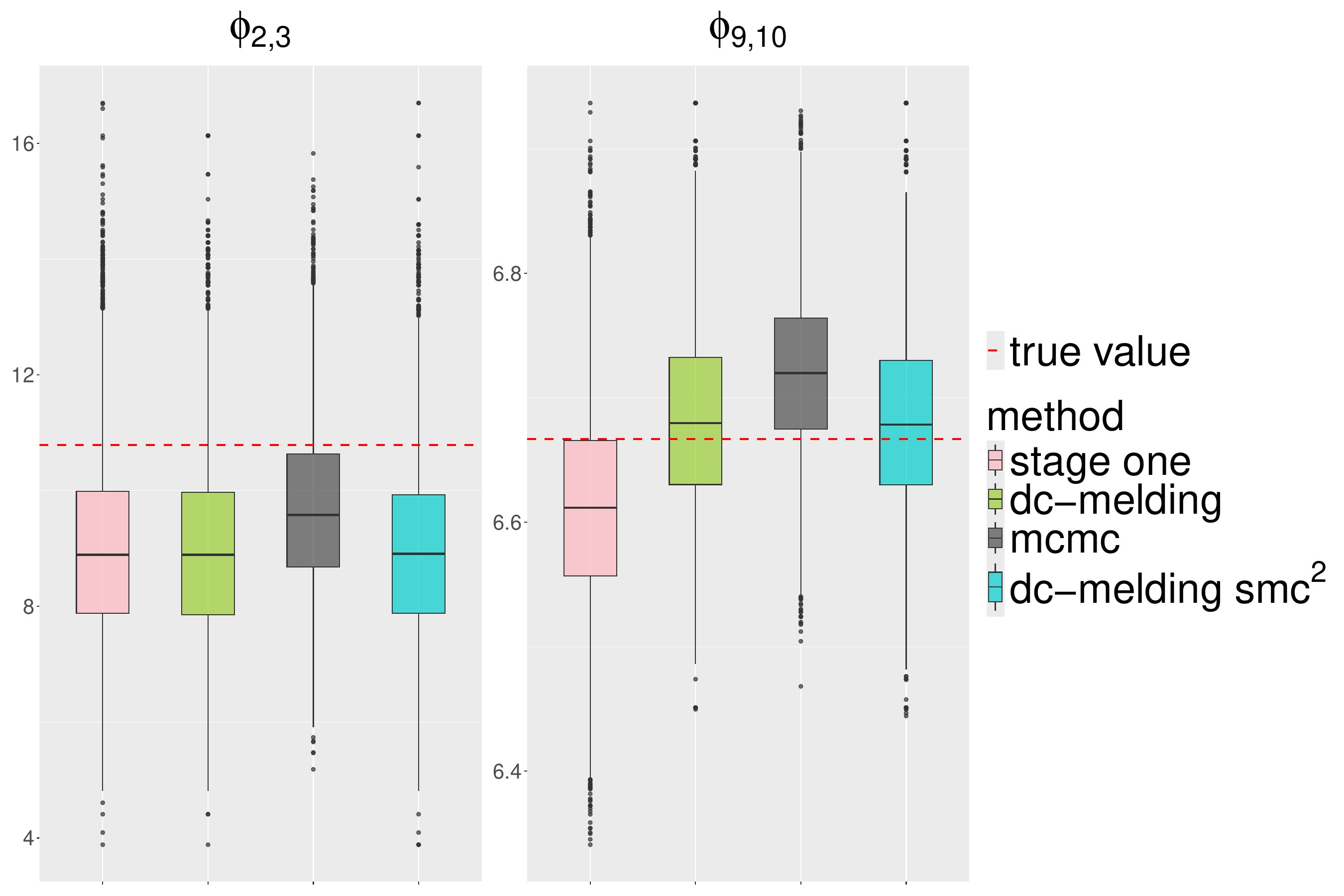}
    \caption{The estimates of $\phi_{2,3}$ and $\phi_{9,10}$ from the same realisation as the one in Figure \ref{Fig_phi45_78}.}
    \label{Fig_phi23_9_10}
\end{figure}

\begin{figure}[tb]
    \centering
    \includegraphics[width=1\textwidth, height=.42\textheight]{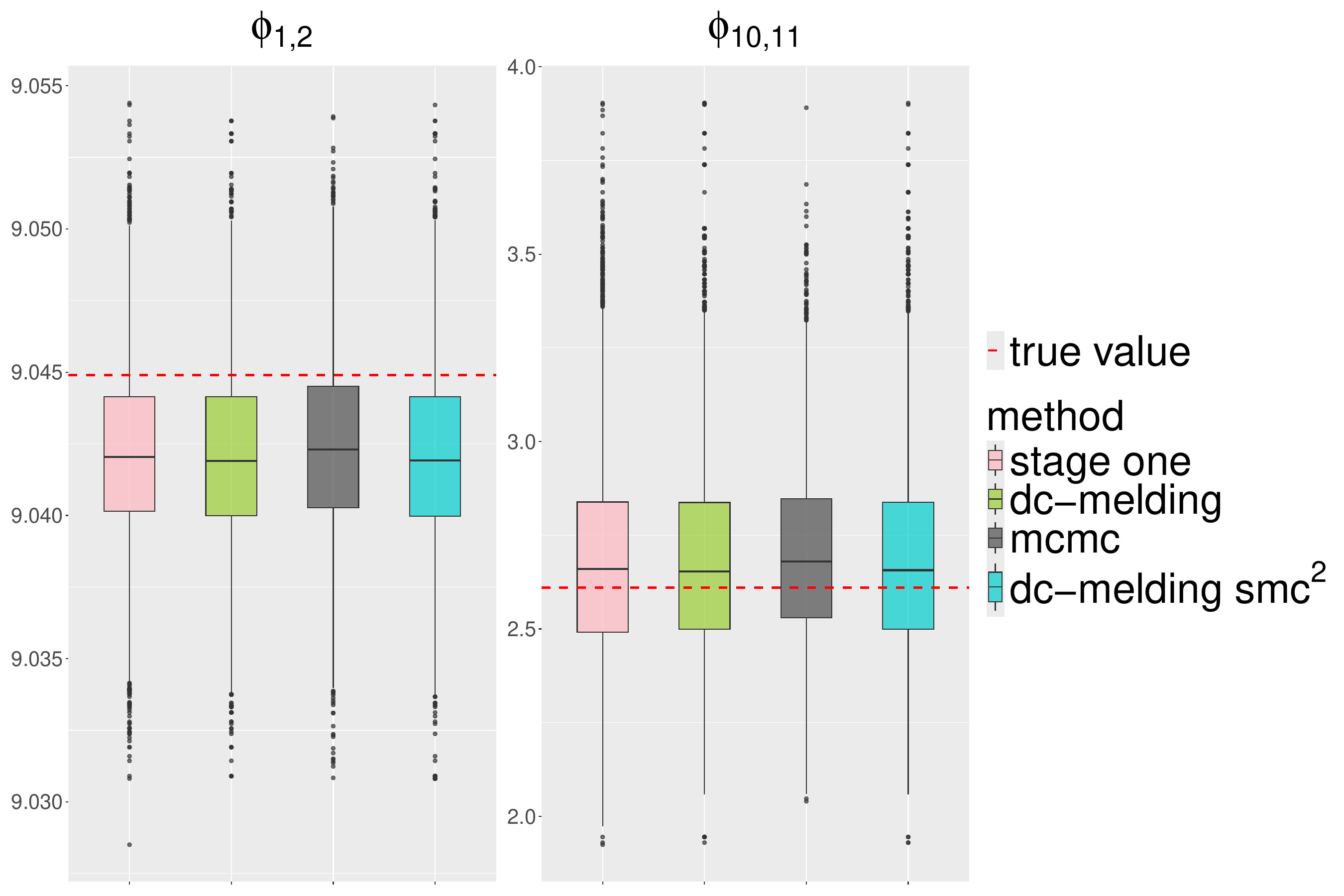}
    \caption{The estimates of $\phi_{1,2}$ and $\phi_{10,11}$ from the same realisation as the one in Figure \ref{Fig_phi45_78}.}
    \label{Fig_phi12_10_11}
\end{figure}

Table \ref{tab:S1} provides the results for the remaining common parameters. Figures \ref{Fig_phi45_78}--\ref{Fig_phi12_10_11} show the estimates for all parameters other than $\phi_{5,6}$ and $\phi_{6,7}$, based on the same replicate considered in the main manuscript. For these parameters, both D\&C-melding samplers exhibit performance comparable to that of the gold-standard MCMC sampler. 

\end{document}